\documentclass[12pt,a4paper]{article}
\pdfoutput=1
\usepackage{color}
\usepackage{amssymb,amsmath,bm,bbold}
\usepackage{epsf}
\usepackage{epsfig}
\usepackage{relsize}
\usepackage[dvipsnames]{xcolor}
\usepackage[linktoc=page,bookmarks=false,colorlinks=false,linkbordercolor=RoyalBlue,citebordercolor=ForestGreen,urlbordercolor=CornflowerBlue]{hyperref}
\usepackage{latexsym,mathrsfs,dsfont}
\usepackage[normalem]{ulem} 
\usepackage[compress]{cite}
\usepackage{graphicx}
\usepackage{slashed}
\usepackage{url}
\usepackage{booktabs}
\usepackage{float}
\usepackage{array}
\usepackage{multirow}
\usepackage{changepage}
\usepackage[hypcap]{caption, subcaption}

\numberwithin{equation}{section}
\usepackage{amsmath}
\usepackage{mathtools}
\usepackage{comment}
\usepackage[dvipsnames]{xcolor}

\usepackage{tabularx,colortbl}

\usepackage{fancyhdr}
\advance \headheight by 3.0truept       

\allowdisplaybreaks[1]

\definecolor{red}{cmyk}{0,1,1,0.4}
\definecolor{darkgreen}{rgb}{0.0,0.6,0.0}
\definecolor{cDarkGrey}{RGB}{91,91,91}
\definecolor{cGrey}{RGB}{245,243,238}
\definecolor{cBlue}{RGB}{0,110,191}
\definecolor{cLightBlue}{RGB}{214,237,252}
\definecolor{cRed}{RGB}{196,0,100}
\definecolor{cLightRed}{RGB}{254,222,237}
\definecolor{cGreen}{RGB}{0,166,80}
\definecolor{cLightGreen}{RGB}{254,222,237}
\definecolor{cOrange}{RGB}{221,74,44}
\definecolor{cLightOrange}{RGB}{255,215,210}
\definecolor{cPurple}{RGB}{93,35,125}
\definecolor{cLightPurple}{RGB}{241,230,252}
\definecolor{cYellow}{RGB}{252,191,10}
\definecolor{cISSRBlue}{RGB}{0,111,174}
\definecolor{cISSRGrey}{RGB}{167,169,172}

\newcommand{\beq}{\begin{equation}}
\newcommand{\eeq}{\end{equation}}
\newcommand{\be}{\begin{equation}}
\newcommand{\ee}{\end{equation}}
\newcommand{\bi}{\begin{itemize}}
\newcommand{\ei}{\end{itemize}}
\newcommand{\ba}{\begin{array}}
\newcommand{\ea}{\end{array}}
\newcommand{\beqa}{\begin{eqnarray}}
\newcommand{\eeqa}{\end{eqnarray}}
\newcommand{\bea}{\begin{eqnarray}}
\newcommand{\eea}{\end{eqnarray}}
\newcommand{\beqn}{\begin{eqnarray}}
\newcommand{\eeqn}{\end{eqnarray}}

\newcounter{TODO}

\newcommand{\GeV}{\,\text{GeV}}

\newcommand{\vcb}{|V_{cb}|}

\def\kpn{K^+\rightarrow\pi^+\nu\bar\nu}
\def\klpn{K_{L}\rightarrow\pi^0\nu\bar\nu}

\begin{document}

\begin{flushleft}
{\em Version of \today}
\end{flushleft}

\vspace{-1.4cm}

\begin{flushright}
    AJB-24-1\\
    MITP-24-049
\end{flushright}

\medskip

\begin{center}
{\Large\bf
\boldmath{Disentangling new physics in $K\to\pi\nu\Bar{\nu}$ and $B\rightarrow K(K^*)\nu\Bar{\nu}$ observables}}
\\[1.2cm]
{\bf
    Andrzej~J.~Buras$^{a,b}$,
  Julia Harz$^{c}$, and Martin A. Mojahed$^{b,c}$
}\\[0.5cm]

{\small
$^a$TUM Institute for Advanced Study,
    Lichtenbergstr. 2a, D-85747 Garching, Germany \\[0.2cm]
    $^b$Physik Department, TUM School of Natural Sciences, TU M\"unchen,\\ James-Franck-Stra{\ss}e, D-85748 Garching, Germany  \\[0.2cm]
    $^c$PRISMA$^+$ Cluster of Excellence \& Mainz Institute for Theoretical Physics,\\ FB 08 - Physics, Mathematics and Computer
Science,\\ 
 Johannes Gutenberg-Universit\"{a}t Mainz, 55099 Mainz, Germany
}
\end{center}

\vskip 0.7cm

\begin{abstract}
  \noindent
 We investigate the possibility of disentangling different new physics contributions to the rare meson decays $K\to\pi+\slashed{E}$ and $B\to K(K^*)+\slashed{E}$ through kinematic distributions in the missing energy $\slashed{E}$. We employ dimension-$6$ operators within the Low-Energy Effective Field Theory (LEFT), identifying the invisible part of the final state as either active or sterile neutrinos. Special emphasis is given to lepton-number violating (LNV) operators with scalar and tensor currents. We show analytically that contributions from vector, scalar, and tensor quark currents can be uniquely determined from experimental data of kinematic distributions. In addition, we present new correlations of branching ratios for $K$ and $B$-decays involving scalar and tensor currents. As there could \textit{a priori} also be new invisible particles in the final states, we include dark-sector operators giving rise to two dark scalars, fermions, or vectors in the final state. In this context, we present new calculations of the inclusive decay rate $B\to X_s+\slashed{E}$ for dark operators.  We show that careful measurements of kinematic distributions make it theoretically possible to disentangle the contribution from LEFT operators from most of the dark-sector operators, even when multiple operators are contributing. We revisit sum rules for vector currents in LEFT and show that the latter are also satisfied in some new dark-physics scenarios that could mimic LEFT. Finally, we point out that an excess in rare meson decays consistent with a LNV  hypothesis would point towards highly flavor non-democratic physics in the UV, and could put high-scale leptogenesis under tension. 
\end{abstract}

\thispagestyle{empty}
\newpage

\setcounter{tocdepth}{2}
\tableofcontents

\newpage

%
%
\section{Introduction}
The rare decay processes $K\to\pi+\slashed{E}$ and $B\to K(K^*)+\slashed{E}$ are exceptionally sensitive to new physics (NP), and have played an important role in the tests of the Standard Model (SM) and its various extensions through the last few decades~\cite{Buras:2004uu,Buras:2020xsm, Colangelo:1996ay,Grossman:1995gt, Melikhov:1998ug,Buchalla:2000sk,Altmannshofer:2009ma,Buras:2014fpa,Kou:2018nap, Felkl2021}. Their sensitivity to NP can be attributed to two facts. First, they belong to the theoretically cleanest decays within the context of flavor-changing neutral current (FCNC) processes in
the SM, with high precision in their theoretical predictions~\cite{Buchalla:1992zm,Buchalla:1993bv,Buchalla:1993wq,Buchalla:1997kz,Misiak:1999yg,Buchalla:1998ba,Buras:2006gb,Brod:2008ss,Brod:2010hi,Brod:2021hsj}.
Second, their predictions in the SM are highly suppressed, due to the Glashow–Iliopoulos–Maiani (GIM) mechanism~\cite{Glashow:1970gm}. With current experimental sensitivity, these processes are already sensitive to NP generated at scales up to $\mathcal{O}(100)$ TeV~\cite{Buras:2014zga}. However, finding experimental evidence for NP would leave open questions about origin and implications for beyond the SM (BSM) model building. This paper aims to provide strategies for addressing such implications in $K\to\pi+\slashed{E}$, $B\to K(K^*)+\slashed{E}$, and the inclusive decay $B\to X_s+\slashed{E}$. We pay particular attention to potential signatures of lepton-number violating (LNV) NP.

The effects of NP appearing at a high scale can be described in a model-independent manner within the framework of effective field theories (EFTs), where only minimal assumptions about particle content and symmetries are imposed. In the EFT approach, experimental results are used to determine or constrain values for the Wilson coefficients (WCs) of the higher-dimensional operators in the EFT. It is then possible to address models of interest by matching them onto the EFT and see if they are compatible with the experimental measurements used to constrain the EFT. A particularly important example is the SM Effective Field Theory (SMEFT)~\cite{Buchmuller:1985jz, Grzadkowski:2010es, Brivio:2017vri,Isidori:2023pyp}, which is used to parametrize NP through higher-dimensional local operators that are built from the SM field content and invariant under the SM gauge group. SMEFT is very useful for model-independent studies of BSM physics above the electroweak scale. For the present analysis, it
is more practical to use low-energy EFT (LEFT)~\cite{Jenkins2017, Jenkins:2017dyc, Aebischer2017}, which can be derived from SMEFT by integrating out the top quark, the Higgs, and the three massive gauge bosons $(W^\pm,Z)$. In LEFT, the operators that are most relevant for rare meson decays are of dimension six.\footnote{Dimension-5 dipole operators can also contribute to rare meson decays. However, as they are strongly constrained by searches for neutrino magnetic dipole moments~\cite{Borexino:2017fbd, Giunti:2014ixa}, their effects are not considered in this work.} 
These operators can be separated into a set of lepton-number conserving (LNC) operators and a set of lepton-number violating (LNV) operators. In this work, we consider both classes of operators with special emphasis on the latter.

A future experimental measurement pointing to nonzero WCs for LNV operators would have major consequences for our understanding of particle physics. It could indicate a Majorana nature of neutrinos and have implications for our understanding of how the matter-antimatter asymmetry of our universe came into existence. The search for neutrinoless double beta decay $(0\nu\beta\beta)$ is by many considered to be the most promising way to probe LNV in general, and to search for Majorana neutrino masses. One of the inherent weaknesses of $0\nu\beta\beta$ is that it can only probe LNV physics in the first generation of SM fermions. This is to be contrasted with meson decays, which can probe physics beyond the first generation of SM fermions. In particular, rare meson decays
$K\to\pi+\slashed{E}$ and $B\to K(K^*)+\slashed{E}$ have quite recently been recognized as a new potential probe of LNV, complementing $0\nu\beta\beta$ experiments and collider searches for LNV~\cite{Fridell:2023rtr}. 

In fact, it was recently pointed out that it is possible to search for LNV using the aforementioned LEFT operators in rare meson decays by analyzing the kinematic distribution of the energy carried away by the invisible final states~\cite{Li2019, Deppisch2020, Felkl2021, Gorbahn:2023juq}. Schematically, distributions $d\Gamma(\text{Meson}_1\rightarrow \text{Meson}_2+\text{\,invisible})/ds$ in the invariant mass of the invisible state $s$ often take different shapes when different effective operators are induced. Hence, kinematic distributions can potentially be utilized in strategies to discriminate between NP scenarios where different effective operators are generated. There has not been a systematic study in the literature of how to quantitatively determine NP contributions from kinematic distributions in rare meson decays so far. In particular, previous studies have only considered NP scenarios where one or at most two WCs are nonzero. The current paper aims to fill this gap and provide a systematic approach to quantifying NP contributions in rare meson decays, where experimental data may \textit{a priori} contain contributions from more than one or two higher-dimensional operators. Moreover, the analysis in this paper does not rely on any assumptions about flavor structure in the UV unless stated explicitly in the text. We hope that the results and strategies presented here and in\cite{Li2019, Deppisch2020, Felkl2021, Gorbahn:2023juq}, along with the 
Belle II experiment \cite{Kou:2018nap}, can be used in future searches for LNV interactions.

An important caveat to the discussion so far is that it is not possible to measure and determine the nature of the final states carrying away the missing energy. Therefore, there could be other particles than neutrinos in the invisible final states. Such scenarios are not fully captured by LEFT, whose validity of description relies on the fact that \textit{only} neutrinos are relevant for this process. To remedy this shortcoming, we also include so-called dark LEFT operators~\cite{Aebischer:2022wnl,He2022} in our study, which allows new states in the invisible decay product. By scrutinizing the effects of dark operators we are able to check the prospects of disentangling various LEFT contributions in an experimental data set that may \textit{a priori} also include contributions from particles that are feebly interacting with the SM. It also enables us to systematically study to what extent different effective operators can mimick LNV and other NP signatures in kinematic distributions, which has not been discussed in previous studies.

In writing the paper, we aimed to make it self-contained and accessible to potential readers from different scientific communities. The organization of our paper is as follows. In Section~\ref{sec:generalities} we present the operator basis and some generalities about lepton-number violation. In Section~\ref{sec:Kaons}, we first review the experimental and theoretical status of $\mathcal{B}(K_L\rightarrow \pi^0\nu\widehat{\nu})$ and $\mathcal{B}(K^+\rightarrow \pi^+\nu\widehat{\nu})$. We proceed by deriving a new result for the correlation of $\mathcal{B}(K_L\rightarrow \pi^0\nu\widehat{\nu})-\mathcal{B}(K^+\rightarrow \pi^+\nu\widehat{\nu})$ and show in detail how different operators contributing to rare Kaon decays can be disentangled through detailed measurements of kinematic distributions. In Section~\ref{sec:Bmeson}, we start by reviewing the experimental bounds on $B\rightarrow K(K^*)\nu\widehat{\nu}$. We proceed by showing how different LEFT and dark-operator contributions to $B$-meson decays can be disentangled. Some simple numerical examples are included to illustrate the underlying concept. We also revisit and reinterpret some sum rules for LEFT operators with vector currents~\cite{Buras:2014fpa}. The section ends with some implications that observations of LNV in rare-meson decays would have for flavor physics in the UV and for leptogenesis. In Section~\ref{sec:conclusions}, we summarize the main results of the paper and conclude. In a number of appendices, we collect information on form factors, dark operators, and long expressions for various distributions.


\section{Generalities}
\label{sec:generalities}
In this section, we present the operator bases utilized by us and discuss the lepton number of relevant LEFT operators. The following discussion provides the foundation for calculations and interpretation of results in later sections. 
 
\subsection{Operator basis}
\label{sec:OperatorBasis}

Adopting the notation in refs.~\cite{Felkl2021,Jenkins2017}, the LEFT Lagrangian with the operators of interest to us reads~\cite{Aebischer2017, Jenkins2017}
\begin{align}
    \label{lagrangian}
    \mathcal{L}_{\rm eff}= \mathcal{L}_{\rm SM} +\sum_{X=L,R}C_{\nu d}^{\text{VLX}}\mathcal{O}_{\nu d}^{\text{VLX}}+\left(\sum_{X=L,R}C^{\text{SLX}}_{\nu d}\mathcal{O}^{\text{SLX}}_{\nu d}+C^{\text{TLL}}_{\nu d}\mathcal{O}^{\text{TLL}}_{\nu d}+\text{h.c.}\right).
\end{align}
This Lagrangian contains operators with vector currents,
\begin{align}
\label{VLL}
\mathcal{O}_{\nu d}^{\text{VLL}}&=(\overline{\nu_L}\gamma^\mu \nu_L)(\overline{d_L}\gamma_\mu d_L)\; , \\
\label{VLR}
\mathcal{O}_{\nu d}^{\text{VLR}}&=(\overline{\nu_L}\gamma^\mu \nu_L)(\overline{d_R}\gamma_\mu d_R)\; ,
\end{align}
scalar currents, 
\begin{align}
\mathcal{O}_{\nu d}^{\text{SLL}}&= (\overline{\nu^c_L} \nu_L)(\overline{d_R} d_L)\;\label{S1} ,\\
\mathcal{O}_{\nu d}^{\text{SLR}}&= (\overline{\nu^c_L} \nu_L)(\overline{d_L} d_R)\; \label{S2},
\end{align}
and tensor currents
\begin{align}
    \mathcal{O}_{\nu d}^{\text{TLL}}&= (\overline{\nu^c_L}\sigma_{\mu\nu} \nu_L)(\overline{d_R} \sigma^{\mu\nu}d_L)\;\label{TLL}.
\end{align}
Here $d_{L(R)}$ and $\nu_L$ are Weyl fermions, where the subscript denotes their chirality. The field $\nu^c_L\equiv C\overline{\nu_L}$ is right-handed, and the charge-conjugation operator is given by $C=i\gamma^2\gamma^0$. The scalar operators $\mathcal{O}_{\nu d}^{\text{SLL}}$ and $\mathcal{O}_{\nu d}^{\text{SLR}}$ are symmetric in the neutrino flavors, the tensor operator $\mathcal{O}_{\nu d}^{\text{TLL}}$ is antisymmetric in the neutrino flavors, while the vector operators $\mathcal{O}_{\nu d}^{\text{VLL}}$ and $\mathcal{O}_{\nu d}^{\text{VLR}}$ do not exhibit a manifest (anti)symmetry in the neutrino flavors~\cite{Felkl2021}. 
All possible flavor combinations are implicitly summed over in the Lagrangian (\ref{lagrangian}).
In particular, (\ref{lagrangian}) can be used for both $K$ and $B$ decays considered by us. This operator basis can also be used in the presence of light sterile neutrinos~\cite{Felkl2021,Felkl2023}. We work in the limit of massless neutrinos, so flavor and mass eigenstates coincide, throughout the paper.

We also consider the possibility of having very light dark-sector particles that generate additional dark-sector LEFT operators, which contribute to an excess in rare meson decays. Operator bases relevant for rare meson decays with two dark-sector particles in the final state have been worked out in Ref.~\cite{He2022} for LEFT augmented with an additional dark-sector particle with either spin $0$, spin $1/2$, or spin $1$. We make use of their results in this work, see also~\cite{Hou:2024vyw} for some very recent applications of (D)SMEFT and (D)LEFT to rare $B$ decays. For the reader's convenience, we have included the full bases of dark operators in Appendix~\ref{appendix:DarkOperators}. From here on, we refer to the operators in (\ref{lagrangian}) as (SM)LEFT to avoid any confusion with operators containing dark-sector particles, which we will refer to as (D)LEFT operators. 

\subsection{Sterile neutrinos and lepton number violation}
\label{sec:OperatorLNV}
The basis in~(\ref{lagrangian}) incorporates the active neutrinos $\nu_{Li}$, $i=1,2,3$, and can also incorporate an arbitrary number of sterile neutrinos $N_{Ri}=\nu_{Li}^c$, $i=4,..$, which both carry lepton number $+1$~\cite{Li2020}.\footnote{Note that with this assignment the Dirac Yukawa term $\overline{L}HN_R$ conserves lepton number as expected.} The operators in (\ref{lagrangian}) could be either LNV or LNC~\cite{Li2020}, depending on whether the neutrinos are sterile and/ or active. The purpose of this section is to clarify under which circumstances the operators violate lepton number. In the following discussion, we will use $i$ and $j$ to denote the generation of the active $i,j\in\{1,2,3\}$ and sterile neutrinos $i,j\geq 4$.

For $i,j\leq 3$, i.e. two active neutrinos, the vector-current operators $\mathcal{O}_{\nu d}^{\text{VLL}}$ and $\mathcal{O}_{\nu d}^{\text{VLR}}$ are LNC while the scalar-currents operators $\mathcal{O}_{\nu d}^{\text{SLL}}$ and $\mathcal{O}_{\nu d}^{\text{SLR}}$ and the tensor-current operators $\mathcal{O}_{\nu d}^{\text{TLL}}$ violate lepton number. Hence, measuring a nonzero WC for one of the scalar-current operators or one of the tensor-current operators would automatically signal LNV in the absence of light sterile neutrinos.

For $i,j\geq 4$, i.e. two sterile neutrinos, the vector-current operators $\mathcal{O}_{\nu d}^{\text{VLL}}$ and $\mathcal{O}_{\nu d}^{\text{VLR}}$ are LNC while the scalar-currents operators $\mathcal{O}_{\nu d}^{\text{SLL}}$ and $\mathcal{O}_{\nu d}^{\text{SLR}}$ and the tensor-current operators $\mathcal{O}_{\nu d}^{\text{TLL}}$ violate lepton number. This result is the same as the result for $i,j\leq 3$.

For $1\leq i\leq 3<j$ and $1\leq j\leq 3<i$, the vector-current operators $\mathcal{O}_{\nu d}^{\text{VLL}}$ and $\mathcal{O}_{\nu d}^{\text{VLR}}$ violate lepton number while the scalar-currents operators $\mathcal{O}_{\nu d}^{\text{SLL}}$ and $\mathcal{O}_{\nu d}^{\text{SLR}}$ and the tensor-current operators $\mathcal{O}_{\nu d}^{\text{TLL}}$ are LNC. Hence, measuring a nonzero WC for a scalar-current or tensor-current operator is \textit{not} a proof of lepton-number violation, unless the associated neutrinos are identified as either two active or two sterile neutrinos. Similarly, measuring a nonzero WC for $\mathcal{O}_{\nu d}^{\text{VLR}}$ or a WC differing from the SM value for $\mathcal{O}_{\nu d}^{\text{VLL}}$ does not automatically imply lepton-number conserving NP. The various scenarios are summarized in Table~\ref{tab:LNV/LNC}.

Finally, it is worth pointing out that it is straightforward to rewrite the basis in (\ref{lagrangian}) if the neutrinos are written in the form of Dirac spinors~\cite{Gorbahn:2023juq}. For the case of Majorana neutrinos, an explicit matching of the basis in (\ref{lagrangian}) to the $S,P,V,A,T$ basis has been carried out and used to compute finite mass corrections to the $B$ meson processes considered in the present paper~\cite{Felkl2021}. Finite neutrino-mass corrections can be relevant if there is at least one additional (light) sterile neutrino present whose mass is non-negligible compared to the experimental sensitivity and other mass-scales in the processes under consideration~\cite{Felkl2021, Felkl2023, Gorbahn:2023juq}.

\setlength{\extrarowheight}{4pt}
\begin{table}[t!]
    \centering
    \begin{tabular}{|>{\raggedright}m{72pt}|>{\raggedright}m{70pt}|>{\raggedright}m{70pt}|>{\raggedright}m{70pt}|>{\centering}m{60pt}|c}
    \cline{1-5}
    $\mathcal{O}$ & $1\leq i,j\leq 3$ & $1\leq i\leq 3<j$ & $1\leq j\leq 3<i$ & $i,j\geq 4$ & \\[2pt]
    \cline{1-5}
    \multirow{1}{*}{$\mathcal{O}_{\nu d}^{\text{SLL}}$ \& $\mathcal{O}_{\nu d}^{\text{SLR}}$} 
     & \centering LNV & \centering LNC & \centering LNC & \centering LNV & \\[2pt]
     \cline{1-5}
    \multirow{1}{*}{$\mathcal{O}_{\nu d}^{\text{VLL}}$ \& $\mathcal{O}_{\nu d}^{\text{VLR}}$} & \centering LNC & \centering LNV & \centering LNV & \centering LNC & \\[2pt]
      \cline{1-5}
    \multirow{1}{*}{\,\,\,\,\,\,\,\,\,\, $\mathcal{O}_{\nu d}^{\text{TLL}}$} & \centering LNV & \centering LNC & \centering LNC & \centering LNV &\\[2pt]
     \cline{1-5}
    \end{tabular}
    \caption{Classification of the (SM)LEFT operators according to LNV/LNC for all possible combinations of neutrino-generation indices $i,j$.}
    \label{tab:LNV/LNC}
\end{table}


\section{Kaon decays}
\label{sec:Kaons}
In this section, we consider strategies to probe various operators contributing to rare Kaon decays. To facilitate the discussion, we first provide a brief review of experimental constraints and analytic relations for the branching ratios $\mathcal{B}(\klpn)$ and $\mathcal{B}(\kpn)$. Then we show how the branching ratios for $\klpn$ and $\kpn$ are correlated with each other in scenarios where scalar-current operators provide the dominant NP contribution and compare this correlation found in NP scenarios dominated by vector currents. The second part of the section concerns kinematic distributions for $\kpn$ and $\klpn$, and how they can be used to disentangle NP contributions. Specifically, we illustrate how dedicated measurements of kinematic distributions for $\kpn$ and $\klpn$ can be used to quantify the contributions from the different operators in Eqs.~(\ref{VLL})-(\ref{TLL}). In the end, we go beyond pure (SM)LEFT to discuss prospects of disentangling (SM)LEFT operators from light NP originating from a dark sector. The latter serves partially as preparation for Section~\ref{sec:Bmeson}, which contains a more general and extended discussion of how to disentangle NP with neutrinos in the final state from dark-sector interactions.

\boldmath
\subsection{Current experimental limits on $K \to \pi \nu\bar{\nu}$}
\unboldmath
The very recent result for $\kpn$ from NA62 \cite{CortinaGil:2021nts,NA62:2024} and the  $90\%$ confidence level (CL) upper bound on $\klpn$  from KOTO  \cite{Ahn:2018mvc} read respectively
\begin{align}
\label{EXP19}
    \mathcal{B}(\kpn)_\text{exp}&=\left(13.0^{+3.3}_{-2.9}\right)\times 10^{-11},\\
\mathcal{B}(\klpn)_\text{exp}&\le 2.0\times 10^{-9}\,.
\end{align}
The four events of $\klpn$ reported by KOTO in \cite{Ahn:2020opg}, that violated the Grossman and Nir (GN) \cite{Grossman:1997sk} upper bound, have been excluded in the 2023 analysis.

It is worth noting that the E949 experiment~\cite{BNL-E949:2009dza} performed a dedicated analysis assuming either only an underlying vector or scalar current leading to the following current-dependent upper limits at $90\%$CL, respectively,
\begin{align}
\label{E949}
    \mathcal{B}(\kpn)^{\mathrm{vec}}_\text{exp}<3.35\times 10^{-10}\,,\qquad
\mathcal{B}(\kpn)^{\mathrm{sc}}_\text{exp}<21\times 10^{-10},
\end{align}
demonstrating the lower experimental sensitivity for scalar currents. 
These experimental results are to be compared with the most recent SM predictions~\cite{Buras:2021nns},
\begin{align}
\label{KSM}
    \mathcal{B}(\kpn)_{\text{SM}}&= (8.60\pm0.42)\times 10^{-11}, \\
    \label{KLSM}
    \mathcal{B}(\klpn)_{\text{SM}}&=(2.94\pm0.15)\times 10^{-11}.
\end{align}

Comparing the current experimental limits with the SM expectations, it is clear that there is still room for NP contributions to these rare processes. With NA62 and KOTO aiming to reach precision in the ballpark of $10\%$ in the future, it becomes of particular interest to understand to what extent scalar and tensor currents can still be accommodated in processes with neutrinos in the final state and how they can be discriminated from operators with vector currents. As a first step towards resolving this question, we proceed to review how the branching ratios in rare Kaon decays change when the SM is augmented by the operators in (\ref{lagrangian}).

\boldmath
\subsection{Branching ratios for $\kpn$ and $\klpn$}
\unboldmath
\label{subsec:kpnandklpn}

The contributions of the operators in (\ref{VLL})-(\ref{TLL}) to the branching ratios $K_L\to \pi^0\nu\widehat{\nu}$ and $K^+\to \pi^+\nu\widehat{\nu}$ can be expressed as~\cite{Li2019} 
\begin{align}
\label{brkl-text}
	\mathcal{B}({K_L\rightarrow\pi^0\nu\widehat{\nu}})&=J_S^{K_L} \sum_{\alpha\leq \beta} \left(1-{1\over2}\delta_{\alpha\beta}\right)  \left|C_{\nu d,\alpha\beta sd}^{\text{SLL}}+C_{\nu d,\alpha\beta sd}^{\text{SLR}}
+C_{\nu d,\alpha\beta ds}^{\text{SLL}}+C_{\nu d,\alpha\beta ds}^{\text{SLR}}\right|^2\nonumber \\
&+J_T^{K_L} \sum_{\alpha<\beta}\left(\left|C^{\text{TLL}}_{\nu d,\alpha\beta ds}+C^{\text{TLL}}_{\nu d,\alpha\beta sd}\right|^2\right)\nonumber \\
&+J_V^{K_L}\sum_{\alpha, \beta}\left(1-{1\over2}\delta_{\alpha\beta}\right)  \left|C_{\nu d,\alpha\beta sd}^{\text{VLL}}+C_{\nu d,\alpha\beta sd}^{\text{VLR}}-C_{\nu d, \alpha\beta ds}^{\text{VLL}}-C_{\nu d,\alpha\beta ds}^{\text{VLR}}\right|^2,
\end{align}
and 
\begin{align}
\label{brkp-text}
&\mathcal{B}({K^+\rightarrow\pi^+\nu\widehat\nu})=J_S^{K^+}\sum_{\alpha\leq \beta} \left(1-{1\over2}\delta_{\alpha\beta}\right)\left(\left|C_{\nu d,\alpha\beta sd}^{\text{SLL}}+C_{\nu d,\alpha\beta sd}^{\text{SLR}}\right|^2+\left|C_{\nu d,\alpha\beta ds}^{\text{SLL}}+C_{\nu d,\alpha\beta ds}^{\text{SLR}}\right|^2\right)\nonumber \\
&+J_T^{K^+} \sum_{\alpha<\beta} \left( \left|C^{\text{TLL}}_{\nu d,\alpha\beta sd}\right|^2+\left|C^{\text{TLL}}_{\nu d,\alpha\beta ds}\right|^2\right)
+J_V^{K^+}\sum_{\alpha, \beta}\left(1-{1\over2}\delta_{\alpha\beta}\right)   \left|C_{\nu d,\alpha\beta sd}^{\text{VLL}}+C_{\nu d,\alpha\beta sd}^{\text{VLR}}\right|^2, 
\end{align}
with 
\begin{align}
	J_S^{K_L}&={1 \over \Gamma_{K_L}^{\text{Exp}}}{B_L^2\over 2^9\pi^3m_{K^0}^3}\int ds\, s\lambda^{1/2}(s,m_{K^0}^2,m_{\pi^0}^2)\, \left| f^{K^0}_0(s) \right|^2=34.6 \,G_F^{-2} ,
\label{J1KL-text}\\
\label{JTKL}
J_T^{K_L}&={1 \over \Gamma_{K_L}^{\text{Exp}}}{1\over 3\cdot2^{5}\pi^3m_{K^0}^3(m_{K^0}+m_{\pi^0})^2}\int ds\,s\, \lambda^{3/2}(s,m_{K^0}^2,m_{\pi^0}^2)\,\left|f^{K^0}_T(s)\right|^2\nonumber \\
&= 0.13\,G_F^{-2},
\\
\label{JVKL}
J_V^{K_L}&={1 \over \Gamma_{K_L}^{\text{Exp}}}{1\over 3\cdot 2^{11}\pi^3m_{K^0}^3}\int ds \lambda^{3/2}(s,m_{K^0}^2,m_{\pi^0}^2)\,\left|f^{K^0}_+(s)\right|^2= 0.25\,G_F^{-2} ,
\\\label{J1KP-text}
J_S^{K^+}&={1 \over \Gamma_{K^+}^{\text{Exp}}}{ B_+^2\over 2^8\pi^3m_{K^+}^3}\int ds\, s
\lambda^{1/2}(s,m_{K^+}^2,m_{\pi^+}^2)\, \left|f^{K^+}_0(s)\right|^2=15.2\,G_F^{-2},
\\
J_T^{K^+}&={1 \over \Gamma_{K^+}^{\text{Exp}}}{1\over 3\cdot2^{4}\pi^3m_{K^+}^3(m_{K^+}+m_{\pi^+})^2}\int ds\,s\, \lambda^{3/2}(s,m_{K^+}^2,m_{\pi^+}^2)\,\left|f^{K^+}_T(s)\right|^2\nonumber \\
&=0.11\,G_F^{-2},
\\\label{J2KP-text}
J_V^{K^+}&={1 \over \Gamma_{K^+}^{\text{Exp}}}{1\over 3\cdot 2^9\pi^3m_{K^+}^3}\int ds   \lambda^{3/2}(s,m_{K^+}^2,m_{\pi^+}^2)\, \left|f^{K^+}_+(s)\right|^2=0.23\,G_F^{-2}\; ,
\end{align}
where
  \be\label{BLB+-text}
  B_L(\mu)=\frac{m^2_{K^0}-m^2_{\pi^0}}{m_s(\mu)-m_d(\mu)},\qquad
  B_+(\mu)=\frac{m^2_{K^+}-m^2_{\pi^+}}{m_s(\mu)-m_d(\mu)}\,.
  \ee
The symbol $\nu\widehat{\nu}$ indicates that both $\nu\bar\nu$ and $\nu\nu$ final states can occur. Here, $\Gamma_{K_L}^{\text{Exp}}\,(\Gamma_{K^+}^{\text{Exp}})$ is the width of $K_L\,(K^+)$, respectively, and $\lambda(a,b,c)= a^2  +b^2 + c^2 - 2 (a b+ b c + a c)$ denotes the usual Källén function. The invariant mass-squared of the neutrinos is given by $s = (k + k')^2$, and is experimentally measured as missing energy.

To our knowledge, the expression for $J_T^{K_L}$ is new, while the results for the vector and scalar currents agree with results in Ref.~\cite{Li2019}, and the result for $J_T^{K^+}$ agrees with the result for $B$-decays in~\cite{Felkl2021}. The results in (\ref{J1KL-text}), (\ref{JVKL}), (\ref{J1KP-text}), and (\ref{J2KP-text}) are refinements of the results in~\cite{Li2019}. In contrast to~\cite{Li2019}, we use form factors in our calculations, which encode additional
$s$-dependence and isospin-breaking effects~\cite{Mescia:2007kn}. The factors {
$B_L=2.57~\mathrm{GeV}$ and $B_+=2.52~\mathrm{GeV}$} also differ from the factor $B$ used in \cite{Li2019}. 
The expressions for the form factors are collected in Appendix~\ref{appendix:formfactors}.

In this context, it should be emphasized that the factors $B_L$  and $B_+$ are dependent on the scale $\mu$. This $\mu$ dependence originates
in the non-vanishing anomalous dimension of the scalar current. It is canceled by the $\mu$ dependence of the corresponding Wilson coefficient so that the branching ratios are scale independent. However, $J_S^{K_L}$ and $J_S^{K^+}$ are scale dependent, and the values given in (\ref{J1KL-text}) and (\ref{J1KP-text}) correspond to $\mu=2\GeV$. They would be smaller for $\mu=1\GeV$. $J_V^{K_L}$ and $J_V^{K^+}$ are $\mu$-independent, due to the conservation of the vector current.

\boldmath
\subsection{Correlation of $\klpn$ and $\kpn$ } \label{sec:KaonCorrelation}
\unboldmath

The branching ratios for $\klpn$ and $\kpn$ are often correlated with each other, where the details of the correlation depend on the NP model considered. Previous analyses of correlations between $\mathcal{B}(\klpn)$ and $\mathcal{B}(\kpn)$ have only considered vector currents~\cite{Blanke:2009pq,Buras:2015yca,Aebischer:2020mkv}. Here, we go beyond these by also discussing the impact of scalar currents on the correlation in question. A first step in this direction has already been made in \cite{Li2019,Deppisch2020}, where it was demonstrated that the GN bound is also satisfied in the presence of scalar-current contributions. 

For illustration purposes, we will assume lepton-flavor universality and that the final state neutrinos have identical flavors for the rest of this {sub}section. The latter automatically excludes tensor-current operators, which are antisymmetric in lepton-flavor indices. In {Sub}section~\ref{subsec:KaonExtraction}, we provide a strategy for testing the absence of tensor currents in future experiments. 

As the vector and scalar contributions do not interfere with each other, the total branching ratios for $\klpn$ and $\kpn$ with three neutrino generations can be parametrized as
\be
\mathcal{B}(\klpn)=\mathcal{B}^V(\klpn)+\mathcal{B}^S(\klpn),
\ee
\be
    \mathcal{B}(\kpn)=\mathcal{B}^V(\kpn)+\mathcal{B}^S(\kpn),
\ee
where $\mathcal{B}^{V(S)}$ denotes the contribution from vector (scalar) current operators. 
Assuming lepton flavor universality, we parameterize Wilson coefficients associated with the vector-current operators as
\be
C_{\nu d,\alpha\beta sd}^{\text{VLL}}+C_{\nu d,\alpha\beta sd}^{\text{VLR}}={(|C_{\rm SM}|e^{i\phi_{\rm SM}}+|C_V|e^{i\phi_V})\delta_{\alpha \beta}},
\ee
and 
\be
C_{\nu d,\alpha\beta ds}^{\text{VLL}}+C_{\nu d, \alpha\beta ds}^{\text{VLR}}= {( |C_{\rm SM}|e^{-i\phi_{\rm SM}}+                  |C_V|e^{-i\phi_V})\delta_{\alpha \beta}},
\ee
which manifestly separates the SM contribution $(C_{\rm SM},\,\phi_{\rm SM})$ from NP $(C_V,\,\phi_V)$. Similarly, we express the WCs associated with the scalar-current operators as
\be
C_{\nu d,sd\alpha\beta}^{\text{SLL}}+C_{\nu d,sd\alpha\beta}^{\text{SLR}}=|C_S|e^{i\phi_S}{\delta_{\alpha \beta}},\qquad 
C_{\nu d,ds\alpha\beta}^{\text{SLL}}+C_{\nu d,ds\alpha\beta}^{\text{SLR}}=|C_S|e^{-i\phi_S}{\delta_{\alpha \beta}}.
\ee
In this notation, the vector and scalar contributions to the branching ratios including the SM contributions read
\begin{align}
\label{BVL+}
    \mathcal{B}^V(\klpn)&=12 J_V^{K_L} T_1, \qquad &&\mathcal{B}^V(\kpn)=3 J_V^{K^+} (T_1+T_2),\\
    \label{B-scalar-kaons}
    \mathcal{B}^S(\klpn)&=6J_S^{K_L} |C_S|^2\cos^2\phi_S, \qquad &&\mathcal{B}^S(\kpn)=3 J_S^{K^+} |C_S|^2,
\end{align}
respectively, where
  \be\label{XYdef}
  T_1=(|C_{\rm SM}|\sin\phi_{\rm SM}+|C_{V}|\sin\phi_{V})^2,\qquad 
  T_2=(|C_{\rm SM}|\cos\phi_{\rm SM}+|C_{V}|\cos\phi_{V})^2\,.
  \ee
By comparing the expressions above with~(\ref{KSM}) and (\ref{KLSM}), we obtain the following values for the SM
\begin{align}
    \mathcal{B}_{\rm SM}(\klpn)&=12 J_V^{K_L}|C_{\rm SM}|^2\sin^2\phi_{\rm SM},
\qquad &&\mathcal{B}_{\rm SM}(\kpn)=3 J_V^{K^+} |C_{\rm SM}|^2,\nonumber \\ 
\label{SM}
|C_{\rm SM}|&= 1.30 \times 10^{-10}~\GeV^{-2}, \qquad &&\phi_{\rm SM}= 0.09~\pi.
\end{align}

We now have all the ingredients necessary to write down two equations relating the branching ratios for $K^+$ and $K_L$. The first is a relation among vector contributions only
    \be\label{BF1}
    \mathcal{B}^V(\klpn)=4\frac{J_V^{K_L}}{J_V^{K^+}}\mathcal{B}^V(\kpn)-12 J_V^{K_L}T_2.     \ee
The second equation relates scalar and vector contributions
\begin{align}
\label{BF2}
\mathcal{B}(\klpn)&=4\frac{J_V^{K_L}}{J_V^{K^+}}\left[\mathcal{B}^V(\kpn)+\mathcal{B}^S(\kpn)\cos^2\phi_S\right]\nonumber \\
&-12 J_V^{K_L} T_2.
\end{align}
The latter was obtained using the following approximation
 \be
    \frac{J_S^{K_L}}{J_S^{K^+}}\approx 2\frac{J_V^{K_L}}{J_V^{K^+}},
    \ee
which follows from
\be
2\frac{J_S^{K_L}}{J_S^{K^+}}=4.55 \qquad \text{and} \qquad 4\frac{J_V^{K_L}}{J_V^{K^+}}=4.35,
\ee
which we identify as the Grossmann-Nir (GN) bound.

\begin{figure}[th]
\centering
\includegraphics[width=0.95\textwidth]{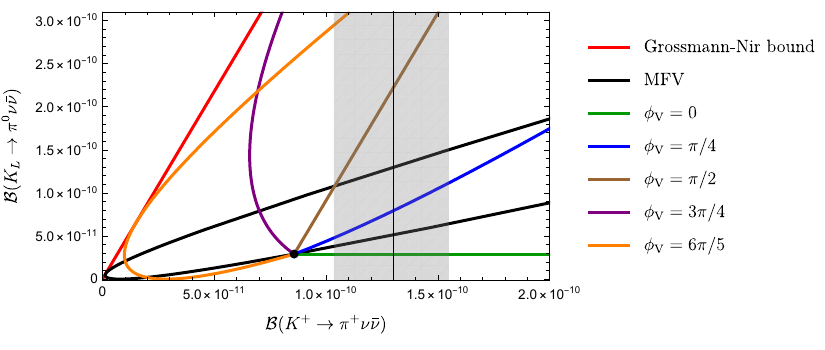}\\
\includegraphics[width=0.95\textwidth]{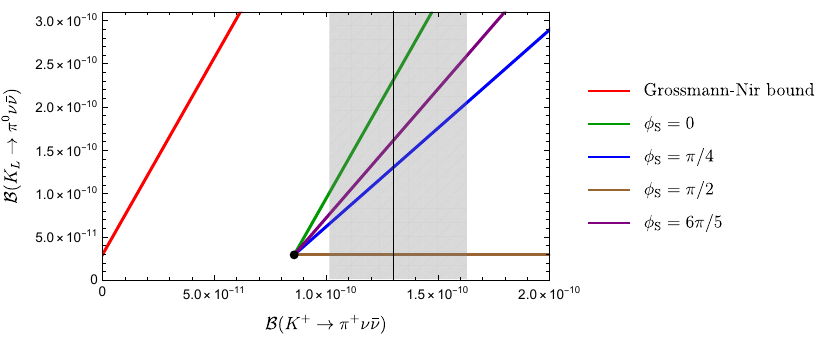}
\caption{$\mathcal{B}(\kpn)-\mathcal{B}(\klpn)$-plane for
  vector-current contributions with $C_S=0$ (top) and scalar-current contributions with $C_V=0$ (bottom) assuming lepton-flavor universality. $\phi_V$ $(\phi_S)$ is fixed to different values (see legend) and $C_V$ $(C_S)$ varied.
    The red line indicates the Grossmann-Nir bound. The SM contribution in (\ref{SM}) is represented by a black point. The grey region represents the present experimental $1\sigma$ range. }
\label{fig:KaonVectorScalar}
\end{figure}

The results of a numerical analysis of equations~(\ref{BF1}) and (\ref{BF2}) are displayed in Fig.~\ref{fig:KaonVectorScalar}. In the upper plot, we show the results of a pure NP vector contribution with $C_S=0$ (by varying $C_V$) in the $\mathcal{B}(\kpn)-\mathcal{B}(\klpn)$-plane. In the lower plot, we show the corresponding results for a pure NP scalar contribution with $C_V=0$ (varying $C_S$). In both cases, the SM contribution in (\ref{SM}) is represented by a black point, the central experimental value of $\mathcal{B}(\kpn)$ by
a thin vertical black line, and the GN bound by a red line. Below, we summarize the main observations concerning the above relations and the resulting plots.

\paragraph{New vector currents}\mbox{}\\ \mbox{}\\
Eq.~(\ref{BF1}) encodes a class of scenarios that have been discussed previously in the literature~\cite{Blanke:2009pq,Buras:2015yca,Aebischer:2020mkv}. For $\phi_V=0$, there is no NP contribution to the $\klpn$ branching ratio, and the r.h.s of (\ref{BF1}) reduces to the SM contribution. This case is depicted by the horizontal green line in the top part of Fig.~\ref{fig:KaonVectorScalar}. For $\phi_V=\pi/2$, depicted by the brown line in the upper part of the same figure, the correlation is parallel to the GN bound. The first term in (\ref{BF1}) represents the GN bound, and the correlation between the two branching ratios takes place on a line parallel to the GN bound due to a strictly negative shift from the SM contribution. As pointed out in \cite{Blanke:2009pq}, such a correlation occurs in NP models where only left-handed NP couplings are present
  and new contributions to the parameter $\varepsilon_K$ are negligible. A recent analysis of such a scenario in the context of $Z^\prime$ models
  has been presented in \cite{Aebischer:2023mbz}.

For $0<\phi_V<\pi/2$, the correlation between the two branching ratios is represented by curves laying between the two branches found above with slopes becoming steeper as $\phi_V$ is increased, as depicted in the top part of Fig.~\ref{fig:KaonVectorScalar}. For $\phi_V>\pi/2$, the curves lay between the GN-line and the branch parallel to it. The black line, representing models with Minimal Flavour Violation, extends to both sides of the $\phi_V=\pi/2$ curve.
These results confirm the findings in \cite{Blanke:2009pq,Buras:2015yca,Aebischer:2020mkv}.

 \paragraph{NP beyond vector currents}\mbox{}\\ \mbox{}\\
The novel feature, presented here for the first time, is the impact of pure {\em scalar} contributions through the correlation relation in (\ref{BF2}), shown in the lower part of Fig.~\ref{fig:KaonVectorScalar}. While the NP vector contribution to $\mathcal{B}(\klpn)$ vanishes for $\phi_V=0$, the NP scalar contribution is maximal for $\phi_S=0$, as the first term on the r.h.s of (\ref{BF2}) reduces to a straight line parallel to the GN line. For non-vanishing $\phi_S$, the correlation between the two branching ratios proceeds on straight lines that are not parallel to the GN line. The slope of the lines decreases with increasing $\phi_S$ for $0\le\phi_S\le\frac{\pi}{2}$.

When comparing the two plots in  Fig.~\ref{fig:KaonVectorScalar},
it becomes clear that a scalar contribution can only increase the branching ratios while a vector contribution can also decrease them compared to the SM model expectation.  When all four NP parameters
\be
    |C_{V}|, \quad \phi_{V}, \quad |C_{S}|, \quad \phi_{S},
\ee
can be nonvanishing at the same time, the full $\mathcal{B}(\kpn)-\mathcal{B}(\klpn)$-plane below the Grossmann-Nir bound (red solid line) opens up. Meanwhile, a pure scalar NP contribution is confined to the area between the green ($\phi_S= 0$) and brown lines ($\phi_S=\pi/2$) in the lower part in  Fig.~\ref{fig:KaonVectorScalar}. Hence, if a deviation from the SM to lower values is experimentally measured in either $\mathcal{B}(\kpn)$ or $\mathcal{B}(\klpn)$, then a scalar current can only be present with an additional vector contribution.

Finally, we want to point out a difference between the scalar and vector contribution to $\klpn$. While there is a relative sign difference between
$C_{\nu d,\alpha\beta sd}^{\text{VLL}}$, $C_{\nu d,\alpha\beta sd}^{\text{VLR}}$ and $C_{\nu d,\alpha\beta ds}^{\text{VLL}}$, $C_{\nu d,\alpha\beta ds}^{\text{VLR}}$ in (\ref{brkl-text}), there are no relative signs between the scalar WCs. This fact can be understood from the relations among matrix elements shown below 
  \be\label{eq:CP}
  \langle\pi^0|\bar d s|\bar K^0\rangle=\langle\pi^0|\bar s d|K^0\rangle,\qquad  \langle\pi^0|\bar d \gamma_\mu s|\bar K^0\rangle=-\langle\pi^0|\bar s \gamma_\mu d| K^0\rangle.
  \ee
  This observation implies that whereas vector contributions can only make an impact on $\klpn$ in the case of a complex Wilson coefficient, a scalar NP contribution can have an impact on $\klpn$ even when it is CP conserving\footnote{To our knowledge this has been first noticed in \cite{Kiyo:1998zm}.}.

The results presented above demonstrate the different imprints that vector and scalar contributions can give rise to in the $\mathcal{B}(\kpn)-\mathcal{B}(\klpn)$-plane. However, the branching ratios alone will not allow us to identify a possible underlying vector or scalar current. Fortunately,
as pointed out already in \cite{Li2019, Deppisch2020}, it is possible to make progress in this direction with the help of kinematic distributions for $\kpn$ and $\klpn$. Dedicated analysis of kinematic distributions allows for the separation of vector current contributions from scalar ones without specifying a NP model. A first look in this direction
has been made in \cite{Li2019, Deppisch2020}. However, in \cite{Deppisch2020} only the possibility of having a LNV  scalar current and its potential to discriminate from the SM current was discussed. In the following, we go significantly beyond previous treatments by proposing a method to discriminate and quantify different NP contributions systematically. We start by demonstrating \emph{analytically} that it is possible to separate vector, scalar, and tensor contributions from each other using kinematic distributions.

\subsection{Extracting different contributions with neutrino final states only}
\label{subsec:KaonExtraction}
{We now relax the simplifying assumption of lepton-flavor conservation that we imposed in the previous subsection. This opens up the possibility of having contributions from tensor currents in addition to the vector and scalar currents considered so far.}
To simplify the notation in the upcoming equations, we introduce the following abbreviation for the differential partial decay widths
\be
\label{Ds}
\mathcal{D}_L^{\rm exp}(s)\equiv{d\Gamma({K_L\rightarrow \pi^0\nu\widehat{\nu}})\over d s}, \qquad
\mathcal{D}_+^{\rm exp}(s)\equiv{d\Gamma({K^+\rightarrow \pi^+\nu\widehat{\nu}})\over d s},
\ee
where $\nu\widehat{\nu}$ again indicates that both $\nu\bar\nu$ and $\nu\nu$ final states can occur.

Due to the absence of interference between vector, scalar, and tensor contributions in \eqref{brkl-text} and~\eqref{brkp-text}, we can parametrize the differential partial widths as 
\begin{align}
 \mathcal{D}_L^{\rm exp}(s) &= C_S^L f^L_S (s) + C_T^Lf^L_T(s)+C_V^L f^L_V (s)\label{eq:DL},\\
 \mathcal{D}_+^{\rm exp}(s) &= C_S^+ f^+_S (s) + C_T^+ f^+_T (s)+C_V^+ f^+_V (s)\label{eq:D+}\,.
\end{align}
Each term on the r.h.s. of (\ref{eq:DL}) and (\ref{eq:D+}) is the product of an effective Wilson coefficient $C$ and a kinematic function $f(s)$. The former is a sum over Wilson coefficients, while the latter only depends on particle masses and the invariant-mass variable $s$. The effective Wilson coefficients in (\ref{eq:DL}) and (\ref{eq:D+}) read,
\begin{eqnarray}
C_S^L&=&
\sum_{\alpha\leq \beta} \left(1-{1\over2}\delta_{\alpha\beta}\right)\left|C_{\nu d,sd\alpha\beta}^{\text{SLL}}+C_{\nu d,sd\alpha\beta}^{\text{SLR}}
+C_{\nu d,ds\alpha\beta}^{\text{SLL}}+C_{\nu d,ds\alpha\beta}^{\text{SLR}}\right|^2\,,
\\
C_T^L&=&\sum_{\alpha<\beta}\left|C^{\text{TLL}}_{\nu d,\alpha\beta ds}+C^{\text{TLL}}_{\nu d,\alpha\beta sd}\right|^2\,,\\
C_V^L&=&
\sum_{\alpha, \beta}\left(1-{1\over2}\delta_{\alpha\beta}\right) \left|C_{\nu d,sd\alpha\beta}^{\text{VLL}}+C_{\nu d,sd\alpha\beta}^{\text{VLR}}-C_{\nu d, ds\alpha\beta}^{\text{VLL}}-C_{\nu d,ds\alpha\beta}^{\text{VLR}}\right|^2\,,
\\
C_S^+&=&
\sum_{\alpha\leq \beta} \left(1-{1\over2}\delta_{\alpha\beta}\right)\left(\left|C_{\nu d,sd\alpha\beta}^{\text{SLL}}+C_{\nu d,sd\alpha\beta}^{\text{SLR}}\right|^2+\left|C_{\nu d,ds\alpha\beta}^{\text{SLL}}+C_{\nu d,ds\alpha\beta}^{\text{SLR}}\right|^2\right)\,,\label{CS+}
\\
\label{CT+}
C_T^+&=&\sum_{\alpha<\beta} \left( \left|C^{\text{TLL}}_{\nu d,\alpha\beta sd}\right|^2+\left|C^{\text{TLL}}_{\nu d,\alpha\beta ds}\right|^2\right)\\
C_V^+&=&
\sum_{\alpha, \beta}\left(1-{1\over2}\delta_{\alpha\beta}\right) \left|C_{\nu d,sd\alpha\beta}^{\text{VLL}}+C_{\nu d,sd\alpha\beta}^{\text{VLR}}\right|^2,
\end{eqnarray}
and the explicit expressions for the kinematic functions can be read out from Eqs.~(\ref{brkl-text})-(\ref{J2KP-text}), 
\begin{align}
	f_S^{L}(s)&={B_L^2 \over 2^9\pi^3m_{K^0}^3}\cdot s\,\lambda^{1/2}(s,m_{K^0}^2,m_{\pi^0}^2)\, \left| f^{K^0}_0 (s)\right|^2,\\
 f_S^{+}(s)&={B_+^2 \over 2^8\pi^3m_{K^+}^3}\cdot s\,
\lambda^{1/2}(s,m_{K^+}^2,m_{\pi^+}^2)\, \left|f^{K^+}_0(s)\right|^2,\\
f_V^{L}(s)&={1\over 3\cdot 2^{11}\pi^3m_{K^0}^3}\cdot \lambda^{3/2}(s,m_{K^0}^2,m_{\pi^0}^2)\,\left|f^{K^0}_+(s)\right|^2, \\
f_V^{+}(s)&={1\over 3\cdot 2^9\pi^3m_{K^+}^3}\cdot \lambda^{3/2}(s,m_{K^+}^2,m_{\pi^+}^2)\, \left|f^{K^+}_+(s)\right|^2 ,\\
f_T^{L}(s)&={1\over 3\cdot2^{5}\pi^3m_{K^0}^3(m_{K^0}+m_{\pi^0})^2}\cdot s\, \lambda^{3/2}(s,m_{K^0}^2,m_{\pi^0}^2)\,\left|f^{K^0}_T(s)\right|^2, \\ 
f_T^{+}(s)&={1\over 3\cdot2^{4}\pi^3m_{K^+}^3(m_{K^+}+m_{\pi^+})^2}\cdot s\, \lambda^{3/2}(s,m_{K^+}^2,m_{\pi^+}^2)\,\left|f^{K^+}_T(s)\right|^2.
\end{align}

Hence, we find only three free \textit{real} parameters $C_S^L, C_T^L, C_V^L$ and $C_S^+, C_T^+,C_V^+$ for $\klpn$ and $\kpn$, respectively. Importantly, these parameters are independent of any kinematics. 
This implies that measuring the value of the $s$-distribution at three different values $s_1$, $s_2$, and $s_3$ allows us to obtain a set of three simple algebraic equations with three unknowns, namely $C_S^L, C_V^L$, $C_T^L$ and $C_S^+, C_V^+$, $C_T^+$ for $\klpn$  and $\kpn$, respectively. Solving them gives us the magnitude of scalar, vector, and tensor contributions,
\begin{align}
    \label{eq:CS+}
    C_{S}^+&=\frac{ D_+^{\rm exp}(s_1)\left[f_V^+(s_2)f_T^+(s_3)-f_V^+(s_3)f_T^+(s_2)\right]+\text{cyclic}}{f_{S}^{+}(s_1)\left[f_V^+(s_2)f_T^+(s_3)-f_V^+(s_3)f_T^+(s_2)\right]+\text{cyclic}},\\
    \label{eq:CV+}
     C_{V}^+&=\frac{ D_+^{\rm exp}(s_1)\left[f_S^+(s_2)f_T^+(s_3)-f_S^+(s_3)f_T^+(s_2)\right]+\text{cyclic}}{f_{V}^{+}(s_1)\left[f_S^+(s_2)f_T^+(s_3)-f_S^+(s_3)f_T^+(s_2)\right]+\text{cyclic}}, \\
     \label{eq:CT+}
     C_{T}^+&=\frac{ D_+^{\rm exp}(s_1)\left[f_S^+(s_2)f_V^+(s_3)-f_S^+(s_3)f_V^+(s_2)\right]+\text{cyclic}}{f_{T}^{+}(s_1)\left[f_S^+(s_2)f_V^+(s_3)-f_S^+(s_3)f_V^+(s_2)\right]+\text{cyclic}},
\end{align}
where \textit{cyclic} refers to the remaining two terms obtained by one and two applications of $s_1\rightarrow s_2$, $s_2\rightarrow s_3$, $s_3\rightarrow s_1$, respectively. Similarly, we obtain
\begin{align}
     \label{eq:CSL}
    C_{S}^L&=\frac{ D_L^{\rm exp}(s_1)\left[f_V^L(s_2)f_{T}^L(s_3)-f_V^L(s_3)f_{T}^L(s_2)\right]+\text{cyclic}}{f_{S}^{L}(s_1)\left[f_V^L(s_2)f_{T}^L(s_3)-f_V^L(s_3)f_{T}^L(s_2)\right]+\text{cyclic}},\\
    \label{eq:CVL}
     C_{V}^L&=\frac{ D_L^{\rm exp}(s_1)\left[f_{S}^L(s_2)f_{T}^L(s_3)-f_{S}^L(s_3)f_{T}^L(s_2)\right]+\text{cyclic}}{f_{V}^{L}(s_1)\left[f_{S}^L(s_2)f_{T}^L(s_3)-f_{S}^L(s_3)f_{T}^L(s_2)\right]+\text{cyclic}}, \\
     C_{T}^L&=\frac{ D_L^{\rm exp}(s_1)\left[f_{S}^L(s_2)f_V^L(s_3)-f_{S}^L(s_3)f_V^L(s_2)\right]+\text{cyclic}}{f_{T}^{L}(s_1)\left[f_{S}^L(s_2)f_V^L(s_3)-f_{S}^L(s_3)f_V^L(s_2)\right]+\text{cyclic}}.
\end{align}
In Appendix~\ref{appendix:WilsonWOTensor}, we show explicitly how these relations simplify in the absence of tensor contributions.

Plenty of information can be extracted from these results. Finding a non-vanishing value for $C^L_S$ or $C_{S}^{+}$ ($C_{T}^{L}$ or $C_T^+$), signals the presence of scalar (tensor) currents. As outlined in ~\ref{sec:OperatorLNV}, this could be a smoking gun for lepton-number violation. In that case, complementary probes of LNV interactions such as neutrinoless double beta decay or collider experiments are important to unambiguously confirm the underlying new LNV physics. Moreover, if any of the tensor operators have non-vanishing Wilson coefficients, then $C_T^+$ is necessarily non-vanishing as there is no room for destructive interference between different Wilson coefficients in (\ref{CT+}). 

Finding $C_V^L$ and/or $C_V^+$ to be different from their SM values
would automatically indicate new vector currents at work. However, the suggested strategy cannot be used to determine whether these are new left-handed or right-handed currents. In contrast, we will demonstrate in Section~\ref{sec:Bmeson} that this is possible in the case of $B$ decays. 

The important point is that vector, scalar, and tensor currents can be separated from experimental data, under the assumption that these are the only contributions present.\footnote{We are aware of the fact that this proposal will be difficult to execute in the near future, but its simplicity will hopefully motivate the experiments to make efforts in this direction.}. To obtain good statistics, different values of $s$ could be required, and suitable values of $s$ need to be chosen to get sufficient sensitivity to the NP dynamics. An important cross-check will be the $s$-independence of the determined values of the (sum of) Wilson coefficients.

\begin{figure}
\centering
\includegraphics[width=0.49\textwidth]{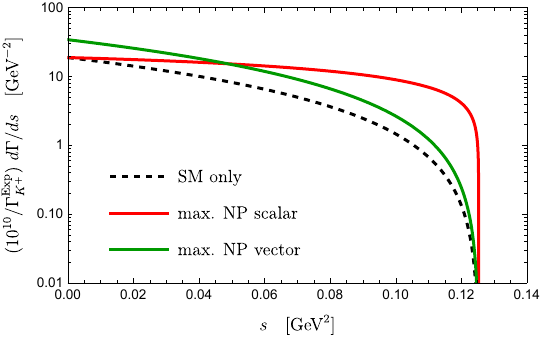}	
\includegraphics[width=0.49\textwidth]{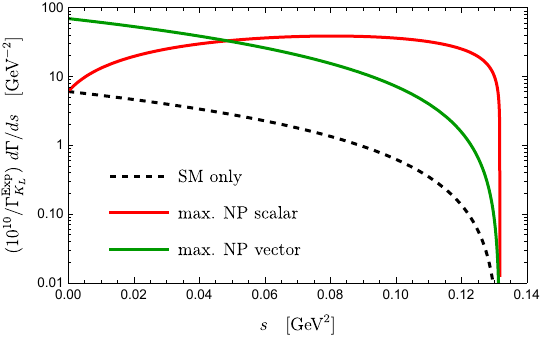}	
\caption{{{Differential distribution of $d\Gamma(\kpn)/ds$ (left plot) and $d\Gamma(\klpn)/ds$ (right plot) normalized to the total experimental decay width $\Gamma_{K^+}^{\mathrm{Exp}}$ and $\Gamma_{K_L}^{\mathrm{Exp}}$, respectively. The SM contribution is depicted by the black dashed curve. For the red (green) curve, we have assumed an additional scalar (vector) contribution on top of the SM contribution, leading to a NP signal around the current experimental upper limit $\mathcal{B}(\kpn)=1.63 \times 10^{-10}$. For the vector (scalar) contribution, a phase of $\phi_V=\pi/2$ $(\phi_S=0)$ was chosen. Lepton-flavor universality is assumed.}}}
\label{fig:DistKLp}
\end{figure}

\paragraph{Probing complex phases with differential distributions}\mbox{}\\ 

For further illustration, we depict $d\Gamma(\kpn)/ds$ (left plot) and $d\Gamma(\klpn)/ds$ (right plot) normalized to the total experimental decay width $\Gamma_{K^+}^{\mathrm{Exp}}$ and $\Gamma_{K_L}^{\mathrm{Exp}}$, respectively, in Fig.~\ref{fig:DistKLp}. The black dashed curve depicts the expected SM contribution, and the red (green) line shows an additional scalar (vector) contribution on top of the SM, leading to a NP signal around the current experimental upper limit $\mathcal{B}(\kpn)=1.63 \times 10^{-10}$.\footnote{For the vector contribution, a phase of $\phi_V=\pi/2$ was chosen. $d\Gamma(\kpn)/ds$ is independent of the scalar phase $\phi_S$, while $\mathcal{B}(\klpn)$ depends on $\phi_S$. For the scalar contribution (red) we fix $\phi_S=0$, leading to the maximal contribution to $\mathcal{B}(\klpn)$.} While the vector contribution follows the shape of the SM contribution, the scalar contribution features a distinct distribution, clearly different from the vector contribution.

{We note that for $K_L$, the NP vector distribution differs significantly
  in magnitude from the SM one for the chosen phase $\phi_V=\pi/2$. The difference between the NP vector distribution and the SM distribution is much less prominent in $K^+$ decays. Similarly,
  the impact of the scalar contribution is much larger in
  $K_L$ decays than in $K^+$ decays for the chosen value of the scalar phase $\phi_S=0$. These observations highlight the importance of precise
  measurements of $s$-distribution for $\klpn$ by KOTO.}

Finally, we have some remarks about opportunities to probe the scalar phase $\phi_S$. While the scalar differential distribution $d\Gamma(\kpn)/ds$ is independent of $\phi_S$,  d$\Gamma(\klpn)/ds$ depends on it, as seen in Eq.~(\ref{B-scalar-kaons}).
Therefore, a combined analysis of the differential distributions for $d\Gamma(\kpn)/ds$ and $d\Gamma(\klpn)/ds$ is very powerful. In case there is a hint towards a new scalar contribution, a comparison of $d\Gamma(\kpn)/ds$ and $d\Gamma(\klpn)/ds$ could tell us if it features a non-zero scalar phase $\phi_S$. Hence, a non-zero scalar phase $\phi_S$ can be probed at the level of branching ratios, as shown in Fig.~\ref{fig:KaonVectorScalar}, and at the level of differential distributions. This is particularly interesting as a non-zero $\phi_S$ would signal a new source of CP violation, which could have interesting consequences for leptogenesis~\cite{Sakharov:1967dj,Fukugita1986, Hagedorn:2017wjy}.

\subsection{Disentangling LEFT from dark-sector physics}
\label{subsec:LNVvsDarkKaons}
{Our analysis thus far assumes that the invisible final states are neutrinos only. A priori, the final state in the NP contribution could also be a single boson, or two or more dark-sector particles~\cite{Fridell:2023ssf,Bolton:2024egx}. The first possibility would simply result in kinematic distributions that follow the SM, but with an additional sharp peak around a fixed value for $s$, where the latter is determined by the mass of the boson. Hence, a NP scenario where the final state is a single boson would in principle be easily distinguishable from NP with two neutrinos in the final state, which in general exhibit kinematic distributions that are nonzero for arbitrary kinematically-allowed values of $s$. The problem of disentangling NP scenarios with two dark-sector particles in the final state from NP induced by the (SM)LEFT operators in Eq.~(\ref{VLL})-(\ref{TLL}) is more subtle, and it is the subject of the following discussion. }

(D)LEFT operators relevant for rare meson decays have been classified in Ref.~\cite{He2022}, and we list them in Appendix.~\ref{appendix:DarkOperators} for the reader's convenience. The classification includes spin $0$, spin $1/2$, and spin $1$ dark particles. In the spin $1$ scenario, the following two cases are considered separately: The dark field is represented by the vector potential $X_{\mu}$ (scenario A), or the field strength tensor $X_{\mu\nu}=\partial_{\mu}A_{\nu}-\partial_{\nu}A_{\mu}$ (scenario B)~\cite{He2022}. In scenario A, the differential decay widths are plagued by divergences in the limit $m\rightarrow 0$, which is an attribute of the longitudinal part in the polarization sum. This is a well-known problem and can be cured operationally by requiring the relevant Wilson coefficients to contain an explicit factor of $m$ to the minimum power necessary to cancel potential divergences in the limit $m\rightarrow 0$, see~\cite{He2022} for details and examples. As the problems originate from the longitudinal polarization sum, a naive approach would be to consider an exactly massless vector field instead, which only has transverse polarization in the first place. It is, however, a well-known result in QFT that massless vector field theories are gauge theories~\cite{Weinberg:1995mt}. Hence, if we assume that the SM quarks are uncharged under the dark gauge symmetry, then the vector enters the effective operators through its field-strength tensor only, which is captured by scenario B. 

The contributions to $\klpn$ and $\kpn$ from (D)LEFT operators have readily been worked out in~\cite{He2022}. 
We have verified and again refined their results by taking into account additional $s$-dependence and isospin-breaking effects. 
We report on these results in detail in Appendix~\ref{appendix:KaonDecays}, where we list all relevant differential decay widths. In the following, we use these results to address differences in (SM)LEFT NP and (D)LEFT NP. All of our results are in the limit of vanishing DM mass, which yields nonzero contributions to kinematic distributions at small values for $s$ analogous to the case of active neutrinos.

The underlying idea for disentangling (SM)LEFT from (D)LEFT in rare meson decays relies on the ability to disentangle the kinematic structures that arise from different operators that contribute to the relevant differential decay widths. For example, from (\ref{J1KL-text}) and (\ref{J1KP-text}), we see that the kinematic functions induced by the scalar-current operators in $d\Gamma(\klpn)/ds$ and $d\Gamma(\kpn)/ds$ are proportional to
\begin{align}
    s\,\lambda^{1/2}(s,m_{K_L}^2,m_{\pi^0}^2)\, \left| f^{K^0}_0(s) \right|^2, \qquad \quad s
\,\lambda^{1/2}(s,m_{K^+}^2,m_{\pi^+}^2)\, \left|f^{K^+}_0(s)\right|^2,
\end{align}
respectively. To be precise, we will call kinematic functions with an overall $s$-independent prefactor stripped off, as in the equation above, {\em kinematic structures}. In Table~\ref{tab:KaonTab}, we list the kinematic structures in $d\Gamma(\klpn)/ds$ and $d\Gamma(\kpn)/ds$ for all (SM)LEFT and (D)LEFT operators. The results were obtained by calculating the contribution from each operator to the differential decay widths, and we provide the full expressions for the latter in Appendix~\ref{appendix:KaonDecays}.\footnote{We have cross-checked our results against available results in the literature when possible~\cite{Li2019,He2022,He2023}.}

\setlength{\extrarowheight}{4pt}
\begin{table}[t!]
    \centering
    \begin{tabular}{|>{\raggedright}m{78pt}|>{\raggedright}m{63pt}|>{\centering}m{220pt}|c}
    \cline{1-3}
    Type & Operator & Kinematic structure in $K\rightarrow \pi\,+\slashed{E}$ & \\[2pt]
    \cline{1-3}
    \multirow{3}{*}{(SM)LEFT} & $\mathcal{O}^{\text{SLL}}_{\nu d}$, $\mathcal{O}^{\text{SLR}}_{\nu d}$ & \cellcolor{blue!25}$s\,\lambda^{1/2}\left|f_0^K\right|^2$& \\[2pt]   
     \cline{2-3}
    & $\mathcal{O}^{\text{VLL}}_{\nu d}$, $\mathcal{O}^{\text{VLR}}_{\nu d}$ & \cellcolor{red!25}$\lambda^{3/2}\left|f_+^K\right|^2$ & \\[2pt]
     \cline{2-3}
     & $\mathcal{O}^{\text{TLL}}_{\nu d}$ & \cellcolor{Green!35}$s\,\lambda^{3/2}\left|f_T^K\right|^2$ & \\[2pt]
     \cline{1-3}
    \multirow{2}{*}{Scalar DM} & $\mathcal{O}_{sd\phi}^S$  & $\lambda^{1/2}\left|f_0^K\right|^2$ & \\[2pt]
     \cline{2-3}
      & $\mathcal{O}_{sd\phi}^V$ & \cellcolor{red!25}$\lambda^{3/2}\left|f_+^K\right|^2$ & \\[2pt]
      \cline{1-3}
    \multirow{3}{*}{Fermion DM} & $\mathcal{O}_{sd\chi1}^{S}, \mathcal{O}_{sd\chi2}^{S}$ & \cellcolor{blue!25}$s\,\lambda^{1/2}\left|f_0^K\right|^2$& \\[2pt]   
     \cline{2-3}
    & $\mathcal{O}_{sd\chi1}^{V}, \mathcal{O}_{sd\chi2}^{V}$ &\cellcolor{red!25} $\lambda^{3/2}\left|f_+^K\right|^2$ & \\[2pt]
     \cline{2-3}
     & $\mathcal{O}_{sd\chi1}^{T}, \mathcal{O}_{sd\chi2}^{T}$ &\cellcolor{Green!35} $s\,\lambda^{3/2}\left|f_T^K\right|^2$ & \\[2pt]
     \cline{1-3}
    \multirow{5}{*}{Vector DM: A} & $\mathcal{O}^S_{sdA}$ & $s^2\,\lambda^{1/2}\left|f_0^K\right|^2$& \\[2pt]   
     \cline{2-3}
    & $\mathcal{O}^V_{sdA2}$ & $s^2\,\lambda^{1/2}\left|f_0^K\right|^2$ & \\[2pt]
     \cline{2-3}
     & $\mathcal{O}^V_{sdA3}, \mathcal{O}^V_{sdA6}$ & $s\,\lambda^{3/2}\left|f_+^K\right|^2$ & \\[2pt]
     \cline{2-3}
    & $\mathcal{O}^V_{sdA4}, \mathcal{O}^V_{sdA5}$ & $s^2\,\lambda^{3/2}\left|f_+^K\right|^2$ & \\[2pt]
     \cline{2-3}
     & $\mathcal{O}^T_{sdA1}$ & $s^2\,\lambda^{3/2}\left|f_T^K\right|^2$ & \\[2pt]
     \cline{1-3}
     \multirow{2}{*}{Vector DM: B} & $\mathcal{O}^S_{sdB1}, \mathcal{O}^S_{sdB2}$ & $s^2\,\lambda^{1/2}\left|f_0^K\right|^2$& \\[2pt]   
     \cline{2-3}
    & $\mathcal{O}^T_{sdB1}, \mathcal{O}^T_{sdB2}$ & $s^2\,\lambda^{3/2}\left|f_T^K\right|^2$ & \\[2pt]
     \cline{1-3}
    \end{tabular}
    \caption{List of operators contributing to $K^{+(L)}\rightarrow \pi^{+(0)}\,+\slashed{E}$ and the kinematic structure they give rise to in $d\Gamma(K^{+(L)}\rightarrow \pi^{+(0)}\,+\slashed{E})/ds$. Here $\lambda$ is shorthand notation for $\lambda(m_K^2,m_{\pi}^2,s)$, and $f_0^K$ is short for $f_0^K(s)$, etc. We have only listed operators that give a contribution to $K^{+(L)}\rightarrow \pi^{+(0)}\,+\slashed{E}$ (see text for details).}
    \label{tab:KaonTab}
\end{table}

From Table~\ref{tab:KaonTab}, we see that in contrast to massless dark scalars and massless dark vectors, only very light or massless dark fermions can generate the same kinematic structure as $\mathcal{O}^{\text{SLL}}_{\nu d}$ and $\mathcal{O}^{\text{SLR}}_{\nu d}$, which we have highlighted in blue. Hence, {in the limit where only one operator can be nonzero at a time,} it is possible to distinguish $\mathcal{O}^{\text{SLL}}_{\nu d}$ and $\mathcal{O}^{\text{SLR}}_{\nu d}$ from any dark LEFT operator with a dark scalar or a dark vector particle. In the case of $\mathcal{O}^{\text{TLL}}_{\nu d}$, we see that only a dark fermion can replicate an identical kinematic structure, as highlighted in green. Finally, we observe that both dark-fermion operators with vector currents and dark-scalar operators with vector currents generate a kinematic structure identical to that of $\mathcal{O}^{\text{VLL}}_{\nu d}$ and $\mathcal{O}^{\text{VLR}}_{\nu d}$, as highlighted in red.\footnote{Only in the case of type $A$ dark vectors, $d\Gamma(K^{+(L)}\rightarrow \pi^{+(0)}\,+\slashed{E})/ds$ receives contributions from interference terms between different WCs. Full expressions for $d\Gamma(K^{+(L)}\rightarrow \pi^{+(0)}\,+\slashed{E})/ds$ are included in Appendix~\ref{appendix:KaonDecays}, but here we note that none of the interference terms have a kinematic structure similar to that of any LEFT operator.} 

Having analyzed kinematic structures from one operator at a time, it is natural to ask how our findings are modified if we allow for multiple operators to be present simultaneously. We will address this question in the next section in the context of $B$-mesons.

\section{$B$-meson decays}
\label{sec:Bmeson}
In the preceding sections, we have shown how powerful kinematic structures can be in discriminating between NP sources to rare Kaon decays. In the following sections, we proceed to the even richer story of $B$-meson decays. Here, the decay channel $B\rightarrow K^*+\slashed{E}$ provides even better opportunities for disentangling NP~\cite{Altmannshofer:2009ma}. We also provide a detailed discussion of whether contributions from (SM)LEFT can be disentangled from contributions from (D)LEFT when any number of (SM)LEFT and (D)LEFT operators have non-vanishing Wilson coefficients. 

\subsection{Current experimental limits and SM predictions}
The current experimental bounds on $b\rightarrow s\Bar{\nu}\nu$, including the most recent results from Belle II \cite{Belle-II:2023esi}, read~\cite{Olive:2016xmw,Grygier:2017tzo}
\begin{align}
\label{BelleIIValue}
{\mathcal{B}}(B^+\to K^+\nu\bar\nu) &={(13\pm4)}\times 10^{-6},\\
{\mathcal{B}}(B^0\to K^0\nu\bar\nu) &\leq 2.6\times 10^{-5}\quad \text{@ 90\% CL}~,\\
{\mathcal{B}}(B^+\to  K^{+*} \nu\bar\nu) &\leq 4.0\times 10^{-5}\quad \text{@ 90\% CL}~,\\
\label{BKstarExLimit}
{\mathcal{B}}(B^0\to K^{0*}\nu\bar\nu) &\leq 1.8\times 10^{-5}\quad \text{@ 90\% CL}\,.
\end{align}
These bounds should be compared with the following SM predictions based on form factors
from the HPQCD-2022 collaboration~\cite{Parrott:2022zte,Parrott:2022rgu,Parrott:2022smq} for $B^+\to K^+\nu\bar\nu$ and BSZ 2015~\cite{Bharucha:2015bzk} for
$B^0\to K^{0*}\nu\bar\nu$
\begin{align}\label{BVNEW}
{\mathcal{B}}(B^+\to K^+\nu\bar\nu)_{\rm SM}^{\rm SD} &={(4.92\pm 0.30)\times 10^{-6}},\\
\label{BVNEW2}
{\mathcal{B}}(B^0\to K^{0*}\nu\bar\nu)_{\rm SM} &= {(10.11\pm 0.96)\times 10^{-6}},
\end{align}
where the superscript ``SD'' in (\ref{BVNEW}) indicates that the tree-level long-distance contribution pointed out in \cite{Kamenik:2009kc} has been left out, as in the Belle II result.\footnote{By keeping the tree-level long-distance contribution pointed out in \cite{Kamenik:2009kc}, one would obtain ${\mathcal{B}}(B^+\to K^+\nu\bar\nu)_{\rm SM}=(5.53\pm 0.30)\times 10^{-6}$ instead. The latter value is due to the HPQCD22 collaboration~\cite{Parrott:2022rgu}, which included the long-distance contribution.}

The SM results quoted above were obtained using the value $\vcb = 42.6(4)\times 10^{-3}$ from \cite{Buras:2022wpw,Buras:2022qip}.
It should be kept in mind that the branching ratios in the SM (\ref{BVNEW})-(\ref{BVNEW2}) depend quadratically on $\vcb$, where the latter is subject to known tensions between its inclusive and exclusive determinations \cite{Bordone:2021oof,FlavourLatticeAveragingGroupFLAG:2021npn, Finauri:2023kte}. This problem has been studied in detail in  \cite{Buras:2021nns,Buras:2022wpw,Buras:2022qip, Stangl:2024} in the context of rare $K$ and $B$ decays. In particular, the dependence of the results (\ref{BVNEW})-(\ref{BVNEW2}) on form factors has been examined very recently in~\cite{Stangl:2024}: As an example, using the average of the $B\to K$ form factors from HPQCD~2013~\cite{Bouchard:2013eph}, FNAL+MILC~2015~\cite{Bailey:2015dka}, and HPQCD~2022 (as presented in GRvDV~2023~\cite{Gubernari:2023puw}), one finds $\mathcal{B}(B^+\to K^+\nu\bar\nu)_{\rm SM}^{\rm SD}=(4.85\pm 0.23)\times 10^{-6}$, in perfect agreement with the result in (\ref{BVNEW}).

Comparing the SM predictions in (\ref{BVNEW})-(\ref{BVNEW2}) with experimental data, it is evident that there is still significant room left for NP contributions. For some recent analyses see \cite{Bause:2021cna,He:2021yoz,Bause:2022rrs,Becirevic:2023aov,Bause:2023mfe,Allwicher:2023syp,Dreiner:2023cms, Altmannshofer:2023hkn, Gabrielli:2024wys, Hou:2024vyw, He:2024iju, Bolton:2024egx, Marzocca:2024hua, Ho:2024cwk, McKeen:2023uzo}. In anticipation of further experimental results in the years to come, we now direct our focus on strategies to disentangle NP contributions in $b\rightarrow s\Bar{\nu}\nu$ observables.

\subsection{Disentangling new physics with neutrino final states only}
\label{sec:BExtraction}
It is straightforward to generalize the procedure of extracting NP contributions put forward for rare Kaon decays in Section~\ref{subsec:KaonExtraction} to the case of $B$-mesons. Although the experimental limits are weaker for the latter, the potential to identify the nature of new contributions in $B$-meson decays is greater than in Kaon decays, due to their angular distributions in $B\rightarrow K^*+\slashed{E}$ ($F_L$ and $F_T$). In particular, below we show how dedicated measurements of angular variables can distinguish between e.g. left-handed quark current contributions and right-handed ones, which is not possible in the case of $K\to\pi\nu\hat{\nu}$. We also consider differential decay distributions for $B\rightarrow K+\slashed{E}$, and $B\rightarrow X_s+\slashed{E}$.

\subsubsection{Disentangling different contributions in $B\rightarrow K^*\widehat{\nu}\nu$}
$B$-mesons can decay into an intermediate $K^*$, which subsequently decays into $K\pi$. The differential decay width with a longitudinally (transversely) polarized $K^*$, denoted as $d\Gamma_{L(T)}/ds$, can be extracted by an angular analysis of the $K^*$ decay products. Hence, it is experimentally meaningful to introduce the following quantities

\begin{align}
    P_L^{\rm exp}(s)\equiv\frac{d\Gamma_L}{ds},
    \qquad
    P_T^{\rm exp}(s)\equiv\frac{d\Gamma_T}{ds},
\end{align}
measuring the differential decay width for given polarization in $K^*$.
Using the notation introduced in Section~\ref{subsec:KaonExtraction}, we parameterize them as follows, 
\begin{align}
    P_T^{\rm exp}(s)&=f_{V+}^T(s)\times\sum_{\alpha,\beta}\left(1-{1\over2}\delta_{\alpha\beta}\right) \lvert C^{\text{VLL}}_{\nu d,\alpha\beta sb}+C^{\text{VLR}}_{\nu d,\alpha\beta sb}\rvert^2\nonumber \\
    &+f_{V-}^T(s)\times\sum_{\alpha,\beta}\left(1-{1\over2}\delta_{\alpha\beta}\right) \lvert C^{\text{VLL}}_{\nu d,\alpha\beta sb}-C^{\text{VLR}}_{\nu d,\alpha\beta sb}\rvert^2\nonumber \\
    &+f_T^T(s)\times\sum_{\alpha<\beta}\left(\lvert C^{\text{TLL}}_{\nu d,\alpha\beta bs}\rvert^2+\lvert C^{\text{TLL}}_{\nu d,\alpha\beta sb}\rvert^2\right), \nonumber \\
    \label{PTWilson}
    &\equiv f_{V+}^T(s) C_{V+}^T+f_{V-}^T(s) C_{V-}^T+f_T^T(s) C_T^T,
\end{align}
and
\begin{align}
     P_L^{\rm exp}(s)&=f_{V-}^L(s)\times\sum_{\alpha,\beta}\left(1-{1\over2}\delta_{\alpha\beta}\right) \lvert C^{\text{VLL}}_{\nu d,\alpha\beta sb}-C^{\text{VLR}}_{\nu d,\alpha\beta sb}\rvert^2\nonumber \\
     &+f_T^L(s)\times\sum_{\alpha<\beta}  \left(\lvert C^{\text{TLL}}_{\nu d,\alpha\beta bs}\rvert^2+\lvert C^{\text{TLL}}_{\nu d,\alpha\beta sb}\rvert^2\right)\nonumber \\
    &+f_S^L(s)\times\sum_{\alpha\leq\beta}\left(1-\frac{1}{2}\delta_{\alpha\beta}\right)\left[\lvert C^{\text{SLR}}_{\nu d,\alpha\beta sb}-C^{\text{SLL}}_{\nu d,\alpha\beta s b}\rvert^2+\lvert C^{\text{SLR}}_{\nu d,\alpha\beta bs}-C^{\text{SLL}}_{\nu d,\alpha\beta bs}\rvert^2\right]\nonumber \\
    \label{PLWilson}
    &\equiv f_{V-}^L(s)C_{V-}^L+f_T^L(s)C^L_T+f_S^L(s)C_S^L,
\end{align}
where the kinematic functions are given by~\cite{Felkl2021}
\begin{align}
    \label{fvPlusT}
    \frac{1}{\Gamma_{B^0}^{\rm{exp}}}\int ds\,f_{V+}^T(s)&=\frac{1}{\Gamma_{B^0}^{\rm{exp}}}\,\frac{1}{768\pi^3m_B^3(m_B+m_{K^*})^2}\int ds\, s\,\lambda^{3/2}(m_B^2,m_{K^*}^2,s)\left|V_0(s)\right|^2 \nonumber \\
    &=0.6\,G_F^{-2},\\
    \frac{1}{\Gamma_{B^0}^{\rm{exp}}}\int ds\,f_{V-}^T(s)&=\frac{1}{\Gamma_{B^0}^{\rm{exp}}}\frac{(m_B+m_{K^*})^2}{768\pi^3m_B^3}\int ds\, s\,\lambda^{1/2}(m_B^2,m_{K^*}^2,s)\left|A_1(s)\right|^2\nonumber \\
    &=1.2\,G_F^{-2},\\
     \frac{1}{\Gamma_{B^0}^{\rm{exp}}}\int ds\,f_T^T(s)&=\frac{1}{\Gamma_{B^0}^{\rm{exp}}}\frac{1}{24\pi^3m_B^3}\int ds\,\lambda^{1/2}(m_B^2,m_{K^*}^2,s)\left[\lambda(m_B^2,m_{K^*}^2,s)\left|T_1(s)\right|^2\right.\nonumber \\
     &\left.+(m_B^2-m_{K^*}^2)^2\left|T_2(s)\right|^2\right]=178\,G_F^{-2}, \\
     \frac{1}{\Gamma_{B^0}^{\rm{exp}}}\int ds\,f_{V-}^L(s)&= \frac{1}{\Gamma_{B^0}^{\rm{exp}}}\frac{m_B^2m_{K^*}^2}{24\pi^3m_B^3}\int ds\, \lambda^{1/2}(m_B^2,m_{K^*}^2,s)\left|A_{12}(s)\right|^2=1.7\,G_F^{-2},\\
     \frac{1}{\Gamma_{B^0}^{\rm{exp}}}\int ds\,f_T^L(s)&= \frac{1}{\Gamma_{B^0}^{\rm{exp}}}\frac{m_B^2m_{K^*}^2}{3\pi^3m_B^3(m_B+m_{K^*})^2}\int ds\, s\,\lambda^{1/2}(m_B^2,m_{K^*}^2,s)\left|T_{23}(s)\right|^2\nonumber \\
     &=22.6\,G_F^{-2},\\
     \label{fsL}
    \frac{1}{\Gamma_{B^0}^{\rm{exp}}}\int ds\,f_S^{L}(s)&=\frac{1}{\Gamma_{B^0}^{\rm{exp}}}\frac{1}{256\pi^3m_B^3(m_b+m_s)^2}\int ds\, s\,\lambda^{3/2}(m_B^2,m_{K^*}^2,s)\left|A_0(s)\right|^2\nonumber \\
    &=\left(\frac{4.19\,\text{GeV}}{m_b}\right)^2 3.5\,G_F^{-2},
\end{align}
and the numerical values are obtained using the parametrization of the form factors $V_0(s),\,A_1(s),$ etc. in Appendix~\ref{appendix:formfactors}.
The combination of Wilson coefficients in (\ref{PTWilson}) are then determined as,
\begin{align}
    \label{CV+T}
    C_{V+}^T&=\frac{ P_T^{\rm exp}(s_1)\left[f_{V-}^T(s_2)f_{T}^T(s_3)-f_{V-}^T(s_3)f_{T}^T(s_2)\right]+\text{cyclic}}{f_{V+}^{T}(s_1)\left[f_{V-}^T(s_2)f_{T}^T(s_3)-f_{V-}^T(s_3)f_{T}^T(s_2)\right]+\text{cyclic}},\\
    \label{CV-T}
     C_{V-}^T&=\frac{ P_T^{\rm exp}(s_1)\left[f_{V+}^T(s_2)f_{T}^T(s_3)-f_{V+}^T(s_3)f_{T}^T(s_2)\right]+\text{cyclic}}{f_{V-}^{T}(s_1)\left[f_{V+}^T(s_2)f_{T}^T(s_3)-f_{V+}^T(s_3)f_{T}^T(s_2)\right]+\text{cyclic}}, \\
     C_{T}^T&=\frac{ P_T^{\rm exp}(s_1)\left[f_{V+}^T(s_2)f_{V-}^T(s_3)-f_{V+}^T(s_3)f_{V-}^T(s_2)\right]+\text{cyclic}}{f_{T}^{T}(s_1)\left[f_{V+}^T(s_2)f_{V-}^T(s_3)-f_{V+}^T(s_3)f_{V-}^T(s_2)\right]+\text{cyclic}},
\end{align}
where \textit{cyclic} refers to the remaining two terms obtained by one and two applications of $s_1\rightarrow s_2$, $s_2\rightarrow s_3$, $s_3\rightarrow s_1$, respectively. Similarly, the combination of Wilson coefficients in (\ref{PLWilson}) are given by
\begin{align}
    \label{CSL}
    C_{S}^L&=\frac{ P_L^{\rm exp}(s_1)\left[f_{V-}^L(s_2)f_{T}^L(s_3)-f_{V-}^L(s_3)f_{T}^L(s_2)\right]+\text{cyclic}}{f_{S}^{L}(s_1)\left[f_{V-}^L(s_2)f_{T}^L(s_3)-f_{V-}^L(s_3)f_{T}^L(s_2)\right]+\text{cyclic}},\\
    \label{CVL}
     C_{V-}^L&=\frac{ P_L^{\rm exp}(s_1)\left[f_{S}^L(s_2)f_{T}^L(s_3)-f_{S}^L(s_3)f_{T}^L(s_2)\right]+\text{cyclic}}{f_{V-}^{L}(s_1)\left[f_{S}^L(s_2)f_{T}^L(s_3)-f_{S}^L(s_3)f_{T}^L(s_2)\right]+\text{cyclic}}, \\
     \label{CTL}
     C_{T}^L&=\frac{ P_L^{\rm exp}(s_1)\left[f_{S}^L(s_2)f_{V-}^L(s_3)-f_{S}^L(s_3)f_{V-}^L(s_2)\right]+\text{cyclic}}{f_{T}^{L}(s_1)\left[f_{S}^L(s_2)f_{V-}^L(s_3)-f_{S}^L(s_3)f_{V-}^L(s_2)\right]+\text{cyclic}}.
\end{align}
In Appendix~\ref{appendix:WilsonWOTensor}, we show how these expressions simplify in the absence of tensor contributions.

The information encoded in the coefficients (\ref{CV+T})-(\ref{CTL}) can also be extracted from dedicated measurements of the differential decay width of $B\rightarrow K^*\nu\widehat{\nu}$, which is the sum of $P_T$ and $P_L$. The advantage of analyzing $P_T$ and $P_L$ separately is that this approach offers additional consistency checks; the measured values of the coefficients $C_{T}^{L}$ and $C_{T}^{T}$ as well as $C_{V-}^L$ and $C_{V-}^T$ should be equal, c.f.~\eqref{PTWilson} and \eqref{PLWilson}.

The important final point is that scalar, tensor, left-handed vector, and right-handed vector currents can be distinguished: In contrast to Kaon decays (and $B\rightarrow K\widehat{\nu}\nu$, as we will see below), left-handed quark current contributions can be distinguished from right-handed quark current contributions by comparing $C_{V-}^{L/T}$ and $C_{V+}^L$, c.f.~\eqref{PTWilson} and \eqref{PLWilson}. 
Finally, a non-vanishing value for $C^L_S$ ($C_{T}^{L}$ and $C_{T}^{T}$) would signal the presence of scalar (tensor) currents, which could be a sign of LNV.

\subsubsection{Complementarity between different decay channels: $B\rightarrow K\nu\widehat{\nu}$}

One of the great outcomes of analyzing the decay channel $B\rightarrow K^*\nu\hat{\nu}$, is the realization that one can separate left-handed from right-handed vector currents. The same is not true for scalar currents, as we only gain access to the combination
\begin{align}
    \sum_{\alpha\leq\beta}\left(1-\frac{1}{2}\delta_{\alpha\beta}\right)\left[\lvert C^{\text{SLR}}_{\nu d,\alpha\beta sb}-C^{\text{SLL}}_{\nu d,\alpha\beta s b}\rvert^2+\lvert C^{\text{SLR}}_{\nu d,\alpha\beta bs}-C^{\text{SLL}}_{\nu d,\alpha\beta bs}\rvert^2\right].
\end{align}
Therefore, to gain further information about the chirality of scalar currents, it turns out to be very useful to consider the decay channel $B\rightarrow K\nu\widehat{\nu}$. Defining 
\begin{align}
    \mathcal{D}_{BK}^{\rm exp}(s)\equiv{d\Gamma({B\rightarrow K\nu\widehat{\nu}})\over d s},
\end{align}
we can parameterize the partial differential decay width of $B\rightarrow K\nu\widehat{\nu}$ as,\footnote{The analysis of $B\rightarrow K\nu\widehat{\nu}$ is completely analogous to the analysis of $K^+\rightarrow \pi^+\nu\widehat{\nu}$, and the results below are obtained directly from (\ref{eq:D+}) and (\ref{CS+})-(\ref{eq:CT+}) by appropriate replacement of form factors and quark flavors.}
\begin{align}
    \mathcal{D}_{BK}^{\rm exp}(s) &= C_S^{BK} f^{BK}_S (s) + C_T^{BK} f^{BK}_T (s)+C_V^{BK} f^{BK}_V (s)\label{eq:B+}\,,
\end{align}
where 
\begin{align}
    C_S^{BK}&=
\sum_{\alpha\leq \beta} \left(1-{1\over2}\delta_{\alpha\beta}\right)\left(\left|C_{\nu d,\alpha\beta bs}^{\text{SLL}}+C_{\nu d,\alpha\beta bs}^{\text{SLR}}\right|^2+\left|C_{\nu d,\alpha\beta sb}^{\text{SLL}}+C_{\nu d,\alpha\beta sb}^{\text{SLR}}\right|^2\right)\,,\label{CSB}
\\
C_T^{BK}&=\sum_{\alpha<\beta} \left( \left|C^{\text{TLL}}_{\nu d,\alpha\beta bs}\right|^2+\left|C^{\text{TLL}}_{\nu d, 
\alpha\beta sb}\right|^2\right),\\
\label{CVB}
C_V^{BK}&=
\sum_{\alpha, \beta}\left(1-{1\over2}\delta_{\alpha\beta}\right) \left|C_{\nu d,\alpha\beta sb}^{\text{VLL}}+C_{\nu d,\alpha\beta sb}^{\text{VLR}}\right|^2,
\end{align}
and the magnitude of the integrated currents using form factors from Appendix~\ref{appendix:formfactors} are given below
\begin{align}
\frac{1}{\Gamma^{\rm{exp}}_{B^+}}\int ds\,f^{BK}_S (s)&=10.4\,G_F^{-2}, \\
    \frac{1}{\Gamma^{\rm{exp}}_{B^+}}\int ds\,f^{BK}_T (s)&=17.7\,G_F^{-2}, \\
    \frac{1}{\Gamma^{\rm{exp}}_{B^+}}\int ds\,f^{BK}_V (s)&=2.2\,G_F^{-2}.
\end{align}
The combinations $C_S^{BK},\,C_T^{BK},\,C_V^{BK}$ of Wilson coefficients can then be extracted as\footnote{It is worth noting that measuring both $B\rightarrow K^*\nu\widehat{\nu}$ and $B\rightarrow K\nu\widehat{\nu}$ would offer the following consistency checks, $C_V^{BK}=C^T_{V+}$ and $C_T^{BK}=C_T^T=C_T^L$.}
\begin{align}
    \label{CSBK}
    C_{S}^{BK}&=\frac{ D_{BK}^{\rm exp}(s_1)\left[f_V^{BK}(s_2)f_T^{BK}(s_3)-f_V^{BK}(s_3)f_T^{BK}(s_2)\right]+\text{cyclic}}{f_{S}^{BK}(s_1)\left[f_V^{BK}(s_2)f_T^{BK}(s_3)-f_V^{BK}(s_3)f_T^{BK}(s_2)\right]+\text{cyclic}},\\
     C_{V}^{BK}&=\frac{ D_{BK}^{\rm exp}(s_1)\left[f_S^{BK}(s_2)f_T^{BK}(s_3)-f_S^{BK}(s_3)f_T^{BK}(s_2)\right]+\text{cyclic}}{f_{V}^{BK}(s_1)\left[f_S^{BK}(s_2)f_T^{BK}(s_3)-f_S^{BK}(s_3)f_T^{BK}(s_2)\right]+\text{cyclic}}, \\
    \label{CTBK}
     C_{T}^{BK}&=\frac{ D_{BK}^{\rm exp}(s_1)\left[f_S^{BK}(s_2)f_V^{BK}(s_3)-f_S^{BK}(s_3)f_V^{BK}(s_2)\right]+\text{cyclic}}{f_{T}^{BK}(s_1)\left[f_S^{BK}(s_2)f_V^{BK}(s_3)-f_S^{BK}(s_3)f_V^{BK}(s_2)\right]+\text{cyclic}}.
\end{align}
In summary, by extracting both the coefficients $C_S^L$ from $B\rightarrow K^*\nu\widehat{\nu}$ and $C_{S}^{BK}$ from $B\rightarrow K\nu\widehat{\nu}$, it is possible to determine if there are both left and right-handed scalar currents present or only currents with one definite handedness, c.f.~\eqref{PLWilson} and \eqref{CSB}.
In the latter case, it is not possible to use kinematic distributions to determine if the scalar NP is left-handed or right-handed. However, finding $C_S^L=0$ would imply
  $C_{\nu d,\alpha\beta bs}^{\text{SLL}}=C_{\nu d,\alpha\beta bs}^{\text{SLR}}$ and purely scalar quark currents (i.e. parity-symmetric quark currents). Measuring $C_{S}^{BK}=0$ and consequently $C_{\nu d,\alpha\beta bs}^{\text{SLL}}=-C_{\nu d,\alpha\beta bs}^{\text{SLR}}$, would imply purely pseudoscalar quark currents (i.e. parity-antisymmetric quark currents).

\subsubsection{Complementarity between different decay channels: $B\rightarrow X_s\nu\widehat{\nu}$}

We have also considered the inclusive decay channel $B\rightarrow X_s\nu\widehat{\nu}$ and give the full result for the differential decay rate in Appendix~\ref{appendix:Inclusive}. The result can be expressed as, 
\begin{align}
    \label{Xs-decaywidth}
    \frac{d\Gamma(B\rightarrow X_s\nu\widehat{\nu})}{ds}=C^S_1J^S_1(s)+C^S_2J^S_2(s)+C^V_1J^V_1(s)+C^V_2J^V_2(s)+C^TJ^T(s),
\end{align}
where the effective WCs read
\begin{align}
    \label{CS1}
    C^S_1&=\sum_{\alpha\leq \beta}\left(1-\frac{\delta_{\alpha\beta}}{2}\right)\left(\left|C^{\text{SLL}}_{\nu d,\alpha\beta bs}\right|^2+\left|C^{\text{SLL}}_{\nu d,\alpha\beta sb}\right|^2+\left|C^{\text{SLR}}_{\nu d,\alpha\beta bs}\right|^2+\left|C^{\text{SLR}}_{\nu d,\alpha\beta sb}\right|^2\right), \\
    \label{CS2}
    C^S_2&=\sum_{\alpha\leq \beta}\left(1-\frac{\delta_{\alpha\beta}}{2}\right){\text{Re}}\left(C_{\nu d,\alpha\beta bs}^{\text{SLL}}C_{\nu d,\alpha\beta bs}^{\text{SLR}*}+C_{\nu d,\alpha\beta sb}^{\text{SLL}}C_{\nu d,\alpha\beta sb}^{\text{SLR}*}\right),\\
    \label{CV1}
    C^V_1&=\sum_{\alpha, \beta}\left(1-{1\over2}\delta_{\alpha\beta}\right) \left(\left|C^{\text{VLL}}_{\nu d,\alpha\beta sb}\right|^2+\left|C^{\text{VLR}}_{\nu d,\alpha\beta sb}\right|^2\right), \\
    C^V_2&=\sum_{\alpha, \beta}\left(1-{1\over2}\delta_{\alpha\beta}\right) {\rm{Re}}\left(C^{\text{VLL}}_{\nu d,\alpha\beta sb}C^{\text{VLR}*}_{\nu d,\alpha\beta sb}\right), \\
    \label{CT}
    C^T&=\sum_{\alpha< \beta}\left(\left|C^{\text{TLL}}_{\nu d,\alpha\beta bs}\right|^2+\left|C^{\text{TLL}}_{\nu d,\alpha\beta sb}\right|^2\right),
\end{align}
and the integrated kinematic functions are given by, 
\begin{align}
    \label{J1S}
    \frac{1}{\Gamma_{B^+}^{\rm{exp}}}\int ds\,J_1^S(s)&=\frac{1}{\Gamma_{B^+}^{\rm{exp}}}\frac{\kappa(0)}{128\pi^3m_b^3}\int ds\, (m_b^2+m_s^2-s)s\,\lambda^{1/2}(m_b^2,m_s^2,s)\nonumber \\
    &=7.6\,\left(\frac{m_b}{4.19\,\text{GeV}}\right)^5\,G_F^{-2},\\
     \frac{1}{\Gamma_{B^+}^{\rm{exp}}}\int ds\,J_2^S(s)&=\frac{1}{\Gamma_{B^+}^{\rm{exp}}}\frac{\kappa(0)m_sm_b}{32\pi^3m_b^3}\int ds\, s\,\lambda^{1/2}(m_b^2,m_s^2,s)=1.3\,\left(\frac{m_b}{4.19\,\text{GeV}}\right)^5\,G_F^{-2},\\
     \frac{1}{\Gamma_{B^+}^{\rm{exp}}}\int ds\,J_1^V(s)&=\frac{1}{\Gamma_{B^+}^{\rm{exp}}}\frac{\kappa(0)}{768\pi^3m_b^3}\int ds\, \lambda^{1/2}(m_b^2,m_s^2,s)\left[\lambda(m_b^2,m_s^2,s)+3s(m_b^2+m_s^2-s)\right]\nonumber \\
     &=7.6\,\left(\frac{m_b}{4.19\,\text{GeV}}\right)^5\,G_F^{-2},\\
    \frac{1}{\Gamma_{B^+}^{\rm{exp}}}\int ds\,J_2^V(s)&=-\frac{1}{\Gamma_{B^+}^{\rm{exp}}}\frac{\kappa(0)m_sm_b}{64\pi^3m_b^3}\int ds\, \lambda^{1/2}(m_b^2,m_s^2,s)\,s\nonumber \\
     &=-0.7\,\left(\frac{m_b}{4.19\,\text{GeV}}\right)^5\,G_F^{-2},\\
    \label{JT}
    \frac{1}{\Gamma_{B^+}^{\rm{exp}}}\int ds\,J^T(s)&=\frac{1}{\Gamma_{B^+}^{\rm{exp}}}\frac{\kappa(0)}{24\pi^3m_b^3}\int ds\, \lambda^{1/2}(m_b^2,m_s^2,s)\left[2\lambda(m_b^2,m_s^2,s)+3s\left(m_s^2+m_b^2-s\right)\right]\nonumber \\
    &=365\,\left(\frac{m_b}{4.19\,\text{GeV}}\right)^5\,G_F^{-2}.
\end{align}
Here $\kappa(0)=0.83$ represents the QCD correction to the $b\to s\nu\bar\nu$ matrix element \cite{Grossman:1995gt,Buchalla:1995vs,Bobeth:2001jm}. Contrary to the previous decay channels, the number of linearly-independent kinematic structures in (\ref{J1S})-(\ref{JT}) is smaller than the number of effective WCs in (\ref{CS1})-(\ref{CT}). All of the currents in (\ref{J1S})-(\ref{JT}) can be constructed from linear combinations of the following three independent kinematic structures
\begin{align}
    s\,\lambda^{1/2}(m_b^2,m_s^2,s), \qquad s^2\,\lambda^{1/2}(m_b^2,m_s^2,s), \qquad \lambda^{3/2}(m_b^2,m_s^2,s),
\end{align}
with their three associated combinations of WCs, 
\begin{align}
    \label{incl1}
    &C_1^V+2C_1^S+32C^T+\frac{4m_sm_b}{m_b^2+m_s^2}\left(2C_2^S-C_2^V\right), \\
    &C_1^V+2C_1^S+32C^T, \\
    \label{incl3}
    &C_1^V+64C^T,
\end{align}
respectively. These combinations of WCs can be determined through dedicated measurements of $B\rightarrow X_s\nu\widehat{\nu}$

In summary, by performing detailed measurements of kinematic distributions for $B\rightarrow K^*\nu\widehat{\nu}$ and $B\rightarrow K\nu\widehat{\nu}$, one can determine the effective coefficients in (\ref{CV+T})-(\ref{CTL}) and (\ref{CSB})-(\ref{CVB}), respectively.
With these results at hand one can further determine the coefficients in~(\ref{CS1})-(\ref{CT}) through simple algebraic manipulations. Hence, rather than providing additional information, by measuring $B\rightarrow X_s\nu\widehat{\nu}$ and extracting the values of (\ref{incl1})-(\ref{incl3}) one can obtain three consistency checks of the results from measurements of $B\rightarrow K\nu\widehat{\nu}$ and $B\rightarrow K^*\nu\widehat{\nu}$.

\subsubsection{Correlations between branching ratios}
\label{subsubsec:BCorrelations}

\begin{figure}[th]
\centering
\includegraphics[width=0.8\textwidth]{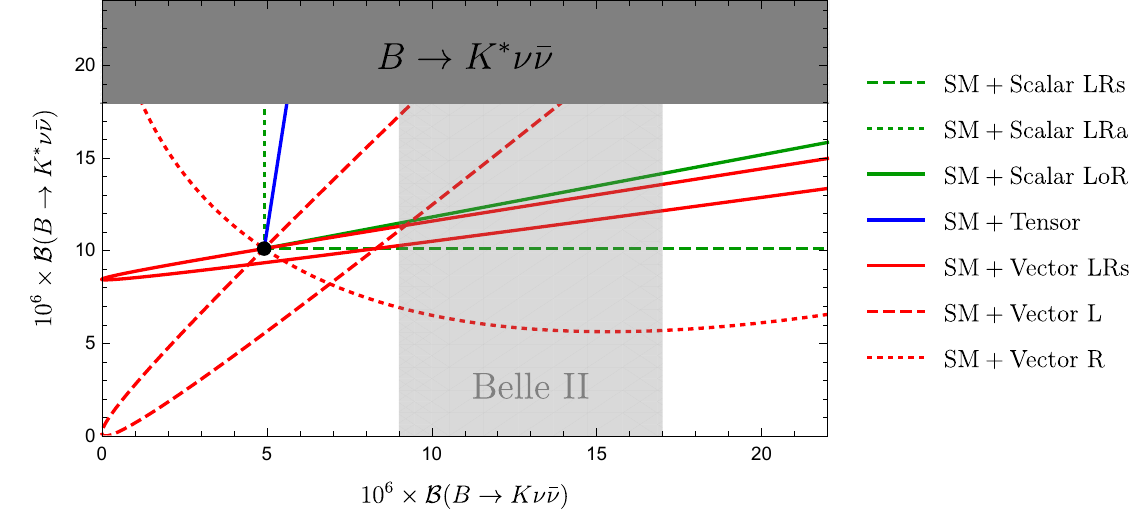}\\
\includegraphics[width=0.8\textwidth]{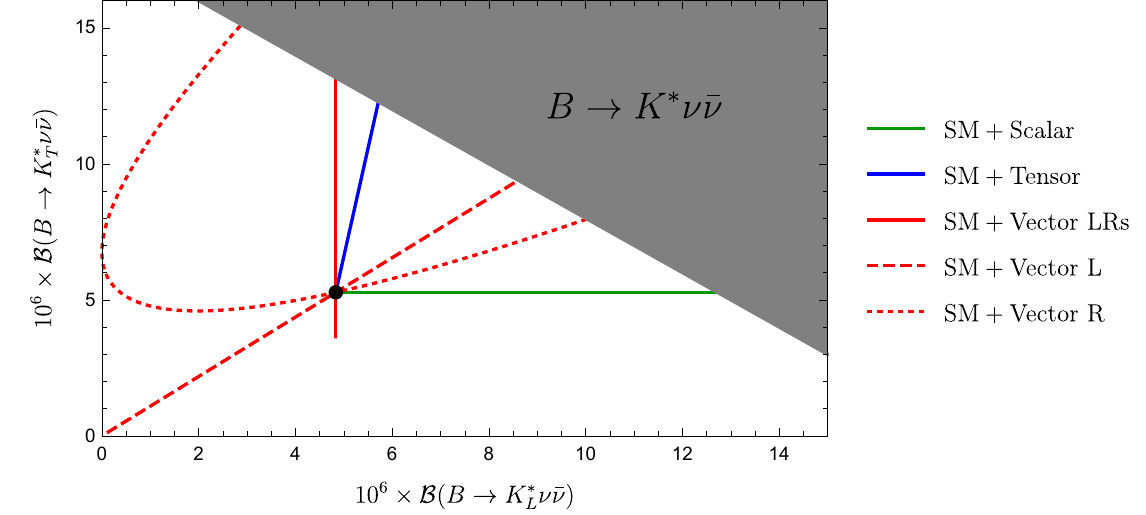}
\caption{The figure displays the $\mathcal{B}(B^+\rightarrow K^+\nu\widehat{\nu})-\mathcal{B}(B\rightarrow K^*\nu\widehat{\nu})$-plane (top) and $\mathcal{B}(B\rightarrow K^*_L\nu\widehat{\nu})-\mathcal{B}(B\rightarrow K^*_T\nu\widehat{\nu})$-plane (bottom) for different NP scenarios. The SM predictions are represented by black points. The light gray region in the upper plot indicates the present experimental range quoted by Belle II~\cite{Belle-II:2023esi}, 
and the dark gray regions are excluded by the experimental limit on $B\rightarrow K^*\widehat{\nu}\nu$ in (\ref{BKstarExLimit}). The green lines show NP scenarios with scalar currents, the blue lines NP scenarios with tensor currents, and the red curves NP scenarios with vector currents (see text for details).}
\label{fig:BPlanes}
\end{figure}

In the previous paragraphs, we have demonstrated that it is theoretically possible to quantify different NP contributions to $B\rightarrow K\nu\widehat{\nu}$ and $B\rightarrow K^*\nu\widehat{\nu}$ through dedicated measurements of kinematic distributions. In anticipation of detailed experimental measurements of the latter, we find it useful to study correlations between different observables, analogously to the analysis conducted for Kaons in Section~\ref{sec:KaonCorrelation}.

In the following, we study correlations in the $\mathcal{B}(B\rightarrow K\nu\widehat{\nu})-\mathcal{B}(B\rightarrow K^*\nu\widehat{\nu})$ and $\mathcal{B}(B\rightarrow K^*_T\nu\widehat{\nu})-\mathcal{B}(B\rightarrow K^*_L\nu\widehat{\nu})$ planes, see the upper part of Fig.~\ref{fig:BPlanes} and the lower part of Fig.~\ref{fig:BPlanes}, respectively.\footnote{See also Ref.~\cite{Rosauro-Alcaraz:2024mvx} for a recent analysis of similar ideas.}
Hereby we consider different benchmark scenarios where only one type of current is generated, i.e. only tensor (blue), scalar (green), or vector (red) currents.

{\bf{Tensor currents:}} Any NP scenario where only tensor currents are generated is confined to the blue lines in Fig.~(\ref{fig:BPlanes}). It is evident from the upper plot that such a scenario is ruled out by the recent results from Belle II~\cite{Belle-II:2023esi}. This result confirms the findings in~\cite{Felkl2023}.

{\bf{Scalar currents:}} The green lines show different scenarios where only scalar currents are generated: The solid line (Scalar LoR) is obtained if only left-handed or only right-handed currents are generated, the dashed line (Scalar LRs) for the symmetric scenario $C^{\text{SLL}}_{\nu d,\alpha\beta q_1q_2}=C^{\text{SLR}}_{\nu d,\alpha\beta q_1q_2}$, and the dotted line (Scalar LRa) for the antisymmetric scenario $C^{\text{SLL}}_{\nu d,\alpha\beta q_1q_2}=-C^{\text{SLR}}_{\nu d,\alpha\beta q_1q_2}$. It is easily seen that all NP scenarios where only scalar currents are generated must be found between the dashed and dotted lines in the upper plot and on the green line in the lower plot.

{\bf{Vector currents:}} Finally, the red curves show different scenarios where only vector currents are generated. In the following, we assume LFU, LFC, and that only three SM neutrinos contribute.\footnote{In the case of vector currents, it is possible to have interference between SM WCs and NP WCs. Assuming absence of LFV implies maximal interference, and assuming LFU fixes the WCs associated with each generation to be equal. It is possible to generate curves different from those depicted in Fig.~\ref{fig:BPlanes} in each of the vector scenarios considered by relaxing either of these two assumptions.}
The solid curve (Vector LRs) is obtained in the symmetric scenario $C^{\text{VLL}}_{\nu d,\alpha\beta q_1q_2}-\left(C^{\text{VLL}}_{\nu d,\alpha\beta q_1q_2}\right)_{\rm{SM}}=C^{\text{VLR}}_{\nu d,\alpha\beta q_1q_2}$, the dashed curve (Vector L) if only left-handed currents are generated, and the dotted curve (Vector R) if only right-handed currents are generated. Here the curves cover a much larger part of the two planes, due to possible interference between the NP WCs and the SM WCs. In particular, interference of WCs can lower the value of observables below their SM predictions, which is impossible in the other cases considered. If a future measurement indicates NP outside the region spanned by the green dashed and dot-dashed curves in the upper plot, and the region spanned by the blue and green lines in the lower plot, then vector currents are generated by the NP. 

It is also worth stressing that only vector currents in (SM)LEFT can access these regions of the plane, while dark operators with vector currents can not, as their WCs do not interfere with the SM WCs. As we will see at the end of Section~\ref{subsec:DisentanglingB}, it is not possible to distinguish vector currents in (SM)LEFT from certain operators in (D)LEFT based on kinematic distributions. However, if future measurements indicate that some observables have lower values than predicted by the SM, then one can be certain that vector currents in (SM)LEFT are at work.

\subsection{Disentangling LEFT from dark-sector physics}
\label{subsec:DisentanglingB}

We have so far assumed that the missing energy produced in the $B\rightarrow K+\slashed{E}$, $B\rightarrow K^*+\slashed{E}$, and $B\rightarrow X_s+\slashed{E}$ channels are carried away by neutrinos only. However,
similar to the case of Kaons, there could a priori be dark operators contributing to the invisible decay.
The only difference between decaying Kaons and $B$-mesons at the operator level is that we replace strange quarks with bottom quarks and down quarks with strange quarks. The (D)LEFT operators have been classified in Ref.~\cite{He2022}, and their contributions to $B$-decays have been partially worked out. Below, we provide a discussion analogous to that in Section~\ref{subsec:LNVvsDarkKaons} on the prospects of disentangling (SM)LEFT from (D)LEFT. We will start with the simplifying assumption that one operator provides the dominant NP contribution, which also captures NP scenarios where several operators are generated and the WC associated with one operator is much larger than the WCs of all other operators. We will then go beyond this simplistic treatment and determine to what extent it is possible to disentangle (SM)LEFT from (D)LEFT, in full generality, from kinematic distributions.

\subsubsection{Single operator analysis}
{\bf{$B\rightarrow K+\slashed{E}$:}}
One can directly read off analytical expressions for $d\Gamma(B\rightarrow K+\slashed{E})/ds$ from the result for $d\Gamma(K^+\rightarrow \pi^++\slashed{E})/ds$ presented in Section~\ref{sec:Kaons}, after replacing masses of mesons and form factors appropriately. In particular, kinematic distributions in the single-operator analysis of $B\rightarrow K+\slashed{E}$ take the same form as those presented for $K^+\rightarrow \pi^++\slashed{E}$ in Table~\ref{tab:KaonTab}. In Table~\ref{tab:KaonTab} we highlighted that, {in the limit where only one operator is contributing at a time,} the only operators that generate the same kinematic distribution as $\mathcal{O}^{\rm{SLL}}_{\nu d}$ or $\mathcal{O}^{\rm{SLR}}_{\nu d}$ are two (D)LEFT operators with fermionic dark-sector particles $\mathcal{O}^{\text{S}}_{sb\chi 1}$ and $\mathcal{O}^{\text{S}}_{sb\chi 2}$ as highlighted in blue.\footnote{The two corresponding dark pseudoscalar (D)LEFT operators $\mathcal{O}^{\text{P}}_{sb\chi 1}$ and $\mathcal{O}^{\text{P}}_{sb\chi 2}$ do not contribute to $B\rightarrow K+\slashed{E}$ due to parity conservation.}
We also found that only two (D)LEFT operators yield the same kinematic distribution as $\mathcal{O}^{\rm{TLL}}_{\nu d}$, namely $\mathcal{O}^{\rm{T}}_{sb\chi 1}$ and $\mathcal{O}^{\rm{T}}_{sb\chi 2}$, as highlighted in green. Finally, we observed that both two dark-fermion operators with vector currents, $\mathcal{O}_{sd\chi1}^{V}$ and $\mathcal{O}_{sd\chi2}^{V}$, and a dark-scalar operator with vector currents, $\mathcal{O}_{sd\phi}^V$, generate a kinematic structure identical to that of $\mathcal{O}^{\text{VLL}}_{\nu d}$ and $\mathcal{O}^{\text{VLR}}_{\nu d}$, as highlighted in red.

\setlength{\extrarowheight}{4pt}
\begin{table}[t!]
    \centering
    \begin{tabular}{|>{\raggedright}m{76pt}|>{\raggedright}m{60pt}|>{\centering}m{284pt}|c}
    \cline{1-3}
    Type & Operator & Kinematic structure in $B\rightarrow K^*\,+\slashed{E}$ & \\[2pt]
    \cline{1-3}
    \multirow{3}{*}{(SM)LEFT} & $\mathcal{O}^{\text{SLL}}_{\nu d}$, $\mathcal{O}^{\text{SLR}}_{\nu d}$ & \cellcolor{blue!25}$s\lambda^{3/2}\left|A_0\right|^2$& \\[2pt]   
     \cline{2-3}
    & $\mathcal{O}^{\text{VLL}}_{\nu d}$, $\mathcal{O}^{\text{VLR}}_{\nu d}$ &\cellcolor{red!25} $\lambda^{1/2}\big[\frac{s\lambda}{(m_B+m_{K^*})^2}\left|V_0\right|^2+s(m_B+m_{K^*})^2\left|A_1\right|^2$\\
    $+32m_B^2m_{K^*}^2\left|A_{12}\right|^2\big]$ & \\[2pt]
     \cline{2-3}
     & $\mathcal{O}^{\text{TLL}}_{\nu d}$ & \cellcolor{Green!35}$\lambda^{1/2}\Big[\lambda\left|T_1\right|^2+(m_B^2-m_{K^*}^2)^2\left|T_2\right|^2+\frac{8m_B^2m_{K^*}^2s}{(m_B+m_{K^*})^2}\left|T_{23}\right|^2\Big]$ & \\[2pt]
     \cline{1-3}
    \multirow{3}{*}{Scalar DM} & $\mathcal{O}_{sb\phi}^P$  & $\lambda^{3/2}\left|A_0\right|^2$ & \\[2pt]
     \cline{2-3}
     & $\mathcal{O}_{sb\phi}^A$  & $\lambda^{1/2}\Big[s(m_B+m_{K^*})^2\left|A_1\right|^2+32m_B^2m_{K^*}^2\left|A_{12}\right|^2\Big]$ & \\[2pt]
     \cline{2-3}
      & $\mathcal{O}_{sb\phi}^V$ & $s\lambda^{3/2}\left|V_0\right|^2$ & \\[2pt]
      \cline{1-3}
    \multirow{4}{*}{Fermion DM} & $\mathcal{O}_{sb\chi1}^{P}, \mathcal{O}_{sb\chi2}^{P}$ & \cellcolor{blue!25}$s\lambda^{3/2}\left|A_{0}\right|^2$& \\[2pt]   
     \cline{2-3}
     &$\mathcal{O}_{sb\chi1}^{A}, \mathcal{O}_{sb\chi2}^{A}$ & $\lambda^{1/2}\Big[s(m_B+m_{K^*})^2\left|A_1\right|^2+32m_B^2m_{K^*}^2\left|A_{12}\right|^2\Big]$& \\[2pt]   
     \cline{2-3}
    & $\mathcal{O}_{sb\chi1}^{V}, \mathcal{O}_{sb\chi2}^{V}$ & $s\lambda^{3/2}\left|V_{0}\right|^2$ & \\[2pt]
     \cline{2-3}
     & $\mathcal{O}_{sb\chi1}^{T}, \mathcal{O}_{sb\chi2}^{T}$ &\cellcolor{Green!35} $\lambda^{1/2}\Big[\lambda\left|T_1\right|^2+(m_B^2-m_{K^*}^2)^2\left|T_2\right|^2+\frac{8m_B^2m_{K^*}^2s}{(m_B+m_{K^*})^2}\left|T_{23}\right|^2\Big]$ & \\[2pt]
     \cline{1-3}
    \multirow{8}{*}{Vector DM: A} & $\mathcal{O}^P_{sbA}$ & $s^2\lambda^{3/2}\left|A_{0}\right|^2$& \\[2pt]   
     \cline{2-3}
     & $\mathcal{O}^T_{sbA1}$ & $s\lambda^{3/2}\left|T_{1}\right|^2$& \\[2pt]   
     \cline{2-3}
    & $\mathcal{O}^T_{sbA2}$ & $s\lambda^{1/2}\Big[(m_B^2-m_{K^*}^2)^2\left|T_2\right|^2+\frac{8m_B^2m_{K^*}^2s}{(m_B+m_{K^*})^2}\left|T_{23}\right|^2\Big]$ & \\[2pt]
     \cline{2-3}
     & $\mathcal{O}^V_{sbA3}, \mathcal{O}^V_{sbA6}$ & $s^2\lambda^{3/2}\left|V_{0}\right|^2$ & \\[2pt]
     \cline{2-3}
    & $\mathcal{O}^V_{sbA4}, \mathcal{O}^V_{sbA5}$ & $s^3\lambda^{3/2}\left|V_{0}\right|^2$ & \\[2pt]
    \cline{2-3}
     & $\mathcal{O}^A_{sbA2}$ & $s^2\lambda^{3/2}\left|A_{0}\right|^2$ & \\[2pt]
     \cline{2-3}
     & $\mathcal{O}^A_{sbA3}, \mathcal{O}^A_{sbA6}$ & $s\lambda^{1/2}\Big[s(m_B+m_{K^*})^2\left|A_1\right|^2+32m_B^2m_{K^*}^2\left|A_{12}\right|^2\Big]$ & \\[2pt]
     \cline{2-3}
     & $\mathcal{O}^A_{sbA4}, \mathcal{O}^A_{sbA5}$ & $s^2\lambda^{1/2}\Big[s(m_B+m_{K^*})^2\left|A_1\right|^2+32m_B^2m_{K^*}^2\left|A_{12}\right|^2\Big]$ & \\[2pt]
     \cline{1-3}
     \multirow{2}{*}{Vector DM: B} & $\mathcal{O}^P_{sbB1}, \mathcal{O}^P_{sbB2}$ & $s^2\lambda^{3/2}\left|A_{0}\right|^2$& \\[2pt]   
     \cline{2-3}
    & $\mathcal{O}^T_{sbB1}, \mathcal{O}^T_{sbB2}$ & $s\lambda^{1/2}\Big[\lambda\left|T_1\right|^2+(m_B^2-m_{K^*}^2)^2\left|T_2\right|^2+\frac{8m_B^2m_{K^*}^2s}{(m_B+m_{K^*})^2}\left|T_{23}\right|^2\Big]$ & \\[2pt]
     \cline{1-3}
    \end{tabular}
    \caption{List of operators contributing to $B\rightarrow K^*\,+\slashed{E}$ and the kinematic structure that they generate in $d\Gamma(B\rightarrow K^{*}\,+\slashed{E})/ds$. Here $\lambda$ is a shorthand notation for $\lambda(m_{B}^2,m_{K^{*}}^2,s)$, and $V_0$ for $V_0(s)$, etc. We have only listed operators that give a contribution to $B\rightarrow K^*\,+\slashed{E}$, and interference terms are neglected under the assumption that only one operator is active (see text for details).}
    \label{tab:KStarTab}
\end{table}

\setlength{\extrarowheight}{4pt}
\begin{table}[t!]
    \centering
    \begin{tabular}{|>{\raggedright}m{78pt}|>{\raggedright}m{63pt}|>{\centering}m{220pt}|c}
    \cline{1-3}
    Type & Operator & Kinematic structure in $B\rightarrow X_s\,+\slashed{E}$ & \\[2pt]
    \cline{1-3}
    \multirow{3}{*}{(SM)LEFT} & $\mathcal{O}^{\text{SLL}}_{\nu d}$, $\mathcal{O}^{\text{SLR}}_{\nu d}$ & \cellcolor{blue!25}$s\lambda^{1/2}(m_b^2+m_s^2-s)$& \\[2pt]   
     \cline{2-3}
    & $\mathcal{O}^{\text{VLL}}_{\nu d}$, $\mathcal{O}^{\text{VLR}}_{\nu d}$ & $\cellcolor{red!25}\lambda^{1/2}\big[\lambda+3s(m_b^2+m_s^2-s)\big]$ & \\[2pt]
     \cline{2-3}
     & $\mathcal{O}^{\text{TLL}}_{\nu d}$ & \cellcolor{Green!35}$\lambda^{1/2}\big[2\lambda+3s(m_b^2+m_s^2-s)\big]$ & \\[2pt]
     \cline{1-3}
    \multirow{4}{*}{Scalar DM} & $\mathcal{O}_{sb\phi}^S$  & $\lambda^{1/2}\big[(m_b+m_s)^2-s\big]$ & \\[2pt]
     \cline{2-3}
     & $\mathcal{O}_{sb\phi}^P$  & $\lambda^{1/2}\big[(m_b-m_s)^2-s\big]$ & \\[2pt]
     \cline{2-3}
     & $\mathcal{O}_{sb\phi}^V$  & $\cellcolor{RedOrange!25}\lambda^{1/2}\big[\lambda+3s\big((m_b-m_s)^2-s\big)\big]$ & \\[2pt]
     \cline{2-3}
      & $\mathcal{O}_{sb\phi}^A$ & $\cellcolor{RedOrange!25}\lambda^{1/2}\big[\lambda+3s\big((m_b+m_s)^2-s\big)\big]$ & \\[2pt]
      \cline{1-3}
    \multirow{5}{*}{Fermion DM} & $\mathcal{O}_{sb\chi1}^{S}, \mathcal{O}_{sb\chi2}^{S}$ & \cellcolor{cyan!25}$s\lambda^{1/2}\big[\big((m_b+m_s)^2-s\big)\big]$& \\[2pt]   
     \cline{2-3}
     & $\mathcal{O}_{sb\chi1}^{P}, \mathcal{O}_{sb\chi2}^{P}$ & \cellcolor{cyan!25}$s\lambda^{1/2}\big[\big((m_b-m_s)^2-s\big)\big]$& \\[2pt]   
     \cline{2-3}
     &$\mathcal{O}_{sb\chi1}^{A}, \mathcal{O}_{sb\chi2}^{A}$ & $\cellcolor{RedOrange!25}\lambda^{1/2}\big[\lambda+3s\big((m_b+m_s)^2-s\big)\big]$& \\[2pt]   
     \cline{2-3}
    & $\mathcal{O}_{sb\chi1}^{V}, \mathcal{O}_{sb\chi2}^{V}$ &\cellcolor{RedOrange!25} $\lambda^{1/2}\big[\lambda+3s\big((m_b-m_s)^2-s\big)\big]$ & \\[2pt]
     \cline{2-3}
     & $\mathcal{O}_{sb\chi1}^{T}, \mathcal{O}_{sb\chi2}^{T}$ & \cellcolor{Green!35}$\lambda^{1/2}\big[2\lambda+3s(m_b^2+m_s^2-s)\big]$ & \\[2pt]
     \cline{1-3}
    \multirow{10}{*}{Vector DM: A} & $\mathcal{O}^S_{sbA}$ & $s^2\lambda^{1/2}\big[(m_b+m_s)^2-s\big]$& \\[2pt]   
     \cline{2-3}
     & $\mathcal{O}^P_{sbA}$ & $s^2\lambda^{1/2}\big[(m_b-m_s)^2-s\big]$& \\[2pt]   
     \cline{2-3}
     & $\mathcal{O}^T_{sbA1}$ & $s\lambda^{1/2}\big[2\lambda+3s\big((m_b-m_s)^2-s\big)\big]$& \\[2pt]   
     \cline{2-3}
    & $\mathcal{O}^T_{sbA2}$ & $s\lambda^{1/2}\big[2\lambda+3s\big((m_b+m_s)^2-s\big)\big]$ & \\[2pt]
     \cline{2-3}
     & $\mathcal{O}^V_{sbA2}$ & $s^2\lambda^{1/2}\big[(m_b+m_s)^2-s\big]$ & \\[2pt]
     \cline{2-3}
     & $\mathcal{O}^V_{sbA3}, \mathcal{O}^V_{sbA6}$ & $s\lambda^{1/2}\big[\lambda+3s\big((m_b-m_s)^2-s\big)\big]$ & \\[2pt]
     \cline{2-3}
    & $\mathcal{O}^V_{sbA4}, \mathcal{O}^V_{sbA5}$ & $s^2\lambda^{1/2}\big[\lambda+3s\big((m_b-m_s)^2-s\big)\big]$ & \\[2pt]
     \cline{2-3}
     & $\mathcal{O}^A_{sbA2}$ & $s^2\lambda^{1/2}\big[(m_b-m_s)^2-s\big]$ & \\[2pt]
     \cline{2-3}
     & $\mathcal{O}^A_{sbA3}, \mathcal{O}^A_{sbA6}$ & $s\lambda^{1/2}\big[\lambda+3s\big((m_b+m_s)^2-s\big)\big]$ & \\[2pt]
     \cline{2-3}
     & $\mathcal{O}^A_{sbA4}, \mathcal{O}^A_{sbA5}$ & $s^2\lambda^{1/2}\big[\lambda+3s\big((m_b+m_s)^2-s\big)\big]$ & \\[2pt]
     \cline{1-3}
     \multirow{3}{*}{Vector DM: B} & $\mathcal{O}^S_{sbB1}, \mathcal{O}^S_{sbB2}$ & $s^2\lambda^{1/2}\big[(m_b+m_s)^2-s\big]$& \\[2pt]   
     \cline{2-3}
     & $\mathcal{O}^P_{sbB1}, \mathcal{O}^P_{sbB2}$ & $s^2\lambda^{1/2}\big[(m_b-m_s)^2-s\big]$& \\[2pt]   
     \cline{2-3}
    & $\mathcal{O}^T_{sbB1}, \mathcal{O}^T_{sbB2}$ & $s^2\lambda^{1/2}\big[2\lambda+3s(m_b^2+m_s^2-s)\big]$ & \\[2pt]
     \cline{1-3}
    \end{tabular}
    \caption{List of operators contributing to $B\rightarrow X_s\,+\slashed{E}$ and the kinematic structure they give rise to in $d\Gamma(B\rightarrow X_s\,+\slashed{E})/ds$. $\lambda$ is a shorthand notation for $\lambda(m_{b}^2,m_s^2,s)$. Interference terms are neglected under the assumption that only one operator is active (see text for details).}
    \label{tab:InclusiveTab}
\end{table}

{\bf{$B\rightarrow K^*+\slashed{E}$ and $B\rightarrow X_s+\slashed{E}$:}} We have carried out a similar analysis for $B\rightarrow K^*+\slashed{E}$, and $B\rightarrow X_s+\slashed{E}$, and our results are summarized in Tables~\ref{tab:KStarTab} and \ref{tab:InclusiveTab}, respectively.\footnote{Out of all (D)LEFT operators, only operators with vector DM of type A give interference terms between different WCs in these decays. Full expressions for the differential decay distributions are included in Appendix~\ref{appendix:B-decays}, but here we note that none of the interference terms have a kinematic structure similar to that of any (SM)LEFT operator.} We comment on these results below.

\begin{itemize}
    \item The results for $B\rightarrow K^*+\slashed{E}$ are summarized in Table~\ref{tab:KStarTab}. We found that only $\mathcal{O}^{\text{P}}_{sb\chi 1}$ and $\mathcal{O}^{\text{P}}_{sb\chi 2}$ generate the same kinematic distribution as $\mathcal{O}^{\rm{SLL}}_{\nu d}$ and $\mathcal{O}^{\rm{SLR}}_{\nu d}$, and we have highlighted these contributions in blue.\footnote{$\mathcal{O}^{\text{S}}_{sb\chi 1}$ and $\mathcal{O}^{\text{S}}_{sb\chi 2}$, which yield the same kinematic distribution as $\mathcal{O}^{\rm{SLL}}_{\nu d}$ and $\mathcal{O}^{\rm{SLR}}_{\nu d}$ in $B\rightarrow K+\slashed{E}$, do not contribute to $B\rightarrow K^*+\slashed{E}$ due to parity conservation.} The only operators that generate the same kinematic distribution as $\mathcal{O}^{\rm{TLL}}_{\nu d}$ are $\mathcal{O}^{\rm{T}}_{sb\chi 1}$, $\mathcal{O}^{\rm{T}}_{sb\chi 2}$, which we have highlighted in green. In the case of (SM)LEFT vector-current operators $\mathcal{O}^{\text{VLL}}_{\nu d}$, $\mathcal{O}^{\text{VLR}}_{\nu d}$, highlighted in red, we see that no single (D)LEFT operator generates a similar kinematic structure in the operator basis considered here. However, the (SM)LEFT vector-current kinematic structure can be reproduced by {going beyond a single-operator analysis and} taking suitable linear combinations of dark-fermion operators or dark-scalar operators. This observation is made more precise in the multiple-operator analysis below.

    \item In Table~\ref{tab:InclusiveTab}, we present our results for the kinematic distributions in $B\rightarrow X_s+\slashed{E}$. The results for dark operators are, to the best of our knowledge, presented here for the first time. 
    
    We find that the kinematic distribution of $\mathcal{O}^{\rm{SLL}}_{\nu d}$ and $\mathcal{O}^{\rm{SLR}}_{\nu d}$, highlighted in dark blue, are unique in this decay channel although they would in practice be very hard to distinguish from $\mathcal{O}^{\text{P}}_{sb\chi 1,2}$ and $\mathcal{O}^{\text{S}}_{sb\chi 1,2}$, highlighted in light blue. 
    
    The (SM)LEFT tensor operator $\mathcal{O}^{\rm{TLL}}_{\nu d}$ has the same kinematic distribution as the (D)LEFT operators $\mathcal{O}^{\rm{T}}_{sb\chi 1}$ and $\mathcal{O}^{\rm{T}}_{sb\chi 2}$. Hence, we find that $\mathcal{O}^{\rm{TLL}}_{\nu d}$ is indistinguishable from $\mathcal{O}^{\rm{T}}_{sb\chi 1}$ and $\mathcal{O}^{\rm{T}}_{sb\chi 2}$ in all three decay channels. Additional efforts would therefore be required to distinguish $\mathcal{O}^{\rm{TLL}}_{\nu d}$ from the fermionic tensorial operators in (D)LEFT. 

    We find that the kinematic distribution of $\mathcal{O}^{\rm{VLL}}_{\nu d}$ and $\mathcal{O}^{\rm{VLR}}_{\nu d}$, highlighted in red, are unique in this decay channel although they would in practice be very hard to distinguish from $\mathcal{O}^{\text{V/A}}_{sb\phi}$ and $\mathcal{O}^{\text{V/A}}_{sb\chi 1,2}$, highlighted in light red. 
\end{itemize}

Above, we found that $\mathcal{O}^{\rm{SLL}}_{\nu d}$ and $\mathcal{O}^{\rm{SLR}}_{\nu d}$ can in principle be distinguished from all (D)LEFT operators by studying kinematic distributions of $B\rightarrow K+\slashed{E}$, and $B\rightarrow K^*+\slashed{E}$, assuming that only one (SM)LEFT or one (D)LEFT operator is present. This is a basis-dependent statement which does not apply in a different (D)LEFT basis where the four operators $\mathcal{O}^{\text{S}}_{sb\chi 1,2}$ and $\mathcal{O}^{\text{P}}_{sb\chi 1,2}$ are combined into four basis operators with similar Dirac structure as $\mathcal{O}^{\rm{SLL}}_{\nu d}$ and $\mathcal{O}^{\rm{SLR}}_{\nu d}$. In Appendix~\ref{appendix:LNVDarkFermion}, we show explicitly how this simple fact manifests itself when comparing contributions from $\{\mathcal{O}^{\text{S}}_{sb\chi 1,2},\,\mathcal{O}^{\text{P}}_{sb\chi 1,2}\}$ and $\{\mathcal{O}^{\rm{SLL}}_{\nu d},\,\mathcal{O}^{\rm{SLR}}_{\nu d}\}$ to the three decay channels considered in this work. Similar considerations also hold for the (SM)LEFT operators $\mathcal{O}^{\rm{VLL}}_{\nu d}$, $\mathcal{O}^{\rm{VLR}}_{\nu d}$ and the (D)LEFT operators $\mathcal{O}^{\text{V/A}}_{sb\phi}$, $\mathcal{O}^{\text{V/A}}_{sb\chi 1,2}$.
These ambiguities regarding operator distinguishability in different operator bases are manifestly seen and resolved in the results of the following multiple-operator analysis.

\subsubsection{Multiple operator analysis}

At this stage, we want to stress that the approach of using kinematic distributions to disentangle different operator contributions to meson decays outlined for LEFT in Sections~\ref{subsec:KaonExtraction} and \ref{sec:BExtraction} is generally applicable to any set of operators. With an increasing number of operators contributing to the meson decays more measurements are required to solve the potentially larger sets of linear equations that arise. For a given decay channel, the number of equations to solve is simply given by the number of linearly independent kinematic structures entering the differential decay distribution.

As the most general case, it is possible to add the (SM)LEFT basis in Section~\ref{sec:OperatorBasis} and all four (D)LEFT bases listed in Appendix~\ref{appendix:DarkOperators} into a combined operator set. A straightforward extension of the procedure in Sections~\ref{subsec:KaonExtraction} and \ref{sec:BExtraction} would then yield a large number of linear equations, whose solution would constrain different combinations of the Wilson coefficients in the set. As a less involved case, we now consider a basis consisting of the (SM)LEFT operators in (\ref{VLL})-(\ref{TLL}) and the dark scalar and dark fermion operators in (\ref{ScalarOperatorStart})-(\ref{FermionOperatorEnd}). The extension to include vector operators is straightforward but arguably less interesting: It is easily seen from Tables~\ref{tab:KaonTab}, \ref{tab:KStarTab} and \ref{tab:InclusiveTab} that no linear combination of vector operators can reproduce the kinematic structures of (SM)LEFT operators in $B\rightarrow K+\slashed{E}$, $B\rightarrow K^*+\slashed{E}$ and $B\rightarrow X_s+\slashed{E}$, respectively. Hence, by performing detailed measurements of kinematic distributions in these decay channels one can completely disentangle (SM)LEFT from dark vectors. As a result, none of the expressions below that involve (SM)LEFT Wilson coefficients would be modified.

Under the assumption that the final state can only be two neutrinos, two dark-sector scalars, or two dark-sector fermions, dedicated measurements of $B\rightarrow K+\slashed{E}$ can determine the value of the following four combinations of Wilson coefficients (we have included the four corresponding kinematic structures for the readers convenience), 
\begin{itemize}
    \item Kinematic structure: 
    \begin{align}
        \lambda^{1/2}\left|f_0^K\right|^2.
    \end{align}
    Associated Wilson coefficient:
    \begin{align}
        \label{scalardmK}
    &\left|C_{sb\phi}^S\right|^2.
    \end{align}
    \item Kinematic structure: 
    \begin{align}
        s\,\lambda^{1/2}\left|f_0^K\right|^2.
    \end{align}
    Associated combination of Wilson coefficients:
    \begin{align}
        \label{LNVSBK}
    &\sum_{\alpha\leq \beta} \left(1-{1\over2}\delta_{\alpha\beta}\right)\left(\left|C_{\nu d,\alpha\beta bs}^{\text{SLL}}+C_{\nu d,\alpha\beta bs}^{\text{SLR}}\right|^2+\left|C_{\nu d,\alpha\beta sb}^{\text{SLL}}+C_{\nu d,\alpha\beta sb}^{\text{SLR}}\right|^2\right) \nonumber \\
    &+2\left(\left|C_{sb\chi 1}^S\right|^2+\left|C_{sb\chi 2}^S\right|^2\right).
    \end{align}
    \item Kinematic structure: 
    \begin{align}
        \lambda^{3/2}\left|f_+^K\right|^2.
    \end{align}
    Associated combination of Wilson coefficients:
    \begin{align}
        \label{LNCBK}
    &2\left|C_{sb\phi}^V\right|^2+8\left(\left|C_{sb\chi 1}^V\right|^2+\left|C_{sb\chi 2}^V\right|^2\right)+\sum_{\alpha, \beta}\left(1-{1\over2}\delta_{\alpha\beta}\right)   \left|C_{\nu d,\alpha\beta bs}^{\text{VLL}}+C_{\nu d,\alpha\beta bs}^{\text{VLR}}\right|^2.
    \end{align}
     \item Kinematic structure: 
    \begin{align}
        s\lambda^{3/2}\left|f_T^K\right|^2.
    \end{align}
    Associated combination of Wilson coefficients:
    \begin{align}
        \label{tensorBK}
    &\left(\left|C_{sb\chi 1}^T\right|^2+\left|C_{sb\chi 2}^T\right|^2\right)+2\sum_{\alpha<\beta} \left( \left|C^{\text{TLL}}_{\nu d,\alpha\beta bs}\right|^2+\left|C^{\text{TLL}}_{\nu d,\alpha\beta sb}\right|^2\right).
    \end{align}
\end{itemize}

Similarly, dedicated measurements of $B\rightarrow K^*+\slashed{E}$ can determine the value of the following five combinations of Wilson coefficients,
\begin{itemize}
    \item Kinematic structure: 
    \begin{align}
        \lambda^{3/2}\left|A_0\right|^2.
    \end{align}
    Associated Wilson coefficient:
    \begin{align}
        \label{scalardmKstar}
    &\left|C_{sb\phi}^P\right|^2.
    \end{align}
    \item Kinematic structure: 
    \begin{align}
        s\,\lambda^{3/2}\left|A_0\right|^2.
    \end{align}
    Associated combination of Wilson coefficients:
    \begin{align}
        \label{LNVSBKstar}
    &\sum_{\alpha\leq\beta}\left(1-\frac{\delta_{\alpha\beta}}{2}\right)\left(\left|C^{\rm{SLL}}_{\nu d,\alpha\beta bs}-C^{\rm{SLR}}_{\nu d,\alpha\beta bs}\right|^2+\left|C^{\rm{SLL}}_{\nu d,\alpha\beta sb}-C^{\rm{SLR}}_{\nu d,\alpha\beta sb}\right|^2\right) \nonumber \\
    &+2\left(\left|C^P_{sb\chi 1}\right|^2+\left|C^P_{sb\chi 2}\right|^2\right).
    \end{align}
    \item Kinematic structure: 
    \begin{align}
        \lambda^{1/2}\Big[s(m_B+m_{K^*})^2\left|A_1\right|^2+32m_B^2m_{K^*}^2\left|A_{12}\right|^2\Big].
    \end{align}
    Associated combination of Wilson coefficients:
    \begin{align}
        \label{LNCBKstar0}
    &2\left|C_{sb\phi}^A\right|^2+ 8\left(\left|C^A_{sb\chi 1}\right|^2+\left|C^A_{sb\chi 2}\right|^2\right)+\sum_{\alpha,\beta}\left(1-{1\over2}\delta_{\alpha\beta}\right) \left|C^{\rm{VLL}}_{\nu d,\alpha\beta bs}-C^{\rm{VLR}}_{\nu d,\alpha\beta bs}\right|^2.
    \end{align}
    \item Kinematic structure: 
    \begin{align}
        s\,\lambda^{3/2}\left|V_0\right|^2.
    \end{align}
    Associated combination of Wilson coefficients:
    \begin{align}
        \label{LNCBKstar}
    &2\left|C_{sb\phi}^V\right|^2+ 8\left(\left|C^V_{sb\chi 1}\right|^2+\left|C^V_{sb\chi 2}\right|^2\right)+\sum_{\alpha,\beta}\left(1-{1\over2}\delta_{\alpha\beta}\right) \left|C^{\rm{VLL}}_{\nu d,\alpha\beta bs}+C^{\rm{VLR}}_{\nu d,\alpha\beta bs}\right|^2.
    \end{align}
    \item Kinematic structure: 
    \begin{align}
        \lambda^{1/2}\Big[\lambda\left|T_1\right|^2+(m_B^2-m_{K^*}^2)^2\left|T_2\right|^2+\frac{8m_B^2m_{K^*}^2s}{(m_B+m_{K^*})^2}\left|T_{23}\right|^2\Big].
    \end{align}
    Associated combination of Wilson coefficients:
    \begin{align}
        \label{LNVTBKstar}
    &\left(\left|C^T_{sb\chi 1}\right|^2+\left|C^T_{sb\chi 2}\right|^2\right)+2\sum_{\alpha<\beta}\left(\left|C^{\rm{TLL}}_{\nu d,\alpha\beta sb}\right|^2+\left|C^{\rm{TLL}}_{\nu d,\alpha\beta bs}\right|^2\right).
    \end{align}
\end{itemize}

For the current analysis of $B\rightarrow X_s+\slashed{E}$, it is easily seen from Table~\ref{tab:InclusiveTab} that only the four linearly independent kinematic structures below are relevant 
\begin{align}
    \lambda^{1/2}(m_b^2,m_s^2,s), \qquad s\,\lambda^{1/2}(m_b^2,m_s^2,s), \qquad s^2\,\lambda^{1/2}(m_b^2,m_s^2,s), \qquad \lambda^{3/2}(m_b^2,m_s^2,s).
\end{align}
The combination of Wilson coefficients associated with the first kinematic structure reads
\begin{align}
    \left|C_{sb\phi}^S\right|^2+\frac{(m_b-m_s)^2}{(m_b+m_s)^2}\left|C_{sb\phi}^P\right|^2,
\end{align}
while the expressions for the remaining three are longer and extend the results in (\ref{incl1})-(\ref{incl3}). As they do not provide information that cannot be extracted from measurements of (\ref{scalardmK})-(\ref{LNVTBKstar}), we do not write down their full expressions here. Instead, we summarize some important points about (\ref{scalardmK})-(\ref{LNVTBKstar}) below.

\begin{itemize}
    \item Eqs.~(\ref{scalardmK}) and (\ref{scalardmKstar}) show that it is possible to determine the magnitude of WCs associated with scalar DM operators with (pseudo)scalar currents.
    \item Eqs.~(\ref{LNVSBK}) and (\ref{LNVSBKstar}) show that it is possible to distinguish scalar currents in (SM)LEFT from all operators under consideration except for fermions with both scalar and pseudo-scalar currents in (D)LEFT. In particular, the combinations of operators in (\ref{LNVSBK}) and (\ref{LNVSBKstar}) have similar Dirac structure. 
    \item We find that vector currents in (SM)LEFT are always distinguishable from dark vectors and dark scalars with (pseudo)scalar currents. However, Eqs.~(\ref{LNCBK}), (\ref{LNCBKstar0}), and (\ref{LNCBKstar}), show that it is not always possible to distinguish vector currents in (SM)LEFT from dark scalars and dark fermions with (axial)vector currents. The combinations of operators entering these equations have similar Dirac structure. 
    \item Eqs.~(\ref{tensorBK}) and (\ref{LNVTBKstar}) show that it is possible to distinguish tensor currents in (SM)LEFT from all operators under consideration except for dark fermions with tensor currents by analyzing kinematic distributions only. 
\end{itemize}

In conclusion, we have shown that on an analytical level it is possible to distinguish contributions from many different operators in (SM)LEFT and (D)LEFT through detailed measurements of kinematic distributions in $B\rightarrow K^*+\slashed{E}$ and $B\rightarrow K+\slashed{E}$. Distinguishing (SM)LEFT operators from fermionic (D)LEFT operators with similar Dirac structure would require analysis beyond kinematic distributions, which goes beyond the scope of the present work.

\subsection{Benchmark examples}
\label{subsec:BelleII}

In the preceding sections, we classified kinematic distributions in the three decay channels $B\rightarrow K+\slashed{E}$, $B\rightarrow K^*+\slashed{E}$, and $B\rightarrow X_s+\slashed{E}$, and demonstrated how experimental measurements of kinematic distributions can set quantitative constraints on the magnitude of some combinations of Wilson coefficients. 
In this section, we provide qualitative illustrations of these results in some simple examples to showcase the usage of kinematic distributions on a less abstract level. 

In light of the recent Belle II measurement~\cite{Belle-II:2023esi},
we use a value of 
\begin{align}
    \mathcal{B}(B\rightarrow K+\slashed{E})=(1.3\pm 0.4)\times 10^{-5},
\end{align}
for the $B\rightarrow K+\slashed{E}$ branching fraction in our examples. We stress that we are not performing an analysis of the recent Belle II data here, but merely use the combination of the inclusive and hadronic tagging results for the $B\rightarrow K+\slashed{E}$ branching fraction as a benchmark value for our numerical illustrations below. A more complete analysis would treat the measured branching ratio as a function of the BSM parameters, as explained in detail in Ref.~\cite{Gartner:2024muk}. We will only analyze a single operator at a time, leaving more sophisticated examples where multiple operators give comparable contributions for future investigations. In particular, we will focus on the two scalar-current (SM)LEFT operators $\mathcal{O}^{\rm{SLL}}_{\nu d}$ and $\mathcal{O}^{\rm{SLR}}_{\nu d}$ below. As already seen in Fig.~\ref{fig:BPlanes}, our results will depend on the chiral structure of the NP inducing the (SM)LEFT operators. More specifically, we study a left-right symmetric (LRs) scenario where $C^{\text{SLR}}_{\alpha\beta bs(sb)}=C^{\text{SLL}}_{\alpha\beta bs(sb)}$, and the maximally left-right violating scenario (LoR) where only one of the two types of scalar (SM)LEFT operators ($C^{\text{SLR}}_{\alpha\beta bs(sb)}$ or $C^{\text{SLL}}_{\alpha\beta bs(sb)}$) is present. To make our plots easy to read, we only consider central values of branching ratios in the calculation of Wilson coefficients.

In the following examples, the difference in the $B\rightarrow K+\slashed{E}$ branching fraction between the experimental value and the SM is attributed to one new operator. Hence, we require the operator to account for a NP branching ratio of roughly, 
\begin{align}
    \mathcal{B}(B\rightarrow K+\slashed{E})_{\text{NP}}&=\mathcal{B}(B\rightarrow K+\slashed{E})_{\text{Exp}}-\mathcal{B}(B\rightarrow K+\slashed{E})_{\text{SM}}\nonumber \\
    &=(0.8\pm 0.4)\times 10^{-5},
\end{align}
where we have used the SM prediction in (\ref{BVNEW}).
In our effective description, the SM prediction  translates into the following central value
\begin{align}
    \label{eq:SMWilson}
    \sum_{\alpha, \beta}\left(1-{1\over2}\delta_{\alpha\beta}\right) \left|C_{\nu d,\alpha\beta bs}^{\text{VLL}}+C_{\nu d,\alpha\beta bs}^{\text{VLR}} \right|^2=3.0\times 10^{-16}(\text{GeV})^{-4}.
\end{align}

Similarly, if the LNV operators $\mathcal{O}_{\nu d}^{\text{SLL}}$ and $\mathcal{O}_{\nu d}^{\text{SLR}}$ are responsible for the full NP contribution to $B\rightarrow K+\slashed{E}$, then the central value of the Wilson coefficient is given by 
\begin{align}
    \label{BoundCSLL}
    \sum_{\alpha\leq\beta}\left(1-\frac{1}{2}\delta_{\alpha\beta}\right)\left[\Big| C^{\text{SLR}}_{\alpha\beta sb}+C^{\text{SLL}}_{\alpha\beta s b}\Big|^2+\Big|C^{\text{SLR}}_{\alpha\beta bs}+C^{\text{SLL}}_{\alpha\beta bs}\Big|^2\right]=1.4\times 10^{-16}(\text{GeV})^{-4}.
\end{align}
The associated effective scale of NP is in this case, 
\begin{align}
    \label{LambdaLNV}
    \Lambda_{\text{LNV}}~\approx~9.3\,\text{TeV},
\end{align}
where the $\approx$ in the equation above reflects that the effective scale depends on the flavor structure of the induced scalar-current operators.

\paragraph{Example 1: scalar DM}\mbox{}\\ \mbox{}\\
In our first example, we compare NP signatures from $\mathcal{O}_{\nu d}^{\text{SLL}}$ and $\mathcal{O}_{\nu d}^{\text{SLR}}$ with the (D)LEFT operators listed in (\ref{ScalarOperatorStart})-(\ref{ScalarOperatorEnd}). 
Out of these four (D)LEFT operators, only the following two contribute to $B^+\to K^++\slashed{E}$,  
\begin{align}
    \mathcal{O}_{sb\phi}^{S}, \qquad \mathcal{O}_{sb\phi}^{V}.
\end{align}
Requiring each of these operators to account for the full NP contribution to ${\mathcal{B}}(B^+\to K^++\slashed{E})$ yields the following central values for their Wilson coefficients, 
\begin{align}
    \Big|C_{sb\phi}^{S}\Big |^2=1.5\times 10^{-15}(\text{GeV})^{-2}, \qquad \Big |C_{sb\phi}^{V}\Big|^2=2.4\times 10^{-16}(\text{GeV})^{-4}.
\end{align}
These values correspond to effective mass scales 
\begin{align}
    \Lambda_S\equiv \Big|C_{sb\phi}^{S}\Big |^{-1}=2.6\times 10^4\text{ TeV}, \qquad \Lambda_V\equiv |C_{sb\phi}^{V}\Big|^{-1/2}=8.0\text{ TeV},
\end{align}
respectively. 

\begin{figure}
\centering
\includegraphics[width=0.7\textwidth]{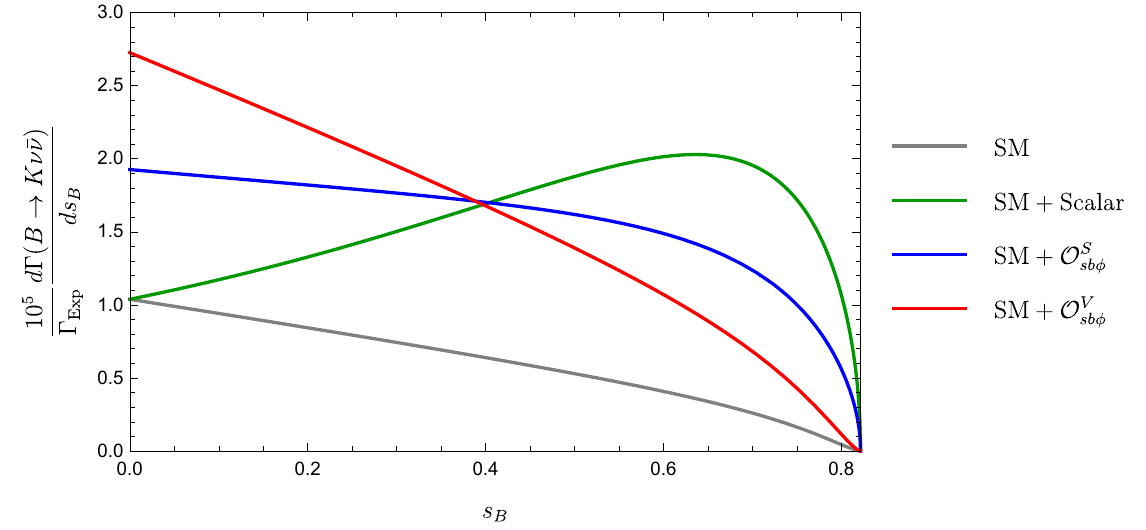}	
\caption{{\bf{Scalar DM.}} Differential decay width of $B\rightarrow K+\slashed{E}$ for the SM (gray), scalar-current (SM)LEFT operators (green), and scalar (D)LEFT operators (blue and red).}
\label{fig:ScalarDMExample}
\end{figure}

With these values for the WCs, we show 
\begin{align}
    &d\Gamma(B\rightarrow K+\slashed{E})_{\text{SM+NP}}/ds_B,
\end{align}
in Fig.~\ref{fig:ScalarDMExample}. Here $s_B=s/m_B^2$. We see that the kinematical distributions associated with the (SM)LEFT operators $\mathcal{O}_{\nu d}^{\text{SLL}}$ and $\mathcal{O}_{\nu d}^{\text{SLR}}$ depicted by the green curve is very different from the kinematical distributions associated with $\mathcal{O}_{sb\phi}^{S}$ ($\mathcal{O}_{sb\phi}^{V}$) shown in blue (red), and the scenarios are easily distinguishable.

\paragraph{Example 2: Vector DM (A)}\mbox{}\\ \mbox{}\\

\begin{figure}[th]
\centering
\includegraphics[width=0.42\textwidth]{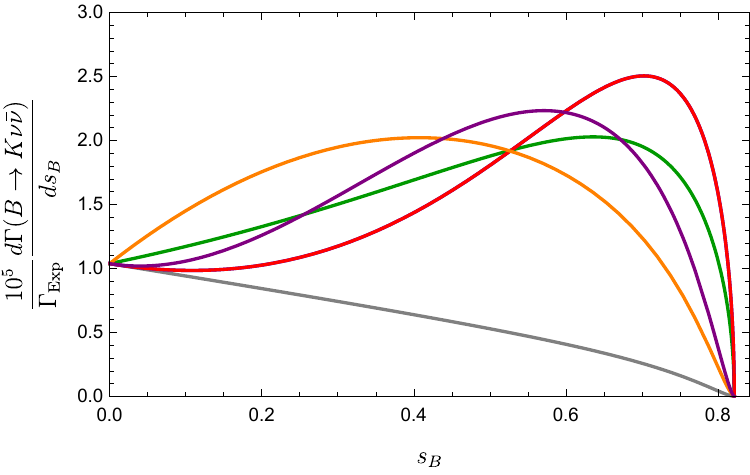}\qquad \quad
\includegraphics[width=0.42\textwidth]{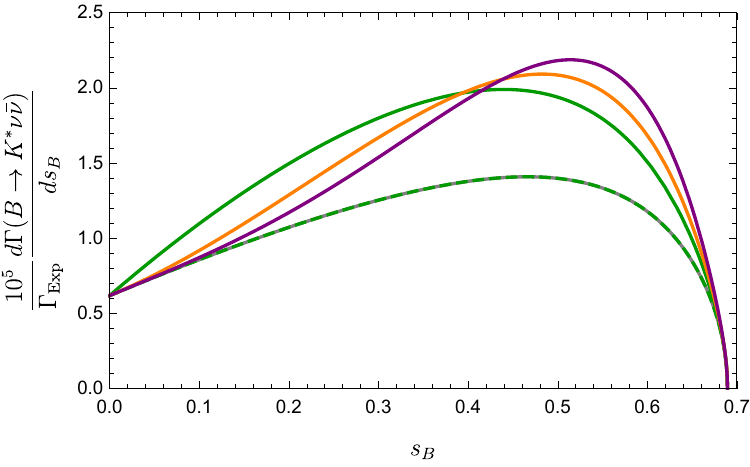}\\
\includegraphics[width=0.65\textwidth]{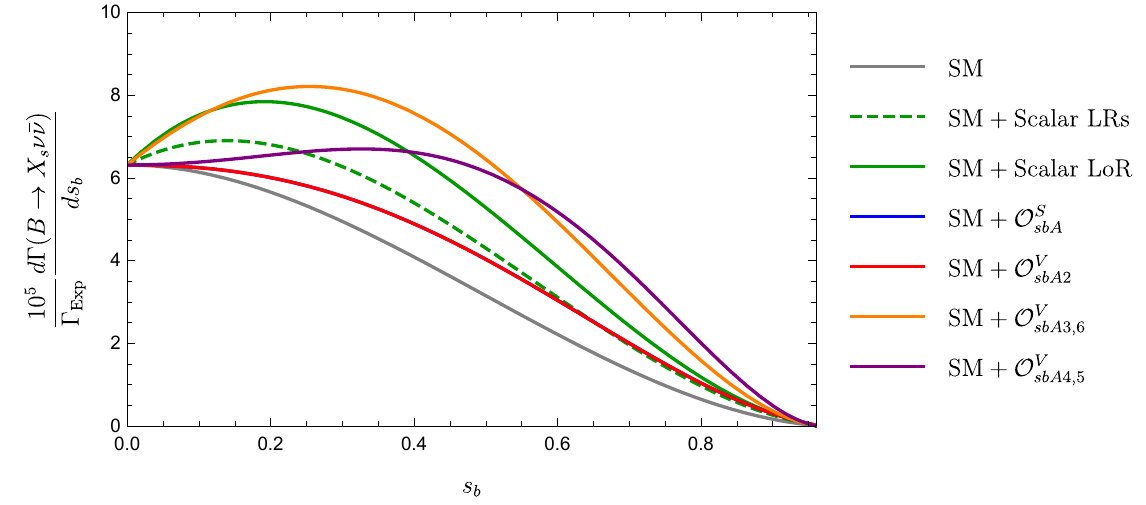}
\caption{{\bf{Vector DM (A).}} Differential decay width of $B\rightarrow K+\slashed{E}$ (top left), $B\rightarrow K^*+\slashed{E}$ (top right), and $B\rightarrow X_s+\slashed{E}$ (bottom) for the SM (gray), scalar-current (SM)LEFT operators (green), and (D)LEFT operators of type A (blue, red, orange, and purple).}
\label{fig:ExampleA1}
\end{figure}

In this example, we consider spin-$1$ dark-sector particles and work in operator scenario A for which the full operator basis is listed in Eqs.~(\ref{Astart})-(\ref{Aend}). In Ref.~\cite{He2023} it was shown that the following subset of operators (defined in Appendix~\ref{appendix:DarkOperators}) can reproduce the Belle II excess without violating experimental bounds on $B\rightarrow K^*\nu\bar{\nu}$,
\begin{align}
    \mathcal{O}_{sbA}^{S},\qquad 
    \mathcal{O}_{sbA2}^{V},\qquad 
    \mathcal{O}_{sbA3}^{V},\qquad 
     \mathcal{O}_{sbA4}^{V},\qquad 
    \mathcal{O}_{sbA5}^{V},\qquad 
    \mathcal{O}_{sbA6}^{V}.
\end{align}
With the contributions of these operators to $d\Gamma(B\rightarrow K+\slashed{E})/ds_B$ at hand, which are listed in Appendix~\ref{appendix:BKinv}, we arrive at the following central values for their corresponding WCs and associated effective mass scales,
\begin{align}
    \left|C_{sbA}^{S}\right|^2&=3.7\times 10^{-17}(\text{GeV})^{-6},  &\Lambda_{sbA}^{S}=550\,\text{GeV}&, \\
    \left|C_{sbA2}^{V}\right|^2&=5.0\times 10^{-17}(\text{GeV})^{-8},  &\Lambda_{sbA2}^{V}=110\,\text{GeV}&, \\
    \left|C_{sbA3}^{V}\right|^2&=2.9\times 10^{-17}(\text{GeV})^{-6},  &\Lambda_{sbA3}^{V}=570\,\text{GeV}&, \\
    \left|C_{sbA4}^{V}\right|^2&=9.6\times 10^{-18}(\text{GeV})^{-8},  &\Lambda_{sbA4}^{V}=130\,\text{GeV}&, \\
    \left|C_{sbA5}^{V}\right|^2&=9.6\times 10^{-18}(\text{GeV})^{-8},  &\Lambda_{sbA5}^{V}=130\,\text{GeV}&, \\
    \left|C_{sbA6}^{V}\right|^2&=2.9\times 10^{-17}(\text{GeV})^{-6},  &\Lambda_{sbA6}^{V}=570\,\text{GeV}&.
\end{align}
With these values for the Wilson coefficients, we show 
\begin{align}
    &d\Gamma(B\rightarrow K+\slashed{E})_{\text{SM+NP}}/ds_B,\\
    &d\Gamma(B\rightarrow K^*+\slashed{E})_{\text{SM+NP}}/ds_B, \\
    &d\Gamma(B\rightarrow X_s+\slashed{E})_{\text{SM+NP}}/ds_b,
\end{align}
in Fig.~(\ref{fig:ExampleA1}). Here $s_b=s/m_b^2$, and we employ the 1S quark-mass value $m_b^{1S}=4.75$ GeV~\cite{Bernlochner:2020jlt}. The blue and red curves overlap, and equal the SM contribution in $B\rightarrow K^*+\slashed{E}$.

Fig.~\ref{fig:ExampleA1} illustrates how different decay channels can be complementary in identifying the source of NP. In particular, it highlights that differences between the (SM)LEFT operators and each of the $\mathcal{O}_{sbA2,3,4,5,6}^{V}$ operators can be most prominent for some values of $s_{B(b)}$ in one decay channel and most prominent for a different range of values of $s_{B(b)}$ in a different decay channel, as seen by comparing e.g. the green and purple curves in the three plots.

\paragraph{Example 3: Vector DM (B)}\mbox{}\\ \mbox{}\\

\begin{figure}[th]
\centering
\includegraphics[width=0.391\textwidth]{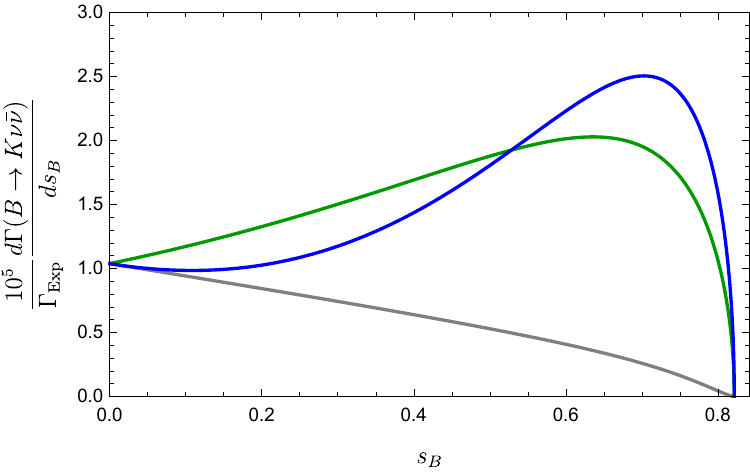}\qquad 
\includegraphics[width=0.55\textwidth]{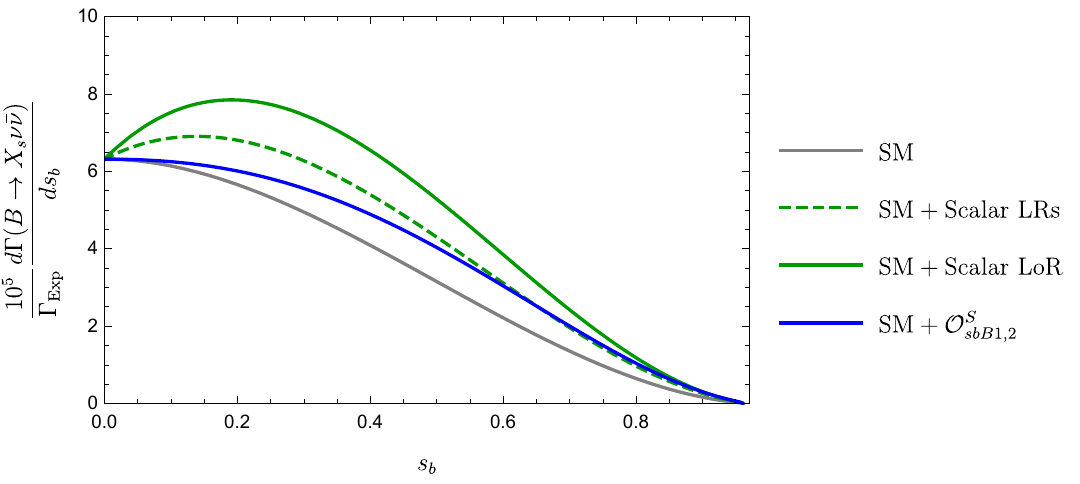}
\caption{{\bf{Vector DM (B).}} Differential decay width of $B\rightarrow K+\slashed{E}$ (left), and $B\rightarrow X_s+\slashed{E}$ (right) for the SM (gray), scalar-current (SM)LEFT operators (green), and (D)LEFT operators of type B (blue).}
\label{fig:ExampleB}
\end{figure}

In the last example, we consider spin-$1$ dark-sector particles and work in operator scenario B for which the full operator basis is listed in Eqs.~(\ref{Bstart})-(\ref{Bend}). The following four operators contribute to $B\rightarrow K\nu\widehat{\nu}$, 
\begin{align}
    \mathcal{O}_{sbB1}^S, \qquad \mathcal{O}_{sbB2}^S,\qquad \mathcal{O}_{sbB1}^T,\qquad \mathcal{O}_{sbB2}^T.
\end{align}
The central values for their corresponding Wilson coefficients and associated effective mass scale are listed below,
\begin{align}
    \left|C_{sbB1,2}^{S}\right|^2&=4.6\times 10^{-18}(\text{GeV})^{-6},  &\Lambda_{sbB1,2}^{S}=780\,\text{GeV}&, \\
    \left|C_{sbB1,2}^{T}\right|^2&=1.5\times 10^{-16}(\text{GeV})^{-6},  &\Lambda_{sbB1,2}^{T}=430\,\text{GeV}&.
\end{align}
We have verified that $\mathcal{O}_{sbB1,2}^T$ are ruled out by experimental bounds on $B\rightarrow K^*\nu\bar{\nu}$. With these values for the WCs, we show $d\Gamma(B\rightarrow K+\slashed{E})_{\text{SM+NP}}/ds_B$ $(d\Gamma(B\rightarrow X_s+\slashed{E})_{\text{SM+NP}}/ds_b)$  in Fig.~\ref{fig:ExampleB} left (right). In this scenario, we see that the low $s_{B(b)}$ region can be particularly useful to distinguish the scalar-current (SM)LEFT operators from the (D)LEFT operators $\mathcal{O}_{sbB1,2}^S$.


\subsection{Sum-rule analysis}
\label{subsec:complementary}
So far, we have succeeded in developing theoretical strategies to disentangle NP contributions through the study of $s$-distributions in $b\to s\nu\widehat{\nu}$ transitions. In particular, we have shown that detailed studies of $s$-distributions can be used to determine the magnitude of certain combinations of Wilson coefficients. The purpose of the following discussion is to highlight some strategies that are less involved from an experimental perspective, analogous to the discussion in Section~\ref{subsubsec:BCorrelations}, but this time emphasizing the utility of sum rules and making some remarks on (D)LEFT.

If only the vector-current (SM)LEFT operators~(\ref{VLL}) and (\ref{VLR}) contribute to rare $B$-decays, then the ratios $\mathcal{R}_K^\nu$, $\mathcal{R}_{K^*}^\nu$, $\mathcal{R}^\nu_{\rm incl}$, and $\mathcal{R}^\nu_{F_L}$ can be expressed in terms of real parameters $\epsilon_{\alpha \beta}>0$ and $\eta_{\alpha \beta} \in [-1/2,1/2]$ as follows~\cite{Buras:2014fpa},
\begin{align}
\label{RK1}
\mathcal{R}_K^\nu&\equiv \frac{\mathcal{B}(B \to K \nu \bar{\nu})}{\mathcal{B}_{\rm SM}(B \to K \nu \bar{\nu})}=\sum_{\alpha \beta} \frac{1}{3}(1-2\eta_{\alpha \beta})\epsilon_{\alpha \beta}^2 \,, \\
\mathcal{R}_{K^*}^\nu&\equiv \frac{\mathcal{B}(B \to K^* \nu \bar{\nu})}{\mathcal{B}_{\rm SM}(B \to K^* \nu \bar{\nu})}=\sum_{\alpha \beta}\frac{1}{3}(1+\kappa_\eta \eta_{\alpha \beta}) \epsilon_{\alpha \beta}^2~,\label{RKRK*1}\\
\label{eq:epseta-incl}
\mathcal{R}^\nu_{\rm incl}&\equiv\frac{\mathcal{B}(B\to X_s\nu\bar\nu)}{\mathcal{B}(B\to X_s\nu\bar\nu)_{\rm{SM}}}=\sum_{\alpha \beta}\frac{1}{3}(1+\kappa_X\eta_{\alpha\beta})\epsilon_{\alpha\beta}^2\, ,
\\
\mathcal{R}^\nu_{F_L} &\equiv \frac{F_L}{F_L^\text{SM}} 
 =  \frac{1}{\mathcal{R}_{K^*}^\nu}
  \sum_{\alpha \beta} \frac{1}{3}   (1+2\eta_{\alpha \beta})\epsilon_{\alpha \beta}^2
 \,,
\label{eq:epseta-R}
\end{align}
with
\be\begin{split}
\epsilon_{\alpha \beta} &=\dfrac{\sqrt{|C_{\nu d,\alpha\beta bs}^{\text{VLL,SM}}\delta_{\alpha \beta}
    +C_{\nu d,\alpha\beta bs}^{\text{VLL}}|^2+|C_{\nu d,\alpha\beta bs}^{\text{VLR}}|^2}}{ |C_{\nu d,\alpha\beta bs}^{\text{VLL,SM}}|} ~, \\
	\eta_{\alpha \beta} &= -\dfrac{\text{Re}\left[ \left( C_{\nu d,\alpha\beta bs}^{\text{VLL,SM}}\delta_{\alpha \beta} +C_{\nu d,\alpha\beta bs}^{\text{VLL}}\right) C_{\nu d,\alpha\beta bs}^{\text{VLR}*}\right]}{|C_{\nu d,\alpha\beta bs}^{\text{VLL,SM}}\delta_{\alpha \beta}
    +C_{\nu d,\alpha\beta bs}^{\text{VLL}}|^2+|C_{\nu d,\alpha\beta bs}^{\text{VLR}}|^2} ~ .
\end{split}\label{eq:epsetadef}
\ee
In the SM one has $\epsilon^{\alpha\beta}=\delta_{\alpha\beta}$, and $\eta^{\alpha\beta}\neq 0$ signals the presence of right-handed currents. The parameter $\kappa_\eta$ depends on the form factors and its explicit form is given in an Appendix in  \cite{Buras:2014fpa}. Presently $\kappa_\eta=1.33\pm0.05$. The parameter $\kappa_X$, which we include for the first time, is given by
\begin{align}
\kappa_X=\frac{12m_bm_s\int_0^{s_{\rm{max}}}ds\,\lambda^{1/2}(s,m_b^2,m_s^2)\cdot s}{\int_0^{s_{\rm{max}}}ds\,\lambda^{1/2}(s,m_b^2,m_s^2)\left[\lambda(s,m_b^2,m_s^2)+3s(m_b^2+m_s^2-s)\right]}\approx 0.08.
\end{align}

The relations presented above are also valid in the case of flavor universality violation and/ or lepton flavor violation. They also imply two model-independent relations, as first pointed out in~\cite{Buras:2014fpa}, which we review below.
The first relation reads
\begin{equation}
\langle F_L \rangle = \langle F_L^\text{SM} \rangle\left[
\left(\frac{(\kappa_\eta-2)\mathcal{R}_K+4\,\mathcal{R}_{K^*}}{(\kappa_\eta+2)\mathcal{R}_{K^*}}\right)+X_1\right]
\,,
\label{eq:FLtest}
\end{equation}
where $X_1$ parametrizes NP that is \textit{not} encoded in vector-current (SM)LEFT operators. 
Eq.~(\ref{eq:FLtest}) is given here in the integrated form, but in principle, it can also be tested experimentally on a bin-by-bin basis. The second relation that we will consider is 
\begin{equation}
\mathcal{B}(B\to X_s\nu\bar\nu)
=
\mathcal{B}(B\to X_s\nu\bar\nu)_\text{SM}\left[\left(
\frac{\kappa_\eta \mathcal{R}_K+2\,\mathcal{R}_{K^*}+\kappa_X(\mathcal{R}_{K^*}-\mathcal{R}_{K})}{\kappa_\eta+2}
\right)+X_2\right]\,,
\label{eq:Xstest}
\end{equation}
\noindent
where $X_2$ parametrizes NP that is \textit{not} encoded in vector-current (SM)LEFT operators. 
This relation improves on the one presented in~\cite{Buras:2014fpa}, where the term involving ${\kappa_X}$ was neglected. Some important comments about (\ref{eq:FLtest}) and (\ref{eq:Xstest}) are listed below.

\begin{itemize}
    \item Experimental results indicating $X_1=X_2=0$ does not uniquely imply that the vector-current (SM)LEFT operators in Eq.~(\ref{VLL})-(\ref{VLR}) are the source of NP. One can show that (\ref{eq:FLtest}) and (\ref{eq:Xstest}) with $X_1=X_2=0$ are also satisfied if the NP is scalar DM satisfying $\left|C^A_{sb\phi}\right|^2=\left|C^V_{sb\phi}\right|^2\neq 0$, with all other NP Wilson coefficients set to zero. The same is true if the NP is fermionic DM with $\left|C^A_{sb\chi1}\right|^2+\left|C^A_{sb\chi2}\right|^2=\left|C^V_{sb\chi1}\right|^2+\left|C^V_{sb\chi2}\right|^2$, with all other NP Wilson coefficients set to zero. This result is consistent with the analysis in Section~(\ref{subsec:DisentanglingB}), where we showed that vector-current interactions in (SM)LEFT cannot always be disentangled from dark fermions or dark scalars by only studying $B\rightarrow K+\slashed{E}$, $B\rightarrow K^*+\slashed{E}$, and inclusive decays.

    \item The relations (\ref{eq:FLtest}) and (\ref{eq:Xstest}) with $X_1=X_2=0$ hold even in the case of lepton flavor non-universality and lepton flavor violation as long as the NP is either attributed to the vector-current LEFT operators in (\ref{VLL})-(\ref{VLR}) or the above-mentioned scenarios with dark particles. Consequently, measuring a nonzero value for $X_1$ or $X_2$ would unambiguously signal either the presence of particles other than neutrinos in the invisible final state or scalar/tensor currents in the (SM)LEFT sector. In other words, measuring a nonzero value for $X_1$ or $X_2$ can be interpreted as a potential sign of LNV, but a conclusive statement would require further evidence. Evidence could be strengthened through dedicated studies of $s$-distributions in $b\to s\nu\widehat{\nu}$ transitions, which have the power to rule out dark scalars and dark vectors as the invisible final states, as shown in Section~\ref{subsec:DisentanglingB}. 
\end{itemize}

{\bf{Limit of LFU and LFC:}} Finally, we recall that in the simplifying limit of LFU and LFC, where $\epsilon_{\alpha \beta}=\epsilon$, $\eta_{\alpha \beta}=\eta$ and
\begin{equation}  \label{eq:epsetadef1}
 \epsilon = \frac{\sqrt{ |C_{\nu d, bs}^{\text{VLL,SM}}+C_{\nu d,bs}^{\text{VLL}}|^2 + |C_{\nu d,bs}^{\text{VLR}}|^2}}{|C_{\nu d, bs}^{\text{VLL,SM}}|}~, \qquad
 \eta = \frac{-\text{Re}\left[\left(C_{\nu d, bs}^{\text{VLL,SM}}+C_{\nu d, bs}^{\text{VLL}}\right) C_{\nu d, bs}^{\text{VLR}*}\right]}{|C_{\nu d, bs}^{\text{VLL,SM}}+C_{\nu d, bs}^{\text{VLL}}|^2 + |C_{\nu d, bs}^{\text{VLR}}|^2}~,
\end{equation}
the relations above simplify significantly~\cite{Buras:2014fpa}:
\begin{align}
 \mathcal{R}^\nu_K   & = (1 - 2\,\eta)\epsilon^2
 \,, &
 \mathcal{R}^\nu_{K^*} 
  & =
  (1 +  \kappa_\eta \eta)\epsilon^2
  \,, &
 \mathcal{R}^\nu_{F_L} \equiv \frac{F_L}{F_L^\text{SM}} 
 & =  
  \frac{1+2\eta}{1+\kappa_\eta\eta}
 \,,
\label{eq:epseta-R1}
\end{align}
\be\label{eq:epseta-incl2}
\mathcal{R}^\nu_{\rm incl}=\frac{\mathcal{B}(B\to X_s\nu\bar\nu)}{\mathcal{B}(B\to X_s\nu\bar\nu)_{\rm{SM}}}=(1+\kappa_X\eta)\epsilon^2\,.
\ee
It is now possible to write down very simple correlations between $\mathcal{R}^\nu_K$, $\mathcal{R}^\nu_{K^*}$, and $\mathcal{R}^\nu_{\rm incl}$,
  \be\label{KK*COR}
  \mathcal{R}^\nu_{K} =\frac{1-2\eta}{1+\kappa_\eta\eta}  \mathcal{R}^\nu_{K^*}=\frac{1-2\eta}{1+\kappa_X\eta} \mathcal{R}^\nu_{\rm incl}.
  \ee

The parameters $\epsilon$ and $\eta$ can be calculated in any neutrino model through (\ref{eq:epsetadef1}), and the results can be presented in the $\epsilon$-$\eta$-plane, with the SM corresponding to
$(\epsilon,\eta)=(1,0)$, and $\eta\not=0$ signals the presence of right-handed currents, see \cite{Buras:2014fpa} for examples. For our purposes, it is important to stress that the presence of right-handed currents would provide non-trivial correlations between $\mathcal{R}^\nu_K$, $\mathcal{R}^\nu_{K^*}$, and $\mathcal{R}^\nu_{\rm incl}$, through (\ref{KK*COR}). Meanwhile, the dark NP scenarios mentioned in the first bullet point on the previous page where either $\left|C^A_{sb\phi}\right|^2=\left|C^V_{sb\phi}\right|^2$ (scalar DM) or $\left|C^A_{sb\chi1}\right|^2+\left|C^A_{sb\chi2}\right|^2=\left|C^V_{sb\chi1}\right|^2+\left|C^V_{sb\chi2}\right|^2$ (fermionic DM), would lead to the following trivial correlation,
\begin{align}
    \mathcal{R}^\nu_{K} =\mathcal{R}^\nu_{K^*}=\mathcal{R}^\nu_{\rm incl}.
\end{align}
Hence, while NP scenarios where only (SM)LEFT vector currents are induced are indistinguishable from these (D)LEFT NP scenarios at the level of the sum rules in (\ref{eq:FLtest}) and (\ref{eq:Xstest}), these (SM)LEFT and (D)LEFT NP scenarios can sometimes be discriminated by considering correlations among $\mathcal{R}^\nu_K$, $\mathcal{R}^\nu_{K^*}$, and $\mathcal{R}^\nu_{\rm incl}$ as summarized below:
\begin{itemize}
    \item A (SM)LEFT NP scenario where only left-handed vector currents are induced can \textit{not} be distinguished from the above-mentioned dark scalar and dark fermion NP scenarios by measuring correlations among $\mathcal{R}^\nu_K$, $\mathcal{R}^\nu_{K^*}$, and $\mathcal{R}^\nu_{\rm incl}$.
    \item All other (SM)LEFT NP scenarios where only vector currents are induced can be distinguished from the above-mentioned dark scalar and dark fermion NP scenarios by measuring correlations among $\mathcal{R}^\nu_K$, $\mathcal{R}^\nu_{K^*}$, and $\mathcal{R}^\nu_{\rm incl}$.
\end{itemize}
The argument extends straightforwardly to the general case described by (\ref{RK1})-(\ref{eq:epseta-R}), where no assumptions about LFV and universality are made, but correlations between $\mathcal{R}^\nu_K$, $\mathcal{R}^\nu_{K^*}$, and $\mathcal{R}^\nu_{\rm incl}$ are generally more involved. 

Having analyzed several strategies for disentangling NP contributions in rare $B$-meson and Kaon decays, we now return to the discussion about lepton-number conservation and lepton-number violation in (SM)LEFT initiated in Section~\ref{sec:OperatorLNV}.


\subsection{Comments on lepton number violation and conservation}
\label{subsec:LNVLNC}
In Section~\ref{sec:OperatorLNV}, we provided an overview of the (SM)LEFT operators relevant for rare meson decays that violate lepton number, as summarized in Table~\ref{tab:LNV/LNC}. The general expectation is that scalar and tensor current operators in (SM)LEFT violate lepton number and that vector current operators in (SM)LEFT conserve lepton number. In other words, if a future measurement points towards a non-zero contribution from (SM)LEFT operators with scalar or tensor currents (vector currents), there could be strong reason to expect LNV (LNC) NP. 

This general expectation fails when the operator contains one active neutrino and one sterile neutrino. In this case, $\mathcal{O}_{\nu d}^{\text{VLL}},\mathcal{O}_{\nu d}^{\text{VLR}}$ are LNV and $\mathcal{O}_{\nu d}^{\text{SLL}},\mathcal{O}_{\nu d}^{\text{SLR}},\mathcal{O}_{\nu d}^{\text{TLL}}$ are LNC. From an experimental perspective, it is generally challenging to draw conclusive statements about the lepton number of operators contributing to rare meson decays, as it requires identification of the invisible final states. However, certain statements about LNV/LNC can be made based on mass considerations, as explained below. 

The contribution from one operator $\mathcal{O}$ to the kinematic distribution $d\Gamma/ds$ is zero for $s<s_{\rm{min}}\equiv(m_1+m_2)^2 $, where $m_1$ and $m_2$ are the masses of the two particles in the invisible final state. With two active neutrinos in the final state, we would expect $0\lesssim s_{\rm{min}}\lesssim (0.2\text{ eV})^2$.\footnote{The lower bound is allowed by neutrino oscillations~\cite{ParticleDataGroup:2018ovx} indicating that at least two active neutrinos are massive, while one could still be massless, and the upper bound is derived from upper limits on absolute neutrino masses from cosmological observations~\cite{Planck:2018vyg} and Tritium decay~\cite{KATRIN:2019yun, KATRIN:2021uub}.} This observation motivates the following two statements: 
\begin{itemize}
    \item If future measurements of NP contributions to kinematic distributions point to $s_{\rm{min}}$ larger than the value compatible with active-neutrino mass measurements, then NP contributions from active neutrinos only are disfavored.\footnote{This statement implicitly assumes that the contribution from active neutrinos to the bin containing the value of $s_{\rm{min}}$ compatible with active neutrino masses is not parametrically smaller than contributions to bins at larger values of $s$.} In practice, this could be realized if the lower momentum bins remain compatible with the SM while an experimental excess is found in bins corresponding to higher values for $s$. 
    \item If future measurements disfavor NP contributions with $s_{\rm{min}}$ beyond a certain threshold $s_{\rm{min}}^{\rm{max}}$, then contributions with one active neutrino and one sterile neutrino with mass $m_N>\sqrt{s_{\rm{min}}^{\rm{max}}}$ in the final state are also disfavored. Contributions with two sterile neutrinos in the final state with mass $m_N>\frac{1}{2}\sqrt{s_{\rm{min}}^{\rm{max}}}$ are similarly disfavored.
\end{itemize}
Similar considerations apply to processes described by (D)LEFT. 
In particular, an experimental excess with $s_{\rm{min}}^{\rm{exp}}$ compatible with the smallness of active neutrino masses and a distribution compatible with scalar or tensor (vector) (SM)LEFT currents, could theoretically have been generated by very light or massless (dark scalars), dark fermions, only active, active and sterile, or only sterile neutrinos, c.f. (\ref{LNVSBK})-(\ref{LNVTBKstar}). This underlines the importance of identifying complementary probes beyond kinematic distributions that can be used to probe the exact nature of the final states. 
The latter goes beyond the scope of the present work.

\subsection{Implications for LNV physics and UV completions}
\label{subsec:implications}
So far, we have considered prospects to search for and disentangle NP in rare meson decays, with special emphasis on contributions that could be lepton-number violating. In this section, we provide perspective on some implications that an observation of LNV in rare meson decays would have for UV physics. Specifically, we discuss the compatibility of $0\nu\beta\beta$ and radiative neutrino mass corrections with observation of lepton-number violation in rare meson decays, and the implications it would have for flavor structures in the UV. We also comment on how the viability of leptogenesis scenarios could be affected if LNV is observed in rare meson decays in the future. The arguments put forward here rely on the assumption of a natural UV completion where the NP contribution to the Wilson coefficients is determined by the mass scale at which NP emerges. A caveat to the following discussion appears if different contributions are canceled in the UV completion, which could relax the NP-scale estimates given below.

\begin{figure}
	\centering
 \raisebox{0.01\height}{\includegraphics[width=0.35\textwidth]{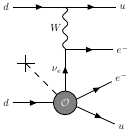}}\,\,\,\,\,\,\,
	\raisebox{0.3\height}{\includegraphics[width=0.35\textwidth]{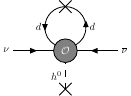}}
	\caption{Contribution to $0\nu\beta\beta$ decay (left) and radiative neutrino-mass diagram (right) induced by $\mathcal{O}_{\overline{d}LQLH1}$.}
	\label{fig:FeynmanDiagrams}
\end{figure}

As a first step, we match scalar-current LEFT operators to SMEFT.
The result, assuming that only active neutrinos are contributing to the NP, reads~\cite{Li2020}
\begin{align}
    \label{LEFT-SMEFT-matching}
    C^{\text{SLL}}_{\nu d, \alpha\beta pr}&=-\frac{v}{4\sqrt{2}}\left(C^{p\alpha r\beta}_{\overline{d}LQLH1}+C^{p\beta r\alpha}_{\overline{d}LQLH1}\right), \qquad C^{\text{SLR}}_{\nu d, \alpha\beta pr}=0, \\
    C^{\text{TLL}}_{\nu d, \alpha\beta pr}&=\frac{v}{16\sqrt{2}}\left(C^{p\alpha r\beta}_{\overline{d}LQLH1}-C^{p\beta r\alpha}_{\overline{d}LQLH1}\right),
\end{align}
where the dimension-$7$ SMEFT operator $\mathcal{O}_{\overline{d}LQLH1}$ is given by, 
\begin{align}
    \label{SMEFTOperator}
\mathcal{O}_{\overline{d}LQLH1}=\epsilon_{ij}\epsilon_{mn}\left(\overline{d}L^i\right)\left(\overline{Q^{Cj}}L^m\right)H^n.
\end{align}
$\mathcal{O}_{\overline{d}LQLH1}$ contributes to $0\nu\beta\beta$ decay, as seen in the Feynman diagram to the left in Fig.~\ref{fig:FeynmanDiagrams}, and the current best constraints from non-observation of $0\nu\beta\beta$ decay reads~\cite{Fridell:2023rtr}, 
\begin{align}
    \label{0vbetabeta-bound}
    C_{\overline{d}LQLH1}\lesssim 7.06\cdot 10^{-8}(\text{TeV})^{-3}, \qquad \Lambda_{\text{NP}}\approx 242\,\text{TeV}.
\end{align}
It is interesting to compare this bound with the NP bounds obtained from rare meson decays. Using the value obtained for the effective Wilson coefficient in (\ref{BoundCSLL}) and the relation (\ref{LEFT-SMEFT-matching}), we obtain the following bounds from rare $B$-meson decays at Belle II 
\begin{align}
    \label{BelleIICutoff}
    C_{\overline{d}LQLH1}\lesssim 0.039\,(\text{TeV})^{-3}, \qquad \Lambda_{\text{NP}}\approx 3.0\,\text{TeV},
\end{align}
where the quoted numerical values are obtained for couplings that are flavor-conserving and universal. The NP bound on $C_{\overline{d}LQLH1}$ from rare $B$ decays is orders of magnitude lower than the corresponding constraint from $0\nu\beta\beta$ in (\ref{0vbetabeta-bound}). However, in the case of $0\nu\beta\beta$ decay only a single flavor combination is tested, and the bound (\ref{0vbetabeta-bound}) only applies to $\mathcal{O}_{\overline{d}LQLH1}$ with the quarks being of first generation and the leptons realized as electron and an electron neutrino. Hence, an observation of LNV in rare-meson decays would provide evidence for the UV physics being flavor non-democratic.

A second interesting aspect of LNV in rare meson decays is its connection to neutrino masses. It is well-known that higher-dimensional LNV operators give radiative contributions to a Majorana neutrino mass~\cite{Cepedello:2017lyo, Cai:2017jrq}, as illustrated to the right in Fig.~\ref{fig:FeynmanDiagrams} for $\mathcal{O}_{\overline{d}LQLH1}$. The radiative-mass contribution from $\mathcal{O}_{\overline{d}LQLH1}$ can be estimated as 
\begin{align}
    \label{delta_m}
    \delta m_{\nu}\approx \frac{y_dv^2}{16\pi^2\Lambda_{\text{NP}}}.
\end{align}
Assuming that the contribution in (\ref{delta_m}) does not exceed $\sim 0.1$ eV implies a lower limit on the NP scale $\Lambda_{\text{NP}}\gtrsim 5\cdot 10^4$ TeV $(10^9\,\text{GeV})$, for the first (second) generation down-type quark Yukawa coupling $y_d=m_d/v$. However, it is imperative to notice that the contribution in (\ref{delta_m}) originates from the flavor-diagonal part of $\mathcal{O}_{\overline{d}LQLH1}$. Meanwhile, the contributions to rare meson decays originate from off-diagonal elements of $\mathcal{O}_{\overline{d}LQLH1}$. Again, the NP scale associated with experimental observation of LNV in rare meson decays in the foreseeable future remains orders of magnitude lower than the lower bound on the NP scale set by neutrino-mass considerations. This provides a second argument for why we would naturally expect non-trivial flavor structures in the UV, if we observe LNV in rare meson decays.\footnote{In particular, in the context of LNV in rare meson decays, an example of a UV setup with leptoquarks yielding sufficiently suppressed neutrino-mass contributions was presented in~\cite{Deppisch2020}.}

Finally, an observation of LNV in rare-meson decays could have implications for high-scale leptogenesis models, such as thermal leptogenesis~\cite{Fukugita1986, Giudice:2003jh}, high-scale resonant leptogenesis~\cite{Covi:1996fm,Pilaftsis:1997dr}, or high-scale flavored leptogenesis~\cite{Abada:2006ea, Domcke:2020quw}. For a review of various specific models, we refer to Refs.~\cite{Chun:2017spz,Elor:2022hpa}. Leptogenesis denotes a class of scenarios for baryogenesis where a lepton asymmetry is generated via CP-violating decays of right-handed neutrinos. The lepton asymmetry is subsequently converted into the observable baryon asymmetry of the universe (BAU) by SM sphaleron processes in the early universe. A necessary condition to achieve a sufficient level of lepton asymmetry required to explain the observed baryon asymmetry of the universe is that processes that diminish the generated lepton asymmetry are sufficiently weak. The latter are commonly referred to as washout processes, and the $\Delta L=2$ processes considered in this paper fall into this category. Hence, the observation of LNV in rare meson decays can be used to constrain mechanisms of leptogenesis, in particular high-scale leptogenesis scenarios where the lepton asymmetry is generated above $\Lambda_{\rm{NP}}$.
See e.g.~\cite{Deppisch2017, Deppisch2020} for further discussion. Specifically, let $n_i$ denote the number density of particle $i$, with $\eta_i \equiv n_i/n_{\gamma}$, where $n_{\gamma}$ is the number density of photons. Even if a very large asymmetry $\eta_L-\eta_{\Bar{L}}\equiv \eta_{\Delta L}\approx 1$ is present at an initial temperature $T$, the asymmetry will be washed out to the extent that it might no longer be able to explain the observed BAU \footnote{Note that for a conclusive falsification of high-scale leptogenesis via low-scale washout processes, equilibration among all flavors has to be ensured, e.g. via the observation of lepton number violation in all flavors, as otherwise the lepton asymmetry could hide in only one flavor~\cite{Cline:1993vv}. Other loopholes exist in scenarios where a decoupled sector shares the baryon asymmetry with the visible sector~\cite{AristizabalSierra:2013lyx} (and references therein). For a general discussion of caveats, see \cite{Deppisch:2013jxa,Deppisch2017}.} in the temperature range~\cite{Deppisch2015,Deppisch2017}
\begin{align}
    \hat{\lambda}\lesssim T\lesssim \Lambda_{\text{NP}}. 
\end{align}
The upper limit is imposed by the EFT description breaking down around $\Lambda_{\text{NP}}$, and the expression for the lower limit can be obtained by solving the Boltzmann equation for $\eta_{\Delta L}$ from the scale where the initial asymmetry $\eta_{\Delta L}$ is created down to the electroweak scale~\cite{Deppisch2017}. For the dimension-$7$ SMEFT operator $\mathcal{O}_{\overline{d}LQLH1}$ relevant to our study, we obtain
\begin{align}
    \hat{\lambda}\sim v,
\end{align}
by using the methods in~\cite{Deppisch2017} and employing the cut-off scale in (\ref{BelleIICutoff}). Hence, observing rare meson decays induced by $\mathcal{O}_{\overline{d}LQLH1}$ would point to a new $\Delta L=2$ washout contribution that is highly efficient in the low TeV range down to the electroweak phase transition. 

\section{Conclusions}
\label{sec:conclusions}
The rare meson decays, $K\rightarrow \pi+\slashed{E}$, $B\rightarrow K+\slashed{E}$, and $B\rightarrow K^*+\slashed{E}$ already place substantial constraints on NP beyond the SM. Yet, there is still significant room left for NP between the SM predictions and currently available experimental data. In anticipation of further experimental data in the years to come, which may show further discrepancies with the SM~\cite{Belle-II:2023esi, CortinaGil:2021nts}, we have proposed a quantitative strategy to disentangle the origin of NP based on careful measurements of kinematic distributions. In particular, our main focus was on prospects of disentangling processes that can be described within (SM)LEFT~\cite{Jenkins2017} or (D)LEFT~\cite{He2022} in the limit of massless neutrinos and massless dark-sector particles. 

Our proposal builds on the observation that different effective operators can give rise to distinguishable kinematic distributions in rare meson decays. In particular, we have shown that it is possible to extract the magnitude of suitable linear combinations of WCs associated with operators of a given Dirac structure. We find that: 

\begin{itemize}
    \item It is possible to separate contributions from (SM)LEFT operators with scalar, vector, and tensor currents in rare meson decays.
    \item It is possible to disentangle left-handed vector currents from right-handed vector currents.
    \item It is possible to determine if scalar currents have a definite handedness,\footnote{Unlike in the case for vector currents, it is \textit{not} possible to determine if the NP current is left-handed or right-handed for scalar currents, only that one is present while the other is not.} or if both left-handed and right-handed scalar currents are present. 
\end{itemize}

A priori, experimental data may contain NP beyond (SM)LEFT. Additional processes where the missing energy is carried away by a single boson will generate a characteristic sharp peak in the kinematic distributions and should be easily distinguishable from (SM)LEFT. However, it is also possible to have two dark particles in the final state, as described by the (D)LEFT framework in~\cite{He2022}. We have employed the latter to investigate to what extent one can distinguish different (SM)LEFT operators from operators that induce rare meson decays with two dark-sector particles in the final state using kinematic distributions alone. In this context, we present new calculations of the inclusive decay rate $B\to X_s+\slashed{E}$ for dark operators. Our treatment goes beyond earlier proposals~\cite{Li2019, Deppisch2020, Gorbahn:2023juq}, as we demonstrate how to systematically quantify NP contributions for an arbitrary number of operators.

More explicitly, we considered the possibility of having all (SM)LEFT and (D)LEFT operators present simultaneously with nonzero Wilson coefficients, and demonstrated the following:
\begin{itemize}
    \item It is possible to disentangle scalar-current (SM)LEFT operators from all (D)LEFT operators with dark scalars and all (D)LEFT operators with dark vectors. 
    \item It is possible to disentangle operators with vector currents in (SM)LEFT from all (D)LEFT operators with spin-$1$ fields, and all (D)LEFT operators with dark scalars involving (pseudo)scalar currents.
    \item It is possible to disentangle tensor-current (SM)LEFT operators from all (D)LEFT operators with dark scalars and all (D)LEFT operators with dark vectors. 
    \item It is \textit{not} possible to distinguish operators with similar Dirac structure, as summarized in Eqs.~(\ref{scalardmK})-(\ref{tensorBK}) for $B\rightarrow K+\slashed{E}$ and (\ref{scalardmKstar})-(\ref{LNVTBKstar}) for $B\rightarrow K^*+\slashed{E}$.
\end{itemize}
In conclusion, analyzing kinematic distributions shows remarkable potential for distinguishing contributions from different operators. However, distinguishing (D)LEFT operators with similar Dirac structure as (SM)LEFT operators is not possible using kinematic distributions \textit{only} and would require complementary probes. We leave this for future work.

We are aware of the fact that our main proposal will be experimentally challenging to execute in the near future. Hence, we also outlined more accessible strategies. Some highlights are: 
\begin{itemize}
    \item We presented novel correlations in the $\mathcal{B}(\kpn)-\mathcal{B}(\klpn)$-plane for pure scalar-current contributions, under the simplifying assumption of lepton-flavor conservation and universality, see Section~\ref{sec:KaonCorrelation}. 
    
    \item We showed that the complex phase in the scalar-current Wilson coefficient can be probed in a combined analysis of $d\Gamma(\kpn)/ds$ and $d\Gamma(\klpn)/ds$, see Section~\ref{subsec:KaonExtraction}.

    \item In the context of $B$-decays, we showed new correlations in the $\mathcal{B}(B^+\rightarrow K^+\nu\bar{\nu})-\mathcal{B}(B\rightarrow K^*\nu\bar{\nu})$-plane and the $\mathcal{B}(B\rightarrow K^*_L\nu\bar{\nu})-\mathcal{B}(B\rightarrow K^*_T\nu\bar{\nu})$-plane for different (SM)LEFT NP scenarios where only scalar currents, tensor currents, or vector currents are generated, see Section~\ref{subsubsec:BCorrelations}.
    
    \item  We revisited and refined two previously known model-independent sum rules that are valid in the presence of flavor universality violation and lepton flavor violation~\cite{Buras:2014fpa}. The first result relates $\langle F_L \rangle$ to $\mathcal{R}_K$ and $\mathcal{R}_{K^*}$, and the second relates $\mathcal{B}(B\to X_s\nu\bar\nu)$ to $\mathcal{R}_K$ and $\mathcal{R}_{K^*}$. We found that these relations only hold for (SM)LEFT operators with vector currents, as was previously known, and for two additional scenarios involving dark scalars and dark fermions. By considering correlations between $\mathcal{R}_K$, $\mathcal{R}_{K^*}$, and $\mathcal{R}^\nu_{\rm{incl}}$, we showed that these dark-sector scenarios are distinguishable from (SM)LEFT scenarios where right-handed vector currents are generated, see Section~\ref{subsec:complementary}.
    
    \item The violation of either of the model-independent sum rules mentioned above would unambiguously signal either the presence of new particles in the invisible final state or scalar/tensor currents in the neutrino sector. Regardless of the source, a violation of the sum rules would motivate a more comprehensive experimental study of kinematic distributions, where the main proposal of this work could be applied. 
\end{itemize}

Finally, we have discussed the intriguing fact that some of the (SM)LEFT considered in this work are lepton-number violating, as summarized in Table~\ref{tab:LNV/LNC}. A general expectation is that the (SM)LEFT with scalar/tensor currents violate lepton number, while the (SM)LEFT operators with vector currents conserve lepton number. This general expectation holds if the two particles in the final state are either both active neutrinos, or both sterile neutrinos. It is violated if the final state consists of one active and one sterile neutrino. In Section~\ref{subsec:LNVLNC}, we commented on how it is possible to discriminate processes with and without sterile neutrinos in the final state using kinematic distributions when the masses of the sterile neutrinos differ significantly from the active-neutrino masses. The discussion of LNV v.s. LNC in rare meson decays highlights that studying kinematic distributions can be useful for gaining insight into NP. However, complementary probes are sometimes required for completely conclusive statements. This should motivate corresponding searches at colliders and beyond.

Finding LNV NP in rare meson decays would have far-reaching consequences for our understanding of questions beyond the SM. If an excess in rare meson decays is in agreement with (SM)LEFT scalar-currents, but in tension with bounds from $0\nu\beta\beta$ decay and radiative neutrino-mass generation, it will either point to highly flavor non-democratic LNV UV physics or LNC physics involving sterile neutrinos or dark-sector fermions. We also discussed the ramifications that observation of a LNV process in rare meson decays would have for leptogenesis. In particular, we showed that a new $\Delta L=2$ washout contribution from scalar and/ or tensor-current operators would emerge. The washout contribution is highly efficient in the low TeV range down to the electroweak phase transition, and has the potential to put high-scale leptogenesis models at tension.

Finally, we are looking forward to accurate data on $B$-decays from Belle II \cite{Kou:2018nap} and on $K$ decays from NA62 and KOTO, which will allow one to use the strategies presented in our paper.

\paragraph{Acknowledgements}\mbox{}\\\mbox{}\\
Financial support of AJB by the Excellence Cluster ORIGINS,
funded by the Deutsche Forschungsgemeinschaft (DFG, German Research
Foundation),
Excellence Strategy, EXC-2094, 390783311 is acknowledged. J. H. acknowledges support from the Emmy Noether grant "Baryogenesis, Dark Matter and Neutrinos: Comprehensive analyses and accurate methods in particle cosmology" (HA 8555/1-1, Project No. 400234416) funded by the Deutsche Forschungsgemeinschaft (DFG, German Research Foundation). M.A.M. acknowledges support from the DFG Collaborative Research Centre “Neutrinos and Dark Matter in Astro- and Particle Physics” (SFB 1258). J.H. and M.A.M. acknowledge support by
the Cluster of Excellence “Precision Physics, Fundamental Interactions, and Structure
of Matter” (PRISMA+ EXC 2118/1) funded by the Deutsche Forschungsgemeinschaft
(DFG, German Research Foundation) within the German Excellence Strategy (Project
No. 390831469).

\begin{appendix}

\section{Form factors}
\label{appendix:formfactors}

\subsection{Kaons}\label{FormfKaons}

The  scalar form factors $f_0^{K^+}(s)$ and $f_0^{K^0}(s)$ are given by
\cite{Mescia:2007kn,Shi:2019gxi,Colangelo:2019axi}
\begin{align}
\label{eq:kaonformfactors}
	f^{K^+}_0(s) &= f^{K^+}_+(0)\left(1 + \lambda_0\frac{s}{m_\pi^2}\right),\quad f^{K^0}_0(s) = f^{K^0}_+(0)\left(1 + \lambda_0\frac{s}{m_\pi^2}\right),
\end{align}
with $\lambda_0 = 13.38\times 10^{-3}$ and,
\begin{align}
\label{eq:fK0}
	f^{K^+}_+(0) = 0.9778, \quad f^{K^0}_+(0) = 0.9544.
\end{align}

The form factor arising from the quark vector current is given by
\begin{align}
\label{eq:kaonformfactors1}
	f^K_+(s) = f^K_+(0)\left(1 + \lambda_+'\frac{s}{m_\pi^2} + \lambda_+''\frac{s^2}{2 m_\pi^4}\right),
\end{align}
where  $\lambda_+' = 24.82\times10^{-3}$, $\lambda_+'' = 1.64\times10^{-3}$.

The form factor arising from the quark tensor current is given by~\cite{Baum:2011rm}
\begin{align}
\label{eq:kaonformfactorstensor}
	f_T^{K^+}(s)=\frac{f_T^{K^+}(0)}{1-\lambda_Ts}, \quad f_T^{K^0}(s)=\frac{f_T^{K^0}(0)}{1-\lambda_Ts}, 
\end{align}
where
\begin{align}
    \frac{1}{\sqrt{2}}f_T^{K^+}(0)=f_T^{K^0}(0)=\frac{0.417}{\sqrt{2}},\quad \lambda_T=1.1\,\text{GeV}^{-2}.
\end{align}
The factor $\frac{1}{\sqrt{2}}$ after the second equality appears because we follow the normalization of form factors in e.g.~\cite{Bordone2019}, which differs from the normalization used in~\cite{Baum:2011rm}.

\subsection{$B$ mesons}\label{FormfB}
For $B\to K\nu\bar\nu$, we need the following three form factors,
\be
f_0^B(s),\qquad  f_+^B(s), \qquad f_T^B(s).
\ee
These form factors have been calculated in \cite{Ball:2004ye} and \cite{Gubernari:2018wyi}. To this end a
fit to the LCSR computation at low $s$ and LQCD \cite{Bouchard:2013eph} at high $s$ has been made.
The results of these papers agree with each other within significant uncertainties. 

To calculate various transversity amplitudes for $B\to K^*\nu\bar\nu$, we need the following form factors
\be
A_0(s), \quad A_1(s), \quad A_{12}(s), \quad V_0(s), \quad T_1(s), \quad T_2(s), \quad T_{23}(s).
\ee
The results for them, obtained using LCSR and LQCD \cite{Horgan:2013hoa,Horgan:2015vla}, can be found
in \cite{Gubernari:2018wyi} and \cite{Bharucha:2015bzk}.

In our numerical calculations, following to some extent \cite{Ball:2004ye},
we have parametrized the form factors as follows 
\be
F(s)=\frac{r_1}{(1-a_1 s/m_B^2)}+\frac{r_2}{(1-a_2  s/m_B^2)}
\ee
and extracted the values of $r_1$, $r_2$, $a_1$ and $a_2$ from
\cite{Ball:2004ye,Bharucha:2015bzk,Gubernari:2018wyi}. To this end, we took
simple averages of the results obtained in these papers and made a fit, which is sufficient for our purposes. The values of these four parameters for the ten form factors in question are collected in Table~\ref{tab:formfactors}.

\begin{table}[ht!]
\begin{center}
\begin{tabular}{||c c c c c||} 
 \hline
 F(s) & $r_1$ & $a_1$ & $r_2$ & $a_2$ \\ [1.5ex] 
 \hline\hline
 $f_0^B(s)$ & $13.84$ & $0.80$ & $-13.52$ & $0.80$ \\ 
 \hline
 $f_+^B(s)$ & $0.18$ & $1.08$ & $0.22$ & $1.08$ \\
 \hline
 $f_T^B(s)$ & $1.19$ & $0.98$ & $-0.87$ & $0.79$ \\
 \hline
  $A_0(s)$ & $0.20$ & $1.15$ & $0.19$ & $1.15$ \\
 \hline
  $A_1(s)$ & $0.66$ & $0.74$ & $-0.39$ & $0.74$ \\
 \hline
  $A_{12}(s)$ & $0.59$ & $0.58$ & $-0.35$ & $0.58$ \\
 \hline
  $V_0(s)$ & $0.07$ & $1.15$ & $0.33$ & $1.15$ \\
 \hline
  $T_1(s)$ & $0.25$ & $1.14$ & $0.11$ & $1.14$ \\
 \hline
  $T_2(s)$ & $0.70$ & $0.73$ & $-0.40$ & $0.73$ \\
 \hline
 $T_{23}(s)$ & $0.22$ & $-0.11$ & $0.41$ & $0.86$ \\[0.5ex] 
 \hline
\end{tabular}
\end{center}
\caption{Fitted form factors for $B$-meson decays.}
\label{tab:formfactors}
\end{table}

\section{Quark-DM operators in LEFT}
\label{appendix:DarkOperators}

For completeness, we also list the dark LEFT operators obtained in Ref.~\cite{He2022}, with a slight change in notation, where $a,b$ denote quark flavors. 

\paragraph{Dark scalar basis:}

\begin{align}
\label{ScalarOperatorStart}
\mathcal{O}_{ab\phi}^S &=  (\overline{q_a} q_b)(\phi^\dagger \phi), 
\\
\mathcal{O}_{ab\phi}^P &=  (\overline{q_a} i \gamma_5 q_b)(\phi^\dagger \phi), 
\\
\mathcal{O}_{ab\phi}^V &=  (\overline{q_a}\gamma^\mu q_b) (\phi^\dagger i \overleftrightarrow{\partial_\mu} \phi), \quad \times_{\rm{rs}} 
\\
\label{ScalarOperatorEnd}
\mathcal{O}_{ab\phi}^A &=  (\overline{q_a}\gamma^\mu\gamma_5 q_b) (\phi^\dagger i \overleftrightarrow{\partial_\mu} \phi),  \quad \times_{\rm{rs}}.  
\end{align}
The $\times_{\rm{rs}}$ indicates that the operator vanishes for a real scalar field~\cite{He2022}. 

\paragraph{Dark fermion basis:}

\begin{align}
\mathcal{O}_{ab\chi1}^{S} &= (\overline{q_a} q_b)(\overline{\chi}\chi),
&
\mathcal{O}_{ab\chi2}^{S}  &= (\overline{q_a} q_b)(\overline{\chi}i \gamma_5\chi), 
\\
\mathcal{O}_{ab\chi1}^{P} &=  (\overline{q_a} i \gamma_5 q_b)(\overline{\chi}\chi),
&
\mathcal{O}_{ab\chi2}^{P} & = (\overline{q_a} \gamma_5 q_b)(\overline{\chi} \gamma_5\chi), 
\\
\label{VdarkStart}
\mathcal{O}_{ab\chi1}^{V} &=  (\overline{q_a}\gamma^\mu  q_b)(\overline{\chi}\gamma_\mu  \chi),
\quad \times_{\rm{M}}
&
\mathcal{O}_{ab\chi2}^{V} &= (\overline{q_a}\gamma^\mu q_b)(\overline{\chi}\gamma_\mu  \gamma_5\chi), 
\\
\label{VdarkEnd}
\mathcal{O}_{ab\chi1}^{A} &=  (\overline{q_a}\gamma^\mu\gamma_5  q_b)(\overline{\chi}\gamma_\mu  \chi),
\quad \times_{\rm{M}}
&
\mathcal{O}_{ab\chi2}^{A} & = (\overline{q_a}\gamma^\mu\gamma_5 q_b)(\overline{\chi}\gamma_\mu  \gamma_5\chi), 
\\
\mathcal{O}_{ab\chi1}^{T} &=  (\overline{q_a}\sigma^{\mu\nu}  q_b)(\overline{\chi}\sigma_{\mu\nu}   \chi),
\quad \times_{\rm{M}}
&
\label{FermionOperatorEnd}
\mathcal{O}_{ab\chi2}^{T} &= (\overline{q_a}\sigma^{\mu\nu} q_b)(\overline{\chi}\sigma_{\mu\nu}  \gamma_5\chi), 
\quad \times_{\rm{M}}
\end{align}
The $\times_{\rm{M}}$ indicates that the operator vanishes if $\chi$ is Majorana~\cite{He2022}.

\paragraph{Dark vector basis case A:}

\begin{align}
\label{Astart}
\mathcal{O}_{ab A}^S &= (\overline{q_a} q_b)(X_\mu^\dagger X^\mu), 
\\
\mathcal{O}_{ab A}^P &= (\overline{q_a}i \gamma_5 q_b)(X_\mu^\dagger X^\mu), 
\\
\mathcal{O}_{ab A1}^T &= {i \over 2} (\overline{q_a}  \sigma^{\mu\nu} q_b) (X_\mu^\dagger X_\nu - X_\nu^\dagger X_\mu),  \quad \times_{\rm{rv}}
\\
\mathcal{O}_{ab A2}^T &= {1\over 2} (\overline{q_a}\sigma^{\mu\nu}\gamma_5 q_b) (X_\mu^\dagger X_\nu - X_\nu^\dagger X_\mu),  \quad \times_{\rm{rv}}
 \\
\mathcal{O}_{ab A1}^V &= {1\over 2} [ \overline{q_a}\gamma_{(\mu} i \overleftrightarrow{D_{\nu)} } q_b] (X^{\mu \dagger} X^\nu + X^{\nu \dagger} X^\mu  ), 
\\
\mathcal{O}_{abA2}^V &= (\overline{q_a}\gamma_\mu q_b)\partial_\nu (X^{\mu \dagger} X^\nu + X^{\nu \dagger} X^\mu  ), 
\\
\mathcal{O}_{ab A3}^V &= (\overline{q_a}\gamma_\mu q_b)( X_\rho^\dagger \overleftrightarrow{\partial_\nu} X_\sigma )\epsilon^{\mu\nu\rho\sigma}, 
\\
\mathcal{O}_{abA4}^V &= (\overline{q_a}\gamma^\mu q_b)(X_\nu^\dagger  i \overleftrightarrow{\partial_\mu} X^\nu), 
\quad \times_{\rm{rv}}
 \\
 \mathcal{O}_{ab A5}^V &= (\overline{q_a}\gamma_\mu q_b)i\partial_\nu (X^{\mu \dagger} X^\nu - X^{\nu \dagger} X^\mu  ),  \quad \times_{\rm{rv}}
 \\
\mathcal{O}_{ab A6}^V &= (\overline{q_a}\gamma_\mu q_b) i \partial_\nu ( X^\dagger_\rho X_\sigma )\epsilon^{\mu\nu\rho\sigma},
\quad \times_{\rm{rv}} 
 \\
\mathcal{O}_{ab A1}^A &= {1\over 2} [\overline{q_a}\gamma_{(\mu} \gamma_5 i \overleftrightarrow{D_{\nu)} }q_b](X^{\mu \dagger} X^\nu + X^{\nu \dagger} X^\mu  ), 
\\
\mathcal{O}_{ab A2}^A &= (\overline{q_a}\gamma_\mu \gamma_5 q_b)\partial_\nu (X^{\mu \dagger} X^\nu + X^{\nu \dagger} X^\mu  ), 
\\ 
\mathcal{O}_{ab A3}^A &= (\overline{q_a}\gamma_\mu\gamma_5 q_b) (X_\rho^\dagger \overleftrightarrow{ \partial_\nu} X_\sigma )\epsilon^{\mu\nu\rho\sigma}, 
\\
\mathcal{O}_{ab A4}^A &= (\overline{q_a}\gamma^\mu\gamma_5 q_b)(X_\nu^\dagger  i \overleftrightarrow{\partial_\mu} X^\nu), 
 \quad \times_{\rm{rv}}
  \\
\mathcal{O}_{ab A5}^A &= (\overline{q_a}\gamma_\mu \gamma_5 q_b)i \partial_\nu (X^{\mu \dagger} X^\nu - X^{\nu \dagger} X^\mu  ),  \quad \times_{\rm{rv}}
 \\
\mathcal{O}_{ab A6}^A &= (\overline{q_a}\gamma_\mu\gamma_5 q_b)i \partial_\nu (  X^\dagger_\rho X_\sigma)\epsilon^{\mu\nu\rho\sigma},
\quad \times_{\rm{rv}}
\label{Aend}
\end{align}
The $\times_{\rm{rv}}$ indicates that the operator vanishes for a real vector field~\cite{He2022}. 

\paragraph{Vector case B:}

\begin{align}
\label{Bstart}
\mathcal{O}_{abB1}^S& = (\overline{q_a}q_b)X_{\mu\nu}^\dagger  X^{\mu\nu},
\\
\mathcal{O}_{abB2}^S& = (\overline{q_a}q_b)X_{\mu\nu}^\dagger \tilde X^{ \mu\nu},
\\
\mathcal{O}_{abB1}^P& = (\overline{q_a}i \gamma_5q_b)X_{\mu\nu}^\dagger X^{ \mu\nu},
\\
\mathcal{O}_{abB2}^P& = (\overline{q_a}i \gamma_5q_b)X_{\mu\nu}^\dagger \tilde X^{ \mu\nu},
\\
\mathcal{O}_{abB1}^T& = {i \over 2} (\overline{q_a}\sigma^{\mu\nu} q_b)(X^{\dagger}_{ \mu\rho} X^{\rho}_{\,\nu}-X^{\dagger}_{ \nu\rho} X^{\rho}_{\,\mu}), \quad \times_{\rm{rv}}
\\
\mathcal{O}_{abB2}^T& = {1\over 2}  (\overline{q_a} \sigma^{\mu\nu}\gamma_5 q_b)(X^{\dagger}_{ \mu\rho} X^{\rho}_{\,\nu}-X^{\dagger}_{ \nu\rho} X^{\rho}_{\,\mu}).\quad \times_{\rm{rv}}
\label{Bend}
\end{align}
The $\times_{\rm{rv}}$ indicates that the operator vanishes for a real vector field~\cite{He2022}. 


\section{Differential decay widths for Kaons}
\label{appendix:KaonDecays}
In this Appendix, we collect explicit expressions for relevant differential decay widths in the limit of vanishing masses for neutrinos and dark-sector particles. The calculations were carried out with the help of \textsc{FeynCalc}~\cite{Shtabovenko2020}. 

\subsection{$K^+\rightarrow \pi^++$inv}
\label{subappendix:K+inv}

\paragraph{Neutrino final state}\mbox{}\\\mbox{}\\
The results are presented in Section~\ref{subsec:kpnandklpn} of our paper.  

\paragraph{Scalar DM final state}\mbox{}\\ \mbox{}\\
The result for a {\em complex} scalar is,
\begin{align}
    \label{KaonScalarDM}
    \frac{d\Gamma}{ds}&=\frac{B_+^2\lambda^{1/2}(m_{K^+}^2,m_{\pi^+}^2,s)}{256\pi^3m^3_{K^+}}\left|f_0^{K^+}(s)\right|^2\left|C_{ds\phi}^S\right|^2\nonumber \\
    &+\frac{\lambda^{3/2}(m_{K^+}^2,m_{\pi^+}^2,s)}{768\pi^3m^3_{K^+}}\left|f_+^{K^+}(s)\right|^2\left|C_{ds\phi}^V\right|^2,
\end{align}
while for {\em real} scalars the term in the second line vanishes, and the term in the first line is a factor $2$ larger. The $s$-dependent functions entering these formulae and the following formulae have been listed for Kaons in Appendix~\ref{FormfKaons}. $B_+$ is given in (\ref{BLB+-text}).
This result agrees with (15) in \cite{He2022}.

\paragraph{Fermion DM final state}\mbox{}\\
\begin{align}
    \label{KaonFermionDM}
    \frac{d\Gamma}{ds}&=\frac{B_+^2\,s\,\lambda^{1/2}(m_{K^+}^2,m_{\pi^+}^2,s)}{128\pi^3m^3_{K^+}}\left|f_0^{K^+}(s)\right|^2\left(\left|C_{ds\chi 1}^S\right|^2+\left|C_{ds\chi 2}^S\right|^2\right)\nonumber \\
    &+\frac{\lambda^{3/2}(m_{K^+}^2,m_{\pi^+}^2,s)}{192\pi^3m^3_{K^+}}\left|f_+^{K^+}(s)\right|^2\left(\left|C_{ds\chi 1}^V\right|^2+\left|C_{ds\chi 2}^V\right|^2\right)\nonumber \\
     &+\frac{s\,\lambda^{3/2}(m_{K^+}^2,m_{\pi^+}^2,s)}{96\pi^3m^3_{K^+}(m_{K^+}+m_{\pi^+})^2}\left|f_T^{K^+}(s)\right|^2\left(\left|C_{ds\chi 1}^T\right|^2+\left|C_{ds\chi 2}^T\right|^2\right).
\end{align}
This result is consistent with the result presented in Appendix A in \cite{He2023}.

\paragraph{Vector DM final state (case A)}\mbox{}\\
The result in the massless limit $m\rightarrow 0$ reads, 
\begin{align}
    \label{KaonVectorDMA1}
    \frac{d\Gamma}{ds}&=\frac{B_+^2\,s^2\,\lambda^{1/2}(m_{K^+}^2,m_{\pi^+}^2,s)}{1024\pi^3m^3_{K^+}}\left|f_0^{K^+}(s)\right|^2\left|\Tilde{C}_{dsA}^S\right|^2\nonumber \\
    &+\frac{s^2(m_{K^+}^2-m_{\pi^+}^2)^2\,\lambda^{1/2}(m_{K^+}^2,m_{\pi^+}^2,s)}{1024\pi^3m^3_{K^+}}\left|f_0^{K^+}(s)\right|^2\left|\Tilde{C}_{dsA2}^V\right|^2\nonumber \\
     &+\frac{s\,\lambda^{3/2}(m_{K^+}^2,m_{\pi^+}^2,s)}{768\pi^3m^3_{K^+}}\left|f_+^{K^+}(s)\right|^2\left(\left|\Tilde{C}_{dsA3}^V\right|^2+\left|\Tilde{C}_{dsA6}^V\right|^2\right)\nonumber \\
     &+\frac{s^2\,\lambda^{3/2}(m_{K^+}^2,m_{\pi^+}^2,s)}{3072\pi^3m^3_{K^+}}\left|f_+^{K^+}(s)\right|^2\left(\left|\Tilde{C}_{dsA4}^V\right|^2+\left|\Tilde{C}_{dsA5}^V\right|^2\right)\nonumber \\
     &+\frac{s^2\,\lambda^{3/2}(m_{K^+}^2,m_{\pi^+}^2,s)}{3072\pi^3m^3_{K^+}(m_{K^+}+m_{\pi^+})^2}\left|f_T^{K^+}(s)\right|^2\left|\Tilde{C}_{dsA1}^T\right|^2\nonumber \\
    &+\mathcal{O}(m^2)\left|\Tilde{C}_{dsA2}^T\right|^2\nonumber \\
    &+\frac{(m_{K^+}^2-m_{\pi^+}^2)^2s^2\,\lambda^{1/2}(m_{K^+}^2,m_{\pi^+}^2,s)}{512\pi^3m_{K^+}^3(m_s-m_d)}\left|f_0^{K^+}(s)\right|^2{\rm{Im}}\left(\Tilde{C}_{dsA}^{S*}\Tilde{C}_{dsA2}^V\right)\nonumber \\
    &+\frac{s^2\lambda^{3/2}(m_{K^+}^2,m_{\pi^+}^2,s)}{1536\pi^3m_{K^+}^3(m_{K^+}+m_{\pi^+})}f_+^{K^+}f_T^{K^+}{\rm{Re}}\left[\Tilde{C}_{dsA1}^T\left(\Tilde{C}_{dsA5}^V-\Tilde{C}_{dsA4}^V\right)^* \right]\nonumber \\
    &-\frac{s^2\lambda^{3/2}(m_{K^+}^2,m_{\pi^+}^2,s)}{1536\pi^3m_{K^+}^3}\left|f_+^{K^+}(s)\right|^2{\rm{Re}}\left(\Tilde{C}_{dsA4}^V\Tilde{C}_{dsA5}^{V*}\right)+\mathcal{O}(m){\rm{Re}}\left(\Tilde{C}_{dsA2}^T\Tilde{C}_{dsA6}^{V*}\right).
\end{align}
The Wilson coefficients with tilde are formally defined as 
\begin{align}
    \label{CTilde}
   C^{S,P}_{dsA}&\equiv m^2 \Tilde{C}^{S,P}_{dsA}, \quad C^{V,A}_{dsA2,4,5}\equiv m^2 \Tilde{C}^{V,A}_{dsA2,4,5},\quad C^{V,A}_{dsA3,6}\equiv m \Tilde{C}^{V,A}_{dsA3,6}, \nonumber \\
   C_{dsA1,2}^T&\equiv m^2\Tilde{C}_{dsA1}^T,
\end{align}
to retain non-diverging results for the decay widths $\Gamma(B\rightarrow K+\slashed{E})$ and $\Gamma(B\rightarrow K^*+\slashed{E})$ in the limit $m\rightarrow 0$~\cite{He2022}. These results agree with (21) and (23) in \cite{He2022}, up to the interference terms, which are presented here for the first time.

\paragraph{Vector DM final state (case B)}\mbox{}\\
\begin{align}
    \label{KaonVectorDMB}
    \frac{d\Gamma}{ds}&=\frac{B_+^2\,s^2\,\lambda^{1/2}(m_{K^+}^2,m_{\pi^+}^2,s)}{128\pi^3m^3_{K^+}}\left|f_0^{K^+}(s)\right|^2\left(\left|C_{dsB1}^S\right|^2+\left|C_{dsB2}^S\right|^2\right)\nonumber \\
    &+\frac{s^2\,\lambda^{3/2}(m_{K^+}^2,m_{\pi^+}^2,s)}{1536\pi^3m^3_{K^+}(m_{K^+}+m_{\pi^+})^2}\left|f_T^{K^+}(s)\right|^2\left(\left|C_{dsB1}^T\right|^2+\left|C_{dsB2}^T\right|^2\right).
\end{align}
This result  agrees with (25) in \cite{He2022}.

\subsection{$K_L\rightarrow \pi^0+$inv}
\label{subsec:KL}

\paragraph{Neutrino final state}\mbox{}\\\mbox{}\\
The result is presented in Section~\ref{subsec:kpnandklpn}.

\paragraph{Scalar DM final state}\mbox{}\\\mbox{}\\
The result
  is readily obtained from (\ref{KaonScalarDM}) by replacing $K^+\rightarrow K^0$, $C_{ds\phi}^S\rightarrow {\text{Re}}\left(C_{ds\phi}^S\right)$, and $C_{ds\phi}^V\rightarrow {\text{Im}}\left(C_{ds\phi}^V\right)$~\cite{He2022}.

\paragraph{Fermion DM final state}\mbox{}\\\mbox{}\\
The result is readily obtained by replacing $K^+\rightarrow K^0$ and 
\begin{align}
    \left|C_{ds\chi1,2}^S\right|^2&\rightarrow \left|{\text{Re}}\left(C_{ds\chi1,2}^S\right)\right|, \qquad \quad\left|C_{ds\chi1,2}^V\right|^2\rightarrow \left|{\text{Im}}\left(C_{ds\chi1,2}^V\right)\right|^2\nonumber \\
    \left|C_{ds\chi1}^T\right|^2&\rightarrow \left|{\text{Re}}\left(C_{ds\chi1}^T\right)\right|, \qquad \quad \quad \,\,\left|C_{ds\chi2}^V\right|^2\rightarrow \left|{\text{Im}}\left(C_{ds\chi2}^V\right)\right|^2
\end{align}
in (\ref{KaonFermionDM}).

\paragraph{Vector DM final state (case A)}\mbox{}\\\mbox{}\\
The result is obtained by replacing $K^+\rightarrow K^0$ and
\begin{align}
    \label{subsA1}
    \left|\Tilde{C}_{dsA}^S\right|^2&\rightarrow \left|{\text{Re}}\left(\Tilde{C}_{dsA}^S\right)\right|^2, \qquad\quad \left|\Tilde{C}_{dsA2-6}^V\right|^2\rightarrow \left|{\text{Im}}\left(\Tilde{C}_{dsA2-6}^V\right)\right|^2, \nonumber \\
    \left|\Tilde{C}_{dsA1}^T\right|^2&\rightarrow \left|{\text{Re}}\left(\Tilde{C}_{dsA1}^T\right)\right|^2, 
    \qquad \quad \,\,\,\,\, \left|\Tilde{C}_{dsA2}^T\right|^2\rightarrow \left|{\text{Im}}\left(\Tilde{C}_{dsA2}^T\right)\right|^2,
\end{align}
in (\ref{KaonVectorDMA1}).

\paragraph{Vector DM final state (case B)}\mbox{}\\\mbox{}\\
The result is obtained by replacing $K^+\rightarrow K^0$ and
\begin{align}
    \label{subsB}
    \left|C_{dsB1,2}^S\right|^2&\rightarrow \left|{\text{Re}}\left(C_{dsB1,2}^S\right)\right|^2, \qquad\quad \left|C_{dsB1}^T\right|^2\rightarrow \left|{\text{Re}}\left(C_{dsB1}^T\right)\right|^2, \nonumber \\
    \left|C_{dsB2}^T\right|^2&\rightarrow \left|{\text{Im}}\left(C_{dsB2}^T\right)\right|^2, 
\end{align}
in (\ref{KaonVectorDMB}).

\section{Differential decay widths for B-mesons}
\label{appendix:B-decays}

\subsection{$B\rightarrow K+$inv}
\label{appendix:BKinv}

All of the results for $B\rightarrow K+$inv can be extracted from Section~\ref{subappendix:K+inv}, by replacing $K\rightarrow B$, $\pi\rightarrow K$, $m_s\rightarrow m_b$, and $m_d\rightarrow m_s$.
This applies in particular to the $s$-dependent functions given in the Appendix~\ref{FormfB} and $B_+$ in (\ref{BLB+-text}).

\subsection{$B\rightarrow K^*+$inv}

\paragraph{Neutrino final state}\mbox{}\\
\begin{align}
    &\frac{d\Gamma(B\rightarrow K^*\nu\widehat{\nu})}{ds}=J_S^{BK^*}\sum_{\alpha\leq\beta}\left(1-\frac{\delta_{\alpha\beta}}{2}\right)\left(\left|C^{\rm{SLL}}_{\nu d,\alpha\beta bs}-C^{\rm{SLR}}_{\nu d,\alpha\beta bs}\right|^2+\left|C^{\rm{SLL}}_{\nu d,\alpha\beta sb}-C^{\rm{SLR}}_{\nu d,\alpha\beta sb}\right|^2\right)\nonumber \\
    &+J_T^{BK^*}\sum_{\alpha<\beta}\left(\left|C^{\rm{TLL}}_{\nu d,\alpha\beta sb}\right|^2+\left|C^{\rm{TLL}}_{\nu d,\alpha\beta bs}\right|^2\right)\nonumber \\
    &+\sum_{\alpha,\beta}\left(1-{1\over2}\delta_{\alpha\beta}\right) \left(J_{V1}^{BK^*}\left|C^{\rm{VLL}}_{\nu d,\alpha\beta bs}-C^{\rm{VLR}}_{\nu d,\alpha\beta bs}\right|^2+J_{V2}^{BK^*}\left|C^{\rm{VLL}}_{\nu d,\alpha\beta bs}+C^{\rm{VLR}}_{\nu d,\alpha\beta bs}\right|^2\right),
\end{align}
where 
\begin{align}
    J_S^{BK^*}&=f_S^L=\frac{s\,\lambda^{3/2}(m_B^2,m_{K^*}^2,s)}{256\pi^3m_B^3(m_b+m_s)^2}\left|A_0(s)\right|^2, \\
    J_T^{BK^*}&=f_T^T+f_T^L=\frac{\lambda^{1/2}(m_B^2,m_{K^*}^2,s)}{24\pi^3m_B^3}\left[\frac{8m_B^2m_{K^*}^2\,s}{(m_B+m_{K^*})^2}\left|T_{23}(s)\right|^2+\lambda(m_B^2,m_{K^*}^2,s)\left|T_1(s)\right|^2\right. \nonumber \\
    &\left.+(m_B^2-m_{K^*}^2)^2\left|T_2(s)\right|^2\right],\\
    J_{V1}^{BK^*}&=f_{V-}^T+f_{V-}^L=\frac{\lambda^{1/2}(m_B^2,m_{K^*}^2,s)}{768\pi^3m_B^3}\left(\left|A_1(s)\right|^2s(m_B+m_{K^*})^2+32\left|A_{12}(s)\right|^2m_B^2m_{K^*}^2\right), \\
    J_{V2}^{BK^*}&=f_{V^+}^T=\frac{\,s\,\lambda^{3/2}(m_B^2,m_{K^*}^2,s)}{768\pi^3m_B^3(m_B+m_{K^*})^2}\left|V_0(s)\right|^2,
\end{align}
where the kinematic functions $f$ and their integrated values were given in~(\ref{fvPlusT})-(\ref{fsL}).
The result is in agreement with~\cite{Altmannshofer:2009ma, Felkl2021}.

\paragraph{Scalar DM final state}\mbox{}\\
\begin{align}
    &\frac{d\Gamma}{ds}=\frac{\lambda^{3/2}(m_B^2,m_{K^*}^2,s)A^2_0(s)}{256\pi^3m_B^3(m_b+m_s)^2}\left|C^P_{sb\phi}\right|^2\nonumber \\
    &+\frac{\lambda^{1/2}(m_B^2,m_{K^*}^2,s)}{384\pi^3m_B^3}\left[s\,(m_B+m_{K^*})^2A^2_1(s)+32m_B^2m_{K^*}^2A^2_{12}(s)\right]\left|C^A_{sb\phi}\right|^2\nonumber \\
    &+\frac{s\,\lambda^{3/2}(m_B^2,m_{K^*}^2,s)V^2_0(s)}{384\pi^3m_B^3(m_B+m_{K^*})^2}\left|C^V_{sb\phi}\right|^2.
\end{align}
This result was first obtained in~\cite{He2022}.

\paragraph{Fermion DM final state}\mbox{}\\
\begin{align}
    &\frac{d\Gamma}{ds}=\frac{\,s\,\lambda^{3/2}(m_B^2,m_{K^*}^2,s)A^2_0(s)}{128\pi^3m_B^3(m_b+m_s)^2}\left(\left|C^P_{sb\chi 1}\right|^2+\left|C^P_{sb\chi 2}\right|^2\right)\nonumber \\
    &+\frac{\lambda^{1/2}(m_B^2,m_{K^*}^2,s)}{96\pi^3m_B^3}\left[A^2_1(s)\,s\,(m_B+m_{K^*})^2+32A^2_{12}(s)m_B^2m_{K^*}^2\right]\left(\left|C^A_{sb\chi 1}\right|^2+\left|C^A_{sb\chi 2}\right|^2\right)\nonumber \\
    &+\frac{s\,\lambda^{3/2}(m_B^2,m_{K^*}^2,s)V^2_0(s)}{96\pi^3m_B^3(m_B+m_{K^*})^2}\left(\left|C^V_{sb\chi 1}\right|^2+\left|C^V_{sb\chi 2}\right|^2\right)\nonumber \\
    &+\frac{\lambda^{1/2}(m_B^2,m_{K^*}^2,s)}{48\pi^3m_B^3}\left[T^2_1(s)\,\lambda(m_B^2,m_{K^*}^2,s)+T^2_2(s)\,(m_B^2-m_{K^*}^2)^2+T^2_{23}\,\frac{8m_B^2m_{K^*}^2\,s}{(m_B+m_{K^*})^2}\right]\nonumber \\
    &\times \left(\left|C^T_{sb\chi 1}\right|^2+\left|C^T_{sb\chi 2}\right|^2\right).
\end{align}
The first two lines are a new result, the rest was first obtained in~\cite{He2023}. 

\paragraph{Vector DM final state (case A)}\mbox{}\\
\begin{align}
    &\frac{d\Gamma}{ds}=\frac{s^2\lambda^{3/2}(m_B^2,m_{K^*}^2,s)\left|A_0(s)\right|^2}{1024\pi^3m_B^3(m_b+m_s)^2}\left|\Tilde{C}^P_{sbA}\right|^2\nonumber \\
    &+\frac{s\lambda^{3/2}(m_B^2,m_{K^*}^2,s)\left|T_1(s)\right|^2}{1536\pi^3m_B^3}\left|\Tilde{C}^T_{sbA1}\right|^2\nonumber \\
    &+\frac{s\lambda^{1/2}(m_B^2,m_{K^*}^2,s)}{1536\pi^3m_B^3}\left[\left(m_B^2-m_{K^*}^2\right)^2\left|T_2(s)\right|^2+\frac{8m_B^2m_{K^*}^2s}{(m_B+m_{K^*})^2}\left|T_{23}(s)\right|^2\right]\left|\Tilde{C}^T_{sbA2}\right|^2\nonumber \\
    &+\mathcal{O}(m^2)\left|\Tilde{C}^V_{sbA2}\right|^2\nonumber \\
    &+\frac{s^2\,\lambda^{3/2}(m_B^2,m_{K^*}^2,s)\left|V_0(s)\right|^2}{384\pi^3m_B^3(m_B+m_{K^*})^2}\left(\left|\Tilde{C}^V_{sbA3}\right|^2+\left|\Tilde{C}^V_{sbA6}\right|^2\right)\nonumber \\
    &+\frac{s^3\,\lambda^{3/2}(m_B^2,m_{K^*}^2,s)\left|V_0(s)\right|^2}{1536\pi^3m_B^3(m_B+m_{K^*})^2}\left(\left|\Tilde{C}^V_{sbA4}\right|^2+\left|\Tilde{C}^V_{sbA5}\right|^2\right)\nonumber \\
    &+\frac{s^2\lambda^{3/2}(m_B^2,m_{K^*}^2,s)\left|A_0(s)\right|^2}{1028\pi^3m_B^3}\left|\Tilde{C}^A_{sbA2}\right|^2\nonumber \\
    &+\frac{s\,\lambda^{1/2}(m_B^2,m_{K^*}^2,s)}{384\pi^3m_B^3}\left[(m_B+m_{K^*})^2s\,\left|A_1(s)\right|^2+32m_B^2m_{K^*}^2\left|A_{12}(s)\right|^2\right]\left(\left|\Tilde{C}^A_{sbA3}\right|^2+\left|\Tilde{C}^A_{sbA6}\right|^2\right)\nonumber \\
     &+\frac{s^2\,\lambda^{1/2}(m_B^2,m_{K^*}^2,s)}{1536\pi^3m_B^3}\left[(m_B+m_{K^*})^2s\,\left|A_1(s)\right|^2+32m_B^2m_{K^*}^2\left|A_{12}(s)\right|^2\right]\left(\left|\Tilde{C}^A_{sbA4}\right|^2+\left|\Tilde{C}^A_{sbA5}\right|^2\right)\nonumber \\
     &+\frac{s^2\lambda^{3/2}(m_B^2,m_{K^*}^2,s)\left|A_0(s)\right|^2}{512\pi^3m_B^3(m_b+m_s)}{\rm{Im}}\left(\Tilde{C}^{P*}_{sbA}\Tilde{C}^A_{sbA2}\right)-\frac{s^3\lambda^{3/2}(m_B^2,m_{K^*}^2,s)\left|V_0(s)\right|^2}{768\pi^3m_B^3(m_B+m_{K^*})^2}{\rm{Re}}\left(\Tilde{C}^{V*}_{sbA4}\Tilde{C}^{V}_{sbA5}\right)\nonumber \\
     &-\frac{s^2\lambda^{1/2}(m_B^2,m_{K^*}^2,s)}{768\pi^3m_B^3}\left[(m_B+m_{K^*})^2s\,\left|A_1(s)\right|^2+32m_B^2m_{K^*}^2\left|A_{12}(s)\right|^2\right]{\rm{Re}}\left(\Tilde{C}^{A*}_{sbA4}\Tilde{C}^{A}_{sbA5}\right)\nonumber \\
     &+\frac{s^2\lambda^{3/2}(m_B^2,m_{K^*}^2,s)T_1(s)V_0(s)}{768\pi^3m_B^3(m_B+m_{K^*})}{\rm{Re}}\left[\left(\Tilde{C}^V_{sbA4}-\Tilde{C}^V_{sbA5}\right)^*\Tilde{C}^T_{sbA1}\right]\nonumber \\
     &+\frac{s^2\lambda^{1/2}(m_B^2,m_{K^*}^2,s)}{768\pi^3m_B^3(m_B+m_{K^*})}\left[(m_B^2-m_{K^*}^2)(m_B+m_{K^*})^2A_1(s)T_2(s)+16m_B^2m_{K^*}^2A_{12}(s)T_{23}(s)\right]\nonumber \\
     &\times{\rm{Im}}\left[\left(\Tilde{C}^A_{sbA4}-\Tilde{C}^A_{sbA5}\right)^*\Tilde{C}^T_{sbA2}\right]+\mathcal{O}(m){\rm{Re}}\left(\Tilde{C}^{V*}_{sbA6}\Tilde{C}^T_{sbA2}\right)+\mathcal{O}(m){\rm{Im}}\left(\Tilde{C}^{A*}_{sbA6}\Tilde{C}^T_{sbA1}\right).
\end{align}
The Wilson coefficients with tilde are defined as in Eq.~(\ref{CTilde}). Our result for the non-interference terms are in agreement with Ref.~\cite{He2022}, while the interference terms are presented here for the first time.

\paragraph{Vector DM final state (case B)}\mbox{}\\
\begin{align}
    &\frac{d\Gamma}{ds}=\frac{s^2\,\lambda^{3/2}(m_B^2,m_{K^*}^2,s)}{128\pi^3m_B^3(m_b+m_s)^2}\left|A_{0}(s)\right|^2\left(\left|C^P_{sbB1}\right|^2+\left|C^P_{sbB2}\right|^2\right)\nonumber \\
    &+\frac{s\,\lambda^{1/2}(m_B^2,m_{K^*}^2,s)}{768\pi^3m_B^3}\left[\lambda(m_B^2,m_{K^*}^2,s)\left|T_{1}(s)\right|^2+(m_B^2-m_{K^*}^2)^2\left|T_{2}(s)\right|^2\right.\nonumber \\
    &\left.+\frac{8m_B^2m_{K^*}^2\,s}{(m_B+m_{K^*})^2}\left|T_{23}(s)\right|^2\right]\times\left(\left|C^T_{sbB1}\right|^2+\left|C^T_{sbB2}\right|^2\right).
\end{align}
The result is in agreement with\cite{He2022}.

\subsection{$B\rightarrow X_s+$inv}
\label{appendix:Inclusive}
The results for $B\rightarrow X_s\nu\widehat{\nu}$ are presented in Section~\ref{sec:BExtraction}.
The results for dark-sector particles in the final state are, to the best of our knowledge, presented here for the first time.

\paragraph{Scalar DM final state}\mbox{}\\
\begin{align}
    \label{InclusiveScalarDM}
    \frac{d\Gamma}{ds}&=\frac{\kappa(0)\lambda^{1/2}(m_b^2,m_s^2,s)}{256\pi^3m_b^3}\left[(m_b+m_s)^2-s\right]\left|C^S_{sb\phi}\right|^2\nonumber \\
    &+\frac{\kappa(0)\lambda^{1/2}(m_b^2,m_s^2,s)}{256\pi^3m_b^3}\left[(m_b-m_s)^2-s\right]\left|C^P_{sb\phi}\right|^2\nonumber \\
    &+\frac{\kappa(0)\lambda^{1/2}(m_b^2,m_s^2,s)}{756\pi^3m_b^3}\left[\lambda(m_b^2,m_s^2,s)+3s\left((m_b-m_s)^2-s\right)\right]\left|C^V_{sb\phi}\right|^2\nonumber \\
    &+\frac{\kappa(0)\lambda^{1/2}(m_b^2,m_s^2,s)}{756\pi^3m_b^3}\left[\lambda(m_b^2,m_s^2,s)+3s\left((m_b+m_s)^2-s\right)\right]\left|C^A_{sb\phi}\right|^2.
\end{align}

\paragraph{Fermion DM final state}\mbox{}\\
\begin{align}
     \label{InclusiveFermionDM}
    &\frac{d\Gamma}{ds}=\frac{\kappa(0)\lambda^{1/2}(m_b^2,m_s^2,s)}{384\pi^3m_b^3}\left[3s\left((m_b+ m_s)^2-s\right)\right]\left(\left|C^S_{sb\chi1}\right|^2+\left|C^S_{sb\chi2}\right|^2\right)\nonumber \\
    &+\frac{\kappa(0)\lambda^{1/2}(m_b^2,m_s^2,s)}{384\pi^3m_b^3}\left[3s\left((m_b-m_s)^2-s\right)\right]\left(\left|C^P_{sb\chi1}\right|^2+\left|C^P_{sb\chi2}\right|^2\right)\nonumber \\
    &+\frac{\kappa(0)\lambda^{1/2}(m_b^2,m_s^2,s)}{192\pi^3m_b^3}\left[\lambda(m_b^2,m_s^2,s)+3s\left((m_b-m_s)^2-s\right)\right]\left(\left|C^V_{sb\chi1}\right|^2+\left|C^V_{sb\chi2}\right|^2\right)\nonumber \\
    &+\frac{\kappa(0)\lambda^{1/2}(m_b^2,m_s^2,s)}{192\pi^3m_b^3}\left[\lambda(m_b^2,m_s^2,s)+3s\left((m_b+m_s)^2-s\right)\right]\left(\left|C^A_{sb\chi1}\right|^2+\left|C^A_{sb\chi2}\right|^2\right)\nonumber \\
    &+\frac{\kappa(0)\lambda^{1/2}(m_b^2,m_s^2,s)}{48\pi^3m_b^3}\left[2\lambda(m_b^2,m_s^2,s)+3s(m_b^2+m_s^2-s)\right]\left(\left|C^T_{sb\chi1}\right|^2+\left|C^T_{sb\chi2}\right|^2\right).
\end{align}

\paragraph{Vector DM final state (case A)}\mbox{}\\

\begin{align}
    &\frac{d\Gamma}{ds}=\frac{\kappa(0)\,s^2\,\lambda^{1/2}(m_b^2,m_s^2,s)}{1024\pi^3m_b^3}\left[\left(m_b+m_s\right)^2-s\right]\left|\Tilde{C}^S_{sbA}\right|^2\nonumber \\
    &+\frac{\kappa(0)\,s^2\,\lambda^{1/2}(m_b^2,m_s^2,s)}{1024\pi^3m_b^3}\left[\left(m_b-m_s\right)^2-s\right]\left|\Tilde{C}^P_{sbA}\right|^2\nonumber \\
    &+\frac{\kappa(0)\,s\,\lambda^{1/2}(m_b^2,m_s^2,s)}{3072\pi^3m_b^3}\left[2\lambda(m_b^2,m_s^2,s)+3s\left((m_b-m_s)^2-s\right)\right]\left|\Tilde{C}^T_{sbA1}\right|^2\nonumber \\
    &+\frac{\kappa(0)\,s\,\lambda^{1/2}(m_b^2,m_s^2,s)}{3072\pi^3m_b^3}\left[2\lambda(m_b^2,m_s^2,s)+3s\left((m_b+m_s)^2-s\right)\right]\left|\Tilde{C}^T_{sbA2}\right|^2\nonumber \\
    &+\frac{\kappa(0)\,s^2\,\lambda^{1/2}(m_b^2,m_s^2,s)}{1024\pi^3m_b^3}(m_b-m_s)^2\left[(m_b+m_s)^2-s\right]\left|\Tilde{C}^V_{sbA2}\right|^2\nonumber \\
    &+\frac{\kappa(0)\,s\,\lambda^{1/2}(m_b^2,m_s^2,s)}{756\pi^3m_b^3}\left[\lambda(m_b^2,m_s^2,s)+3s\left((m_b-m_s)^2-s\right)\right]\left(\left|\Tilde{C}^V_{sbA3}\right|^2+\left|\Tilde{C}^V_{sbA6}\right|^2\right)\nonumber \\
    &+\frac{\kappa(0)\,s^2\,\lambda^{1/2}(m_b^2,m_s^2,s)}{3072\pi^3m_b^3}\left[\lambda(m_b^2,m_s^2,s)+3s\left((m_b-m_s)^2-s\right)\right]\left(\left|\Tilde{C}^V_{sbA4}\right|^2+\left|\Tilde{C}^V_{sbA5}\right|^2\right)\nonumber \\
    &+\frac{\kappa(0)\,s^2\,\lambda^{1/2}(m_b^2,m_s^2,s)}{1024\pi^3m_b^3}(m_b+m_s)^2\left[(m_b-m_s)^2-s\right]\left|\Tilde{C}^A_{sbA2}\right|^2\nonumber \\
    &+\frac{\kappa(0)\,s\,\lambda^{1/2}(m_b^2,m_s^2,s)}{756\pi^3m_b^3}\left[\lambda(m_b^2,m_s^2,s)+3s\left((m_b+m_s)^2-s\right)\right]\left(\left|\Tilde{C}^A_{sbA3}\right|^2+\left|\Tilde{C}^A_{sbA6}\right|^2\right)\nonumber \\
    &+\frac{\kappa(0)\,s^2\,\lambda^{1/2}(m_b^2,m_s^2,s)}{3072\pi^3m_b^3}\left[\lambda(m_b^2,m_s^2,s)+3s\left((m_b+m_s)^2-s\right)\right]\left(\left|\Tilde{C}^A_{sbA4}\right|^2+\left|\Tilde{C}^A_{sbA5}\right|^2\right)\nonumber \\
    &-\frac{\kappa(0)\,s^2\,\lambda^{1/2}(m_b^2,m_s^2,s)}{1536\pi^3m_b^3}\left[\lambda(m_b^2,m_s^2,s)+3s\left((m_b-m_s)^2-s\right)\right]{\rm{Re}}\left(\Tilde{C}^{V*}_{sbA4}\Tilde{C}^{V}_{sbA5}\right)\nonumber \\
    &-\frac{\kappa(0)\,s^2\,\lambda^{1/2}(m_b^2,m_s^2,s)}{1536\pi^3m_b^3}\left[\lambda(m_b^2,m_s^2,s)+3s\left((m_b+m_s)^2-s\right)\right]{\rm{Re}}\left(\Tilde{C}^{A*}_{sbA4}\Tilde{C}^{A}_{sbA5}\right)\nonumber \\
    &-\frac{\kappa(0)\,s^2\,\lambda^{1/2}(m_b^2,m_s^2,s)}{512\pi^3m_b^3}(m_b-m_s)\left((m_b+m_s)^2-s\right){\rm{Im}}\left(\Tilde{C}^{V*}_{sbA2}\Tilde{C}^{S}_{sbA}\right)\nonumber \\
    &-\frac{\kappa(0)\,s^2\,\lambda^{1/2}(m_b^2,m_s^2,s)}{512\pi^3m_b^3}(m_b+m_s)\left((m_b-m_s)^2-s\right){\rm{Re}}\left(\Tilde{C}^{A*}_{sbA2}\Tilde{C}^{P}_{sbA}\right)\nonumber \\
    &+\frac{\kappa(0)\,s^2\,\lambda^{1/2}(m_b^2,m_s^2,s)}{512\pi^3m_b^3}(m_b+m_s)\left((m_b-m_s)^2-s\right){\rm{Im}}\left[\left(\Tilde{C}^{V}_{sbA5}-\Tilde{C}^{V}_{sbA4}\right)^*\Tilde{C}^{T}_{sbA1}\right]\nonumber \\
    &+\frac{\kappa(0)\,s^2\,\lambda^{1/2}(m_b^2,m_s^2,s)}{512\pi^3m_b^3}(m_b-m_s)\left((m_b+m_s)^2-s\right){\rm{Im}}\left[\left(\Tilde{C}^{A}_{sbA5}-\Tilde{C}^{A}_{sbA4}\right)^*\Tilde{C}^{T}_{sbA2}\right]\nonumber \\
    &+\mathcal{O}(m){\rm{Im}}\left(\Tilde{C}^{V*}_{sbA6}\Tilde{C}^{T}_{sbA2}\right)++\mathcal{O}(m){\rm{Re}}\left(\Tilde{C}^{A*}_{sbA6}\Tilde{C}^{T}_{sbA1}\right).
\end{align}

\paragraph{Vector DM final state (case B)}\mbox{}\\
\begin{align}
    &\frac{d\Gamma}{ds}=\frac{\kappa(0)\,s^2\,\lambda^{1/2}(m_b^2,m_s^2,s)}{128\pi^3m_b^3}\left[\left(m_b+m_s\right)^2-s\right]\left(\left|C^S_{sbB2}\right|^2+\left|C^S_{sbB1}\right|^2\right)\nonumber \\
    &+\frac{\kappa(0)\,s^2\,\lambda^{1/2}(m_b^2,m_s^2,s)}{128\pi^3m_b^3}\left[\left(m_b-m_s\right)^2-s\right]\left(\left|C^P_{sbB2}\right|^2+\left|C^P_{sbB1}\right|^2\right)\nonumber \\
    &+\frac{\kappa(0)\,s^2\,\lambda^{1/2}(m_b^2,m_s^2,s)}{756\pi^3m_b^3}\left[2\lambda(m_b^2,m_s^2,s)+3s\left(m_b^2+m_s^2-s\right)\right]\left(\left|C^T_{sbB2}\right|^2+\left|C^T_{sbB1}\right|^2\right).
\end{align}


\section{Simplified relations for Wilson coefficients}
\label{appendix:WilsonWOTensor}
In Subsections~\ref{subsec:KaonExtraction} and \ref{sec:BExtraction}, we presented explicit formulae for the magnitude of effective Wilson coefficients in terms of kinematic functions assuming that NP is fully described by LEFT. These formulae are valid in generic NP scenarios where there can be contributions from (pseudo)scalar, (axial-)vector, and tensor quark currents. In this Appendix, we show how these formulae simplify when there is no tensor quark-current contribution. 

In the case of $K^+\rightarrow \pi^+\nu\widehat{\nu}$, the relations~(\ref{eq:CS+}) and~(\ref{eq:CV+}) describing the magnitude of the scalar and vector contributions, respectively, simplify to
\begin{align}
    C_{S}^+&=\frac{D_+^{\rm exp}(s_1)f^+_V(s_2)-D_+^{\rm exp}(s_2)f^+_V(s_1)}{f^+_S(s_1)f^+_V(s_2)-f^+_S(s_2)f^+_V(s_1)}\,, \\
    C_{V}^+&=\frac{D_+^{\rm exp}(s_1)f^+_S(s_2)-D_+^{\rm exp}(s_2)f^+_S(s_1)}{f^+_V(s_1)f^+_S(s_2)-f^+_V(s_2)f^+_S(s_1)}\,. 
\end{align}
Similarly, the effective WCs describing $K_L\rightarrow \pi^0\nu\widehat{\nu}$ in~(\ref{eq:CSL}) and~(\ref{eq:CVL}) reduce to
\begin{align}
    C_{S}^L&=\frac{D_L^{\rm exp}(s_1)f^L_V(s_2)-D_L^{\rm exp}(s_2)f^L_V(s_1)}{f^L_S(s_1)f^L_V(s_2)-f^L_S(s_2)f^L_V(s_1)}\,, \\
    C_{V}^L&=\frac{D_L^{\rm exp}(s_1)f^L_S(s_2)-D_L^{\rm exp}(s_2)f^L_S(s_1)}{f^L_V(s_1)f^L_S(s_2)-f^L_V(s_2)f^L_S(s_1)}\,. 
\end{align}

For the case of $B$-mesons, the relations~(\ref{CV+T})-(\ref{CV-T}) and (\ref{CSL})-(\ref{CVL}) parametrizing NP in $F_T$ and $F_L$, respectively, simplify to
\begin{align}
    C_{V+}^T&=\frac{P_T^{\rm exp}(s_1)f^T_{V-}(s_2)-P_T^{\rm exp}(s_2)f^T_{V-}(s_1)}{f^T_{V+}(s_1)f^T_{V-}(s_2)-f^T_{V+}(s_2)f^T_{V-}(s_1)}\,, \\
    C_{V-}^T&=\frac{P_T^{\rm exp}(s_1)f^T_{V+}(s_2)-P_T^{\rm exp}(s_2)f^T_{V+}(s_1)}{f^T_{V-}(s_1)f^T_{V+}(s_2)-f^T_{V-}(s_2)f^T_{V+}(s_1)}\,, \\
    C_{S}^L&=\frac{P_L^{\rm exp}(s_1)f^L_{V-}(s_2)-P_L^{\rm exp}(s_2)f^L_{V-}(s_1)}{f^L_{S}(s_1)f^L_{V-}(s_2)-f^L_{V+}(s_2)f^L_{V-}(s_1)}\,, \\
    C_{V-}^L&=\frac{P_L^{\rm exp}(s_1)f^L_{S}(s_2)-P_L^{\rm exp}(s_2)f^L_{S}(s_1)}{f^L_{V-}(s_1)f^L_{S}(s_2)-f^L_{V-}(s_2)f^L_{S}(s_1)}\,.
\end{align}
Similarly, the effective WCs describing $B^+\rightarrow \pi^+\nu\widehat{\nu}$ in~(\ref{CSB}) and~(\ref{CVB}) reduce to
\begin{align}
    C_{S}^{BK}&=\frac{D_{BK}^{\rm exp}(s_1)f^{BK}_V(s_2)-D_{BK}^{\rm exp}(s_2)f^{BK}_V(s_1)}{f^{BK}_S(s_1)f^{BK}_V(s_2)-f^{BK}_S(s_2)f^{BK}_V(s_1)}\,, \\
    C_{V}^{BK}&=\frac{D_{BK}^{\rm exp}(s_1)f^{BK}_S(s_2)-D_{BK}^{\rm exp}(s_2)f^{BK}_S(s_1)}{f^{BK}_V(s_1)f^{BK}_S(s_2)-f^{BK}_V(s_2)f^{BK}_S(s_1)}. 
\end{align}


\section[(In)Distinguishability of dark fermions from (SM)LEFT scalar currents]{(In)Distinguishability of dark fermions from\\ (SM)LEFT scalar currents}
\label{appendix:LNVDarkFermion}

In this Appendix, we explicitly show to what extent $\{\mathcal{O}^{\text{S}}_{sb\chi 1,2},\,\mathcal{O}^{\text{P}}_{sb\chi 1,2}\}$ is (in)distinguishable from $\{\mathcal{O}^{\rm{SLL}}_{\nu d},\,\mathcal{O}^{\rm{SLR}}_{\nu d}\}$ in our treatment. Although the results can be trivially obtained from a redefinition of the dark basis, we have included a short discussion for completeness. The discussion is separated into four cases, which capture all possible scenarios.

\paragraph{1. Two LEFT operators v.s. two dark LEFT operators:}

Starting with the most generic setup, we investigate whether it is always possible to distinguish a scenario where both $\mathcal{O}^{\rm{SLL}}_{\nu d}$ and $\mathcal{O}^{\rm{SLR}}_{\nu d}$ have nonzero Wilson coefficients from a scenario where both $\mathcal{O}^{\text{S}}_{sb\chi 1,2}$ and $\mathcal{O}^{\text{P}}_{sb\chi 1,2}$ have nonzero Wilson coefficients. 

For any nonzero values of $C^{\rm{SLL}}_{\nu d}$, and $C^{\rm{SLR}}_{\nu d}$, one can choose dark-fermion WCs satisfying
\begin{align}
    \label{BKE-condition}
    2\left(\left|C_{sb\chi 1}^S\right|^2+\left|C_{sb\chi 2}^S\right|^2\right)&=\sum_{\alpha\leq \beta} \left(1-{1\over2}\delta_{\alpha\beta}\right)\left(\left|C_{\nu d,\alpha\beta bs}^{\text{SLL}}+C_{\nu d,\alpha\beta bs}^{\text{SLR}}\right|^2\right.\nonumber \\
    &\left.+\left|C_{\nu d,\alpha\beta sb}^{\text{SLL}}+C_{\nu d,\alpha\beta sb}^{\text{SLR}}\right|^2\right)\,,\\
    \label{BKStarE-condition}
    2\left(\left|C^P_{sb\chi 1}\right|^2+\left|C^P_{sb\chi 2}\right|^2\right)&=\sum_{\alpha\leq\beta}\left(1-\frac{1}{2}\delta_{\alpha\beta}\right)\left(\lvert C^{\text{SLR}}_{\nu d,\alpha\beta sb}-C^{\text{SLL}}_{\nu d,\alpha\beta s b}\rvert^2\right. \nonumber \\
    &\left.+\lvert C^{\text{SLR}}_{\nu d,\alpha\beta bs}-C^{\text{SLL}}_{\nu d,\alpha\beta bs}\rvert^2\right),
\end{align}
to reproduce the resulting kinematic distributions and branching ratios of $B\rightarrow K+\slashed{E}$ and $B\rightarrow K^*+\slashed{E}$, respectively.
For scalar currents in LEFT to be indistinguishable from dark fermions with the framework and observables considered in this paper, their predictions for $B\rightarrow X_s+\slashed{E}$ must also match. By imposing the latter, and using results for $B\rightarrow X_s+\slashed{E}$ presented in Appendix~\ref{appendix:Inclusive}, we obtain the following two conditions
\begin{align}
    \label{BXE-condition1}
    &\left(\left|C_{sb\chi 1}^S\right|^2+\left|C_{sb\chi 2}^S\right|^2+\left|C_{sb\chi 1}^P\right|^2+\left|C_{sb\chi 2}^P\right|^2\right)=\sum_{\alpha\leq \beta}\left(1-\frac{\delta_{\alpha\beta}}{2}\right)\nonumber \\
    &\times\left(\left|C^{\text{SLL}}_{\nu d,\alpha\beta bs}\right|^2+\left|C^{\text{SLL}}_{\nu d,\alpha\beta sb}\right|^2+\left|C^{\text{SLR}}_{\nu d,\alpha\beta bs}\right|^2+\left|C^{\text{SLR}}_{\nu d,\alpha\beta sb}\right|^2\right), \\
    \label{BXE-condition2}
    &\left(\left|C_{sb\chi 1}^S\right|^2+\left|C_{sb\chi 2}^S\right|^2\right)-\left(\left|C_{sb\chi 1}^P\right|^2+\left|C_{sb\chi 2}^P\right|^2\right)=\sum_{\alpha\leq \beta}\left(1-\frac{\delta_{\alpha\beta}}{2}\right)\nonumber \\
    &\times2\,{\text{Re}}\left(C_{\nu d,\alpha\beta bs}^{\text{SLL}}C_{\nu d,\alpha\beta bs}^{\text{SLR}*}+C_{\nu d,\alpha\beta sb}^{\text{SLL}}C_{\nu d,\alpha\beta sb}^{\text{SLR}*}\right).
\end{align}
It is easy to verify that the conditions (\ref{BXE-condition1}) and (\ref{BXE-condition2}) are automatically satisfied when (\ref{BKE-condition}) and (\ref{BKStarE-condition}) hold. As a result, we can not distinguish scalar LEFT operators from dark fermion operators using the techniques and observables considered in this paper. We proceed to illustrate how the argument is altered in simpler scenarios. 

\paragraph{2. One LEFT operator v.s. one dark LEFT operator:}
Here we investigate whether it is always possible to distinguish a scenario where one of $\mathcal{O}^{\rm{SLL}}_{\nu d}$ and $\mathcal{O}^{\rm{SLR}}_{\nu d}$ has a nonzero Wilson coefficient, from a scenario where one of $\mathcal{O}^{\text{S}}_{sb\chi 1,2}$ and $\mathcal{O}^{\text{P}}_{sb\chi 1,2}$ has a nonzero Wilson coefficient. 

We have shown above that a LEFT operator can always be distinguished from a dark LEFT operator by considering $B\rightarrow K+\slashed{E}$ and $B\rightarrow K^*+\slashed{E}$. More precisely, it is impossible to satisfy both (\ref{BKE-condition}) and (\ref{BKStarE-condition}), as the r.h.s of the two equations will be nonzero while the l.h.s. of one of them will vanish in the scenario under consideration. Hence, one scalar-current LEFT operator is distinguishable from one dark LEFT operator in the basis we are considering.

 \paragraph{3. One LEFT operator v.s. two dark LEFT operators:}

We now investigate whether it is always possible to distinguish a scenario where only one of $\mathcal{O}^{\rm{SLL}}_{\nu d}$ and $\mathcal{O}^{\rm{SLR}}_{\nu d}$ has a nonzero Wilson coefficient, from a scenario where both $\mathcal{O}^{\text{S}}_{sb\chi 1,2}$ and $\mathcal{O}^{\text{P}}_{sb\chi 1,2}$ have nonzero Wilson coefficients. 

It is easily seen that the conditions (\ref{BKE-condition})-(\ref{BXE-condition2}) are satisfied for 
\begin{align}
    \label{1-2-condition}
    \left|C_{sb\chi 1}^S\right|^2+\left|C_{sb\chi 2}^S\right|^2=\left|C_{sb\chi 1}^P\right|^2+\left|C_{sb\chi 2}^P\right|^2,
\end{align}
if either all $C_{\nu d}^{\text{SLL}}$ or all $C_{\nu d}^{\text{SLR}}$ coefficients are zero. 
In conclusion, distinguishing a scenario where only left-handed or right-handed scalar LEFT WCs are nonzero from a scenario where scalar and pseudo-scalar fermionic LEFT WCs are nonzero and satisfy (\ref{1-2-condition}) would require analysis going beyond what we have presented in this paper.

\paragraph{4. Two LEFT operators v.s. one dark LEFT operator:}

Finally, we investigate whether it is always possible to distinguish a scenario where both $\mathcal{O}^{\rm{SLL}}_{\nu d}$ and $\mathcal{O}^{\rm{SLR}}_{\nu d}$ have nonzero Wilson coefficients from a scenario where only one of $\mathcal{O}^{\text{S}}_{sb\chi 1,2}$ and $\mathcal{O}^{\text{P}}_{sb\chi 1,2}$ has a nonzero Wilson coefficient. 

It follows immediately from (\ref{BKE-condition}) and (\ref{BKStarE-condition}) that the two scenarios are distinguishable if $C_{\nu d, \alpha\beta sb(bs)}^{\text{SLL}}\neq \pm C_{\nu d, \alpha\beta sb(bs)}^{\text{SLR}}$. However, if $C_{\nu d, \alpha\beta sb(bs)}^{\text{SLL}}= \pm C_{\nu d, \alpha\beta sb(bs)}^{\text{SLR}}$, then the r.h.s. of either (\ref{BKE-condition}) or (\ref{BKStarE-condition}) vanishes, and we cannot distinguish the two scenarios without going beyond the analysis considered here.

\end{appendix}

\addcontentsline{toc}{section}{References}

\small

\bibliographystyle{JHEP}
\bibliography{references}

\end{document}